\documentclass[sigconf,nonacm]{acmart}

\AtBeginDocument{%
  \providecommand\BibTeX{{%
    \normalfont B\kern-0.5em{\scshape i\kern-0.25em b}\kern-0.8em\TeX}}}

\setcopyright{none}
\renewcommand\footnotetextcopyrightpermission[1]{}  

\usepackage{amsmath,amsthm}
\usepackage{bm}
\usepackage{breqn}

\usepackage{algorithm}
\usepackage{algorithmic}

\usepackage{booktabs}
\usepackage{array}
\usepackage{makecell}
\usepackage{multirow}
\usepackage{tablefootnote}

\usepackage{graphicx}
\usepackage{adjustbox}
\usepackage{wrapfig}
\usepackage{float}
\usepackage{caption}
\usepackage{subcaption}

\usepackage{hyperref}
\usepackage{hyperxmp}

\usepackage{xcolor}

\begin{document}\citestyle{acmauthoryear}
\title{\textcolor[HTML]{a07bcc}{DiffPhD}: A Unified \textcolor[HTML]{a07bcc}{Diff}erentiable Solver for \textcolor[HTML]{a07bcc}{P}rojective \textcolor[HTML]{a07bcc}{H}eterogeneous Materials in Elasto\textcolor[HTML]{a07bcc}{d}ynamics with Contact-Rich GPU-Acceleration}


\author{%
  Shih-Yu Lai\textsuperscript{1,2$*$},~
  Sung-Han Tien\textsuperscript{1$*$},~
  Jui-I Huang\textsuperscript{1},~
  Yen-Chen Tseng\textsuperscript{1},~
  Yi-Ting Chiu\textsuperscript{1},~
  Siyuan Luo\textsuperscript{3},~\\
  Ziqiu Zeng\textsuperscript{3},~
  Fan Shi\textsuperscript{3},~
  Peter Yichen Chen\textsuperscript{4},~
  Tiantian Liu\textsuperscript{5},~
  Yu-Lun Liu\textsuperscript{6},~
  Bing-Yu Chen\textsuperscript{1$\dagger$}%
}
\authornote{$^*$Equal contribution.\quad $^\dagger$Corresponding author.\\
  \textit{Emails:}~
  \texttt{akinesia112@gmail.com},~
  \texttt{jamie920619@gmail.com},~
  \texttt{rayhuang@cmlab.csie.ntu.edu.tw},~
  \texttt{ajean9388@gmail.com},~
  \texttt{austin030606@gmail.com},~
  \texttt{sy.luo@nus.edu.sg},~ \\
  \texttt{zzeng@nus.edu.sg},~
  \texttt{fan.shi@nus.edu.sg},~
  \texttt{pyc@csail.mit.edu},~
  \texttt{ltt1598@gmail.com},~
  \texttt{yulunliu@cmlab.csie.ntu.edu.tw},~
  \texttt{robin@ntu.edu.tw}}
\affiliation{%
  \institution{%
    \textsuperscript{1}National Taiwan University\quad
    \textsuperscript{2}MoonShine Animation Studio\quad
    \textsuperscript{3}National University of Singapore\\
    \textsuperscript{4}The University of British Columbia\quad
    \textsuperscript{5}Independent Researcher\quad
    \textsuperscript{6}National Yang Ming Chiao Tung University%
  }%
  \country{}
  \city{} 
}

\renewcommand{\shortauthors}{Lai, et al.}



\begin{abstract}
Differentiable simulation of soft bodies is a foundation for system identification, trajectory optimization, and Real2Sim transfer. Yet, existing methods such as the differentiable Projective Dynamics (DiffPD)  struggle when faced with heterogeneous materials with extreme stiffness contrasts, hyperelasticity under large deformations, and contact-rich interactions, which are common scenarios in the real world. We present DiffPhD, a unified GPU-accelerated differentiable Projective Dynamics framework for heterogeneous materials that tackles these intertwined challenges simultaneously. Our key insight is a careful integration of: (i) stiffness-aware projective weights to embed heterogeneity into the global system; (ii) trust-region eigenvalue filtering lifted to the backward pass for stable hyperelastic gradients and a type-II Anderson Acceleration scheme with dual-gate convergence to stabilize forward iteration under large stiffness contrasts; and (iii) a unified GPU pipeline that reuses a single sparse factor across forward, backward, and contact computations, with stiffness-amplified Rayleigh damping folded into the same factor for heterogeneity-aware dissipation at zero recurring cost. DiffPhD achieves strict gradient accuracy while delivering up to an order-of-magnitude speedup over prior differentiable solvers on heterogeneous, hyperelastic, contact-rich benchmarks. Crucially, this speedup does not come at the cost of stability: DiffPhD remains convergent on stiffness contrasts up to 100$\times$ where prior PD solvers degrade. This unlocks end-to-end gradient-based optimization on regimes previously bottlenecked by either solver fragility or per-iteration cost---shell--joint composite creatures, soft characters wielding stiff weapons, and soft-gripper robotic manipulation---all handled within a single forward--backward pass.
\end{abstract}




\begin{teaserfigure}
\centering
\vspace{-1.0em}
\includegraphics[width=1.0\textwidth]
{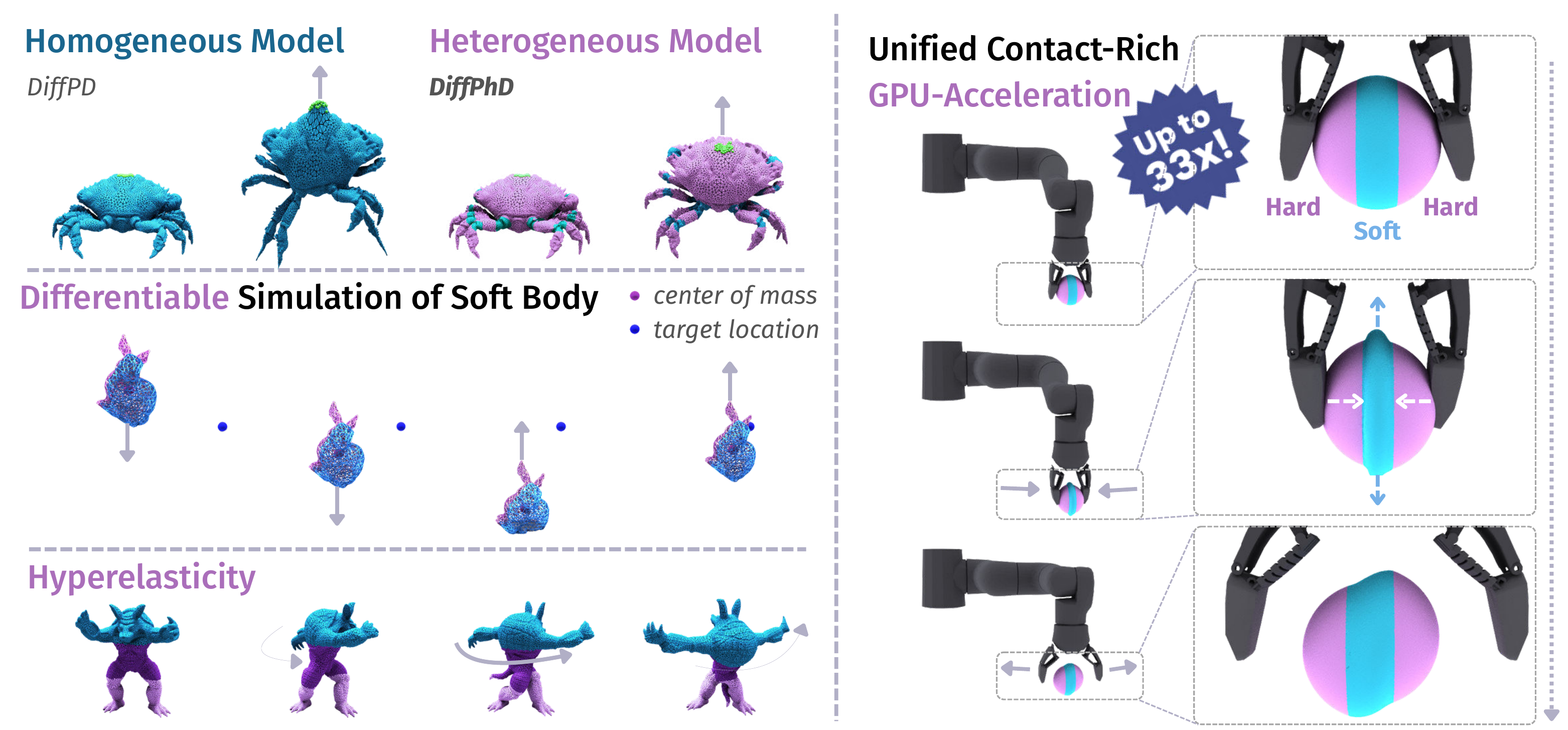}
\vspace{-3.0em}
\caption{We present \textbf{DiffPhD}, a GPU-accelerated differentiable Projective Dynamics (PD) solver for heterogeneous hyperelastic materials: \textbf{1.~Heterogeneous Materials}---stiffness-aware projective weights baked directly into the global matrix handle extreme stiffness contrasts (crab); \textbf{2.~Differentiable Simulation}---trust-region eigenvalue filtering, adaptively switching between absolute-value and clamping projections, is lifted onto the proximal-map Hessian inside the backward pass to yield accurate Neo-Hookean gradients at high Poisson's ratios and large deformations (bunny, armadillo); \textbf{3.~Unified Contact-Rich GPU Acceleration}---a single sparse inverse factor reused across forward, backward, and frictional-contact computations preserves gradient accuracy (robot manipulator). \textcolor{gray}{Gray} arrows: applied forces; \textcolor{green}{green} vertices: loaded DoFs; \textcolor{violet}{dark purple} regions are $100\times$ stiffer than \textcolor{cyan}{cyan}.}
\Description{Four panels: (top) a homogeneous cyan crab and a heterogeneous crab (purple shell, cyan legs) deforming differently under the same lift; (middle) a tetrahedralized bunny in multiple frames with purple centre-of-mass and blue target dots; (bottom-left) an armadillo in four large-deformation poses with motion arrows; (right) a 7-DoF robot arm with a soft compliant gripper, with zoomed insets showing the gripper interacting with a small blue cube.}
\label{fig:teaser}
\end{teaserfigure}
\vspace{-5.5em}

\maketitle

\section{INTRODUCTION}

Differentiable simulation of soft bodies has emerged as a foundation 
for inverse problems in computer graphics and robotics, enabling 
system identification, trajectory optimization, and sim-to-real 
transfer through end-to-end gradient 
flow~\cite{DiffPD,FBA,aquatic2022icml,nclaw2023icml,contactpoints2022iclr}. 
As applications push toward realistic scenarios---characters with 
bones embedded in soft tissue, contacting with rigid bodies, 
or robotic graspers manipulating heterogeneous objects---two 
ingredients become indispensable: (i) \emph{material heterogeneity}, 
where Young's modulus varies by orders of magnitude across a single 
mesh and the parameters are themselves often the design target; and 
(ii) \emph{hyperelasticity} with frictional contact, capturing the 
large, non-linear, volume-preserving deformations of real-world 
flesh, rubber, and tissue. Heterogeneity is not a way to make a body stiffer---it is the 
structural property that makes \emph{spatially varying} response, 
region-specific identification, and per-zone inverse design 
possible at all. Yet their combination 
is precisely where existing differentiable solvers break down.

\textit{\textbf{Why heterogeneity and hyperelasticity are hard for 
differentiable PD?}}
Projective Dynamics (PD)~\cite{PD2014} and its differentiable 
extension DiffPD~\cite{DiffPD} are attractive precisely because the 
global stiffness matrix $\bm{A}$ is constant across iterations, 
admitting a one-time factorization that amortizes every forward and 
backward solve---an advantage that depends on $\bm{A}$ being 
well-conditioned. When stiffness contrasts span $10\times$ or 
more---routine in shell--joint composites and layered biomechanical 
models~\cite{SubspaceMixedFEM,EMU2020}---the condition number of 
$\bm{A}$ explodes, iterative solves stall, and gradients become 
unreliable. Compounding this, Neo-Hookean energies are non-convex: 
the local step requires eigenvalue projection, and the choice 
between absolute-value filtering~\cite{Chen2024Stabler} and 
clamping~\cite{Teran2005Robust} is itself problem-dependent under 
high Poisson's ratios and large volume 
changes~\cite{Trust-Region}---a wrong choice destabilizes forward 
iteration, a fixed choice forfeits convergence speed across the 
state diversity a differentiable rollout traverses. Quasi-Newton variants~\cite{DiffQN} inherit the same pathology, and 
contact-rich rollouts force the backward to repeatedly apply 
$\bm{A}^{-1}$ against contact Jacobians---turning the factorization 
win into a per-iteration bottleneck. Existing tools 
each cover only a slice: FBA~\cite{FBA} delivers a nested-dissection 
sparse inverse $\bm{A}^{-1}\!=\!\bm{S}^{T}\bm{S}$ with unified 
Signorini--Coulomb NCP contact but is forward-only; the IPC-based 
differentiable solver~\cite{Huang2024DiffIPC} delivers end-to-end 
gradients at an order-of-magnitude wall-clock cost above PD; DiffPD 
itself retains gradient flow only in the homogeneous, corotated 
regime, falling back to a non-PD direct solver under Neo-Hookean.

\textit{\textbf{Why is integration the real difficulty?}}
The components we draw on---trust-region filtering~\cite{Trust-Region}, Type-II Anderson Acceleration~\cite{PengAA2018}, FBA's sparse factor~\cite{FBA}---were each designed under assumptions the others violate: trust-region filtering targets the forward \emph{element elastic} Hessian and must be re-targeted onto the \emph{prox-map} Hessian inverted by the backward IFT, with an adaptive rule avoiding per-step Hessian re-assembly and confined to the backward pass alone, since any forward prox-map modification corrupts the AA fixed-point; Anderson Acceleration assumes an autonomous fixed-point map and uses a wide history window, both of which heterogeneous curvature breaks across regions, demanding bounded history and a convergence test immune to AA-induced plateaus; and FBA's forward-only factor must serve both passes under a single sparsity pattern, accommodating contact Jacobians and a per-element prox-map differential absent from the forward NCP without re-factorization that would destroy PD's amortization. Resolving these tensions without sacrificing convergence speed, gradient accuracy, or factor reuse is the core issue.

\textit{Our approach.}
We present \textbf{DiffPhD}, a unified GPU-accelerated differentiable 
solver for heterogeneous, hyperelastic, contact-rich elastodynamics 
within a single gradient-consistent pipeline 
(Fig.~\ref{fig:teaser}). Our design rests on one observation: PD's 
matrix $\bm{A}$ is the shared backbone of every expensive 
operation---forward global step, Delassus contact compliance, 
backward adjoint---so we organize the solver around making $\bm{A}$ 
cheap to invert and robust under contrast.
(i) We encode heterogeneity \emph{structurally} at assembly through 
stiffness-aware projective weights, so contrast lives inside 
$\bm{A}$ rather than in any per-element local solve.
(ii) Trust-region filtering is lifted onto the prox-map Hessian in 
the backward pass with a state-adaptive rule whose quadratic model 
reuses $\bm{A}$ at zero extra cost; the forward fixed-point is 
stabilized by bounded-history Type-II AA and a \emph{dual-gate} 
convergence rule.
(iii) FBA's nested-dissection pipeline~\cite{FBA} is lifted into a 
single GPU-resident loop sharing one persistent factor across 
forward, backward, and contact stages.

Our work makes three core contributions:
\begin{itemize}
    \vspace{-0.5em}
    \item \textbf{Heterogeneity via Stiffness-Aware Projective 
    Assembly.} Stable, gradient-consistent simulation across 
    orders-of-magnitude stiffness 
    contrasts---surpassing FEM, Newton, and 
    Mixed-FEM~\cite{SubspaceMixedFEM} baselines in forward speed 
    and backward conditioning.
    \item \textbf{Differentiable Hyperelasticity via Proximal-Map Trust-Region Filtering.} The first lift of trust-region filtering onto PD's prox-map Hessian in the \emph{backward} pass, with closed-form differential and a state-adaptive rule reusing the forward PD matrix as its quadratic model---complemented by bounded-history Type-II Anderson Acceleration and a \emph{dual-gate} convergence criterion---yielding stable Neo-Hookean gradients beyond regimes prior differentiable PD covers.
    \item \textbf{Unified Contact-Rich GPU Forward$/$Backward 
    Loop.} A single persistent sparse-inverse factor shared across 
    forward, backward, and contact stages---an order-of-magnitude 
    wall-clock improvement over prior end-to-end-differentiable 
    contact solvers on contact-rich heterogeneous Neo-Hookean 
    rollouts.
    \vspace{-0.5em}
\end{itemize}

We validate DiffPhD on heterogeneous, hyperelastic, contact-rich 
benchmarks---spanning system identification, trajectory and 
material-design optimization, and Real2Sim 
manipulation---where prior differentiable solvers cannot stably 
handle, demonstrating strict gradient accuracy and substantial 
speedups over state-of-the-art baselines.
\section{RELATED WORK}
\label{sec:RelatedWorks}
\subsection{Differentiable Simulation of Deformable Solids}
Differentiable physics splits along the time-integration scheme 
(explicit vs.\ implicit) and the spatial discretization (particle, 
FEM, or learned). Explicit schemes pair MPM with autodiff 
(ChainQueen~\cite{Hu2019ChainQueen}, DiffTaichi~\cite{Hu2020DiffTaichi}) 
but require tiny timesteps and aggressive checkpointing for long 
rollouts; meshless alternatives include projective 
peridynamics~\cite{Lu2024PPM} and Simplicits~\cite{simplicits2024siggraph}. 
Implicit solvers obtain gradients via the adjoint method on a 
linearised dynamics---used for Real2Sim viscoelastic 
identification~\cite{Hahn2019Real2Sim}, multi-body frictional contact 
(ADD~\cite{Geilinger2020ADD}), and differentiable 
cloth~\cite{Liang2019DiffClothInverse,Qiao2020Scalable,Li2022diffcloth}---where 
the linear solve dominates per-step cost. Neural 
surrogates~\cite{Li2019Learning,SanchezGonzalez2020,nclaw2023icml} 
bypass the solver, trading accuracy for differentiability by 
construction.

DiffPD~\cite{DiffPD} occupies a distinct middle ground: implicit-FEM 
accuracy with the prefactorised linear backbone of Projective 
Dynamics~\cite{PD2014,Liu2017QuasiNewton} reused in the backward 
pass, extended to aquatic locomotion with neural 
hydrodynamics~\cite{aquatic2022icml} and contact-point discovery for 
soft-body manipulation~\cite{contactpoints2022iclr}. 
DiffQN~\cite{DiffQN} recasts the same elastodynamics in a 
quasi-Newton framework with low-rank Hessian updates, cheaper per 
step but with no remedy for Neo-Hookean indefiniteness. Adjacent 
soft-robotics efforts span gradient-free 
VoxCAD~\cite{Hiller2014VoxCAD}, modal SOFA~\cite{Allard2007SOFA}, 
learning-in-the-loop co-design~\cite{Spielberg2019Learning}, 
biomimetic swimmers~\cite{Min2019SoftCon}, fabricated foam 
quadrupeds~\cite{Bern2019Trajectory}, reduced-order soft-character 
control~\cite{Thieffry2018Control,Barbic2005Real}, dexterous 
volumetric manipulation~\cite{Lee2018Dexterous}, evolutionary 
design~\cite{Cheney2013Unshackling,Bongard2016Material,Corucci2016Evolving}, 
and differentiable IPC~\cite{Huang2024DiffIPC}.

A complementary line compresses the deformation space rather than 
the solver: subspace integration~\cite{Barbic2005Real}, modal 
actuation~\cite{Benchekroun2024ActuatorsALaMode}, and adaptive 
subspaces for contact-rich 
elastodynamics~\cite{Sharp2024TradingSpaces} with stochastic 
force-dual bases~\cite{Sharp2025ForceDualModes}. Also, Subspace 
Mixed-FEM~\cite{SubspaceMixedFEM} is closest in spirit to our 
heterogeneity contribution but is forward-only; gradient flow under 
simultaneous heterogeneity, Neo-Hookean stiffening, and frictional 
contact has not been established by any of the above, which our 
pipeline addresses end-to-end.

\vspace{-0.5em}
\subsection{Hyperelastic Energies and Hessian Stabilization}
Hyperelastic 
models~\cite{Ogden1997Nonlinear,Mooney1940,Rivlin1947,Kumar2016} 
achieve visual realism through non-convex potentials whose 
element-wise Hessians are routinely indefinite under large strain. 
Stable Neo-Hookean~\cite{Smith2018StableNH} addresses rest-state 
stability and inversion robustness, with adaptations to 
dynamic-production variants~\cite{Kim2022DynamicDeformables}, 
Cauchy--Green ARAP~\cite{Lin2022Isotropic}, edge-based 
SVK~\cite{Kikuuwe2009SVK}, and fibre-aligned anisotropic 
muscles~\cite{EMU2020}.

Projected-Newton solvers restore convergence by replacing the 
indefinite Hessian with a PSD surrogate. Per-element eigenvalue 
clamping~\cite{Teran2005Robust} sets non-positive eigenvalues to a 
floor; absolute-value 
filtering~\cite{Chen2024Stabler,Gill1981Practical,Nocedal2006Numerical} 
flips their sign instead, dramatically more efficient on stiff 
Neo-Hookean materials but prone to over-damping near the optimum. 
Diagonal additions~\cite{Fu2016Computing} and Tikhonov 
schemes~\cite{Paternain2019Newton} share this family but introduce 
problem-specific thresholds; 
projection-on-demand~\cite{Longva2023Pitfalls} sidesteps eigenanalysis 
by inflating the mass term until Cholesky succeeds. Beyond 
Newton-type schemes, second-order stencil 
descent~\cite{Chen2023StencilDescent} and preconditioned 
NCG~\cite{Shen2024PNCG} exploit locality and matrix-free structure 
for real-time interior-point hyperelasticity. Our adaptive 
eigenvalue filter follows trust-region and regularized-Newton 
theory~\cite{Trust-Region,Sorensen1982Newton,More1983Computing,More1993Generalizations,PongWolkowicz2014,Steck2023Regularization,Ueda2014Regularized}, 
conceptually related to saddle-free 
Newton~\cite{Dauphin2014Identifying}. 
ADMM~\cite{Narain2016ADMM,Overby2017ADMM} and 
position-based~\cite{Macklin2016XPBD,Mueller2007PBD} reformulations 
offer complementary optimization views but do not address gradient 
flow via the per-element proximal map---the property our hyperelasticity contributes.

\vspace{-0.5em}
\subsection{Linear Solvers, Preconditioning, and Frictional Contact}
For large-scale elastodynamics, the global linear solve dominates 
wall-clock cost. Within PD, acceleration strategies include 
Chebyshev semi-iteration~\cite{Wang2015Chebyshev}, GPU 
descent~\cite{Wang2016Descent}, parallel Gauss--Seidel with 
randomized graph coloring~\cite{Fratarcangeli2016Vivace}, 
reduced-subspace formulations~\cite{Brandt2018Hyper}, and a 
domain-decomposed CPU cloth solver~\cite{Lu2025DDPDCloth}. For the 
SPD system, multigrid---geometric~\cite{Zhu2010Efficient,DickGeorgiiWestermann2011,Wang2018Parallel,Xian2019Scalable} 
and algebraic~\cite{Naumov2015AmgX,Tamstorf2015Smoothed}---is the 
standard tool, joined by multilevel preconditioners for 
Laplacians~\cite{Krishnan2013Efficient} and ill-conditioned matrices 
via multiscale Cholesky~\cite{Chen2021Multiscale}. Direct 
factorizations stay competitive when topology is 
fixed~\cite{HerholzSorkine2020}, with adaptive algebraic 
reuse~\cite{Cheshmi2024PARTH} for dynamic sparsity; both scale 
poorly with problem size and re-meshing.

Domain decomposition offers an orthogonal axis. In Additive Schwarz ~\cite{Schwarz1870,Cai1996Overlapping,DryjaWidlund1989,Frommer1999Weighted} 
and its multilevel 
extensions~\cite{Dryja1996Multilevel,DryjaWidlund1991,Zhang1992Multilevel,Dolean2015Introduction} 
add coarse-space corrections for inter-domain coupling; restricted 
variants~\cite{Cai1999Restricted}, balancing 
decomposition~\cite{Mandel1993Balancing}, 
FETI~\cite{Farhat1991Method,Farhat2001FETIDP}, and Schur-complement 
preconditioners~\cite{Haase1991Approximate,Li2017LowRank} occupy 
related niches. In graphics, decomposition has driven reduced-space 
character 
simulation~\cite{Barbic2011RealTime,Kim2011PhysicsBased,Wu2015Unified,Yang2013Boundary}, 
decomposed implicit elastodynamics~\cite{Li2019Decomposed}, 
discontinuous-Galerkin coarse grids~\cite{Edwards2015Discretely}, and 
high-resolution cloth wrinkles~\cite{Wang2021GPU}. The GPU MAS 
preconditioner~\cite{MAS} shows that small non-overlapping subdomains 
with Nicolaides-style coarse 
spaces~\cite{Nicolaides1987Deflation} outperform multigrid for cloth 
and deformables; spectral coarse 
spaces~\cite{Spillane2013Automatic,Willems2013Spectral} could improve 
heterogeneity handling but are hard to assemble on-the-fly.

Frictional contact divides into two camps. Penalty 
formulations~\cite{Bridson2002Robust,Macklin2020PrimalDual,Wu2020Safe} 
introduce a fictitious repulsive energy with per-scene stiffness 
tuning. Complementarity formulations enforce non-penetration and 
Coulomb friction directly: IPC~\cite{Li2020IPC} and its 
non-distance-barrier GPU extension~\cite{Lan2024EGC} provide 
intersection- and inversion-free dynamics through a smoothed barrier, 
forming the basis for the differentiable 
extension~\cite{Huang2024DiffIPC} and the unified geometric-contact 
potential~\cite{Huang2025GeometricContact}; 
\cite{Ly2020Dry} integrate Signorini--Coulomb into PD's local-global 
loop for cloth, and \cite{FBA} extend this to GPU-resident 
hyperelastic solids via a unified NCP. In-timestep 
remeshing~\cite{Ferguson2023InTimestepRemeshing} reformulates 
contact-rich elastodynamics around adaptive resolution rather than 
solver design; dynamics-aware coarsening~\cite{Chen2017Dynamics} and 
parallel spatial-hashing collision detection~\cite{Tang2018PSCC} 
address adjacent concerns. Yet none co-design contact with material 
heterogeneity and end-to-end differentiability---the gap our unified 
pipeline (Sec.~\ref{sec:method}) closes.

\section{BACKGROUND}

\begin{table}[h]
  \centering\small
  \caption{Principal notation.}
  \label{tab:notation}
  \setlength{\tabcolsep}{4pt}
  \begin{tabular}{ll}
    \toprule
    Symbol & Meaning \\
    \midrule
    $\bm{M}\in\mathbb{R}^{dn_v\times dn_v}$ & lumped mass matrix \\
    $h$ & time-step size \\
    $\bm{A}\in\mathbb{R}^{dn_v\times dn_v}$ & PD global stiffness matrix (SPD, het+damp) \\
    $\bm{G}_e\in\mathbb{R}^{d^2\times dn_v}$ & element deformation-gradient operator \\
    $w_e$, $V_e$ & per-element PD surrogate weight ($w_e\propto\mu_e$), rest volume \\
    $\bm{p}_e\in\mathbb{R}^{d^2}$ & PD local auxiliary (projection) variable \\
    $\mathcal{M}_e$ & constraint manifold of element $e$ \\
    $\mu_e,\lambda_e$ & per-element Lam\'e parameters \\
    $\bar{\mu},\bar{\lambda},\bar{k}$ & mesh-wide scalar means used in local prox \\
    $\mu_{\mathrm{ref}}\!=\!\max_e\mu_e$ & damping normalisation reference \\
    $\alpha$ & scalar Rayleigh mass-damping coefficient \\
    $\beta_e$ & per-element stiffness-damping coefficient ($\propto\mu_e$) \\
    $\bm{F}_e\in\mathbb{R}^{d\times d}$ & deformation gradient of element $e$ \\
    $J_e\!=\!\det\bm{F}_e$ & volume-change ratio \\
    $\bm{\sigma}_F,\bm{\sigma}^{*}$ & singular values of $\bm{F}_e$, converged prox stretches \\
    $\bm{H}_\psi(\bm{\sigma})$ & NH stretch-space Hessian, Eq.~\eqref{eq:nh_hessian_stretch} \\
    $n_e^v$ & nodes per element ($n_e^v=4$ for tetrahedra, $d=3$) \\
    $\bm{H}_e$ & raw element elastic Hessian, $\nabla^{2}_{\bm{q}}(V_{e}\psi_{e})$ \\
    $\bm{H}_e^{\mathrm{prox}}$ & PD prox-map Hessian, $\bm{H}_\psi+\bar{k}\bm{I}$ \\
    $\tau\in[0,1]$ & backward TR blend parameter \\
    $\rho$ & TR ratio, Eq.~\eqref{eq:tr_ratio} \\
    $\varepsilon_{\mathrm{TR}}$ & TR tolerance ($=0.1$) \\
    $\Delta\bm{q}^{*}$ & last converged fixed-point increment, Eq.~\eqref{eq:tr_step} \\
    $m$ & AA window size (default $m\!=\!1$ on het, $m\!=\!5$ on homo) \\
    $\Delta\bm{G},\Delta\bm{Q}$ & AA residual / iterate difference matrices \\
    $\bm{\gamma}^{*}$ & AA mixing coefficients \\
    $\bm{P}$ & METIS nested-dissection permutation \\
    $\bm{S}$ & FBA sparse inverse factor: $\bm{A}^{-1}\!=\!\bm{S}^T\bm{S}$ \\
    $\bm{J}_n,\bm{J}_b,\bm{J}_f$ & contact / bilateral / friction Jacobians \\
    $\bm{\lambda}_n,\bm{\lambda}_b,\bm{\lambda}_f$ & corresponding multipliers \\
    $\bm{W}_{ab}\!=\!\bm{J}_a\bm{A}^{-1}\bm{J}_b^T$ & Delassus compliance sub-block \\
    $\bm{\Omega}_n,\bm{\Omega}_f,\bm{E}_n,\bm{E}_f$ & NCP weight / regulariser matrices \\
    $\delta_{n,c},\delta_{f,c},\mu_j$ & normal gap, tangential slip, Coulomb coefficient \\
    $\Phi(\bm{q})$ & primal objective, Eq.~\eqref{eq:primal} \\
    \bottomrule
  \end{tabular}
\end{table}

\subsection{Problem Formulation}
\label{sec:setup}

A deformable solid with $n_v$ vertices, $n_e$ finite elements, and
spatial dimension $d\!\in\!\{2,3\}$ has generalised position
$\bm{q}\!\in\!\mathbb{R}^{dn_v}$ and velocity $\bm{v}\!=\!\dot{\bm{q}}$.
Each element carries an independent Young's modulus $E_e$ and a shared
Poisson ratio $\nu$:
\begin{equation}
  \mu_e \;=\; \frac{E_e}{2(1+\nu)},
  \qquad
  \lambda_e \;=\; \frac{E_e\,\nu}{(1+\nu)(1-2\nu)}.
  \label{eq:lame}
\end{equation}
The solid interacts with $K_n$ unilateral contacts, $K_b$ bilateral
attachments, and $K_f$ frictional pairs. Implicit Euler gives
\begin{equation}
  \frac{\bm{M}}{h^{2}}\!\left(\bm{q}_{t+h}-\tilde{\bm{q}}\right)
  \;=\;\bm{f}_{\mathrm{ela}}(\bm{q}_{t+h})
  -\bigl(\alpha\bm{M}+\bm{B}_{\bm{\beta}}\bigr)\bm{v}_{t+h}
  +\bm{J}_n^{T}\bm{\lambda}_n+\bm{J}_b^{T}\bm{\lambda}_b+\bm{J}_f^{T}\bm{\lambda}_f,
  \label{eq:dynamics}
\end{equation}
with $\bm{B}_{\bm{\beta}}\!=\!\sum_e\beta_e V_e\bm{G}_e^{T}\bm{G}_e$ the
heterogeneity-weighted stiffness-damping operator
(Sec.~\ref{sec:rayleigh}) and inertial free-fall target
\begin{equation}
  \tilde{\bm{q}}\;=\;\bm{q}_t+h\bm{v}_t
    +h^{2}\bm{M}^{-1}\bigl(\bm{f}_{\mathrm{ext}}+\bm{f}_{\mathrm{state}}(\bm{q}_t,\bm{v}_t)\bigr).
  \label{eq:freefall}
\end{equation}
The contact-free case is the stationarity condition of the primal
objective \cite{Martin2011ExampleBased,PD2014}
\begin{equation}
  \Phi(\bm{q})\;=\;\frac{1}{2h^{2}}\lVert\bm{q}-\tilde{\bm{q}}\rVert_{\bm{M}}^{2}+\Psi(\bm{q}),
  \qquad
  \Psi(\bm{q})\!=\!\sum_eV_e\psi_e(\bm{F}_e(\bm{q})),
  \label{eq:primal}
\end{equation}
analysed for its forward/backward implications in
Sec.~\ref{sec:variational}.

\subsection{Heterogeneous Hyperelastic Energy}
\label{sec:neohookean}

We adopt a compressible Neo-Hookean energy density at element $e$,
\begin{equation}
  \psi_e(\bm{F}_e)
  =\frac{\mu_e}{2}\!\left(\lVert\bm{F}_e\rVert_F^{2}-d\right)
  -\mu_e\ln J_e
  +\frac{\lambda_e}{2}(\ln J_e)^{2},
  \label{eq:neohookean}
\end{equation}
chosen because (i) the $\ln J_e$ barrier resists inversion
($J_e\!\to\!0^{+}\!\Rightarrow\!\psi_e\!\to\!+\infty$); (ii) it
reduces to linear elasticity in the small-strain limit, preserving
the shear/bulk interpretation of $(\mu_e,\lambda_e)$; and (iii) it is
$C^{2}$ on $J_e\!>\!0$, supporting closed-form derivatives required by
the forward Newton solve and the backward IFT. With elastic force
$\bm{f}_{\mathrm{ela}}(\bm{q})\!=\!-\nabla_{\bm{q}}\Psi$, the first
Piola--Kirchhoff stress is
\begin{equation}
  \bm{P}_e(\bm{F}_e)
  =\mu_e\!\left(\bm{F}_e-\bm{F}_e^{-T}\right)
  +\lambda_e\ln(J_e)\,\bm{F}_e^{-T}.
  \label{eq:pkstress}
\end{equation}

\begin{figure}[t]
  \centering
  \includegraphics[width=\columnwidth]{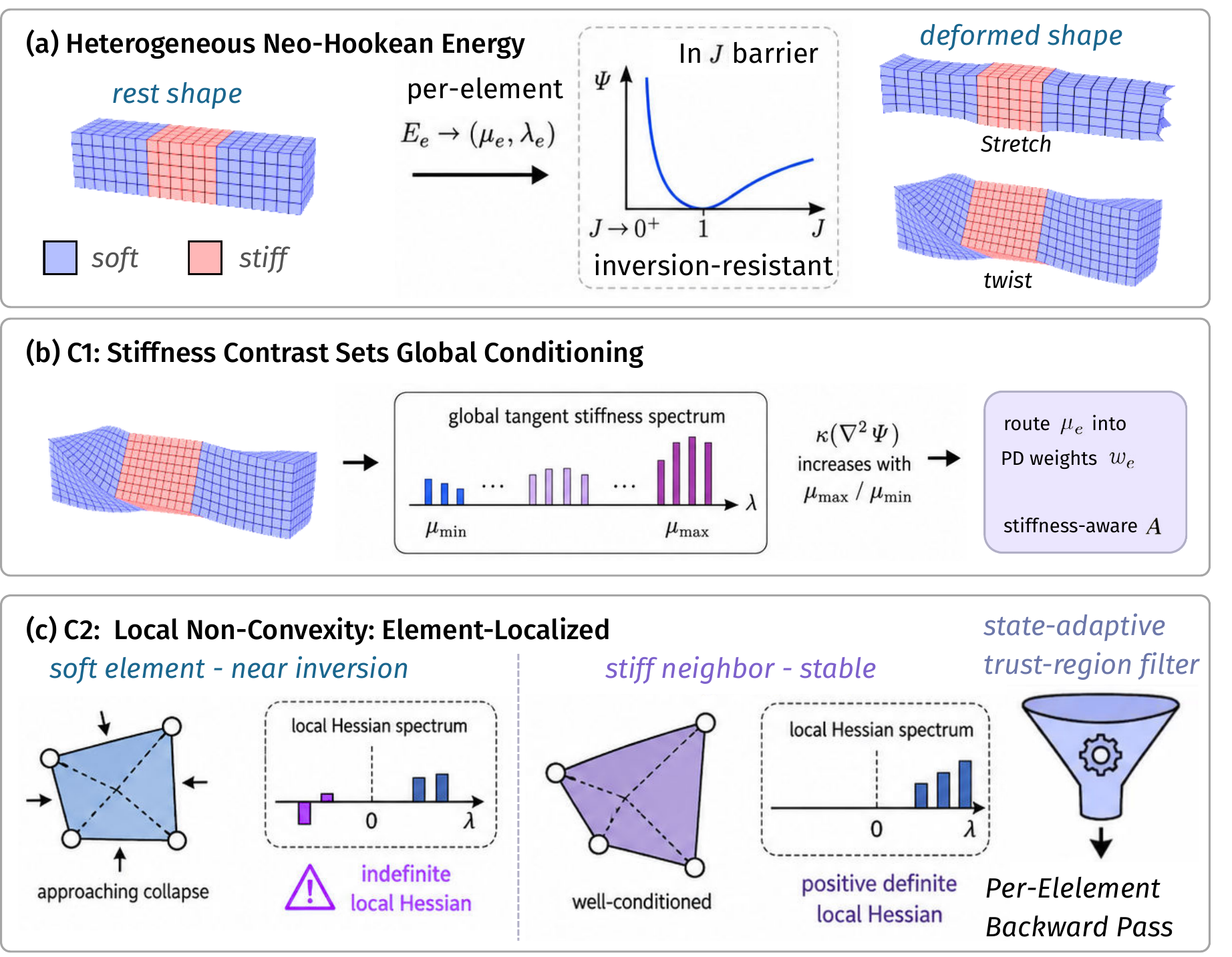}
  \caption{\textbf{Two consequences of material heterogeneity in
  Neo-Hookean PD.}
  \textit{(a) Heterogeneous Neo-Hookean energy.} A bar with stiff
  (red) and soft (blue) regions deforms qualitatively differently
  under stretch and twist, and the energy's logarithmic barrier
  prevents element inversion.
  \textit{(b) Consequence C1 (global).} The spectrum of the global
  tangent stiffness widens directly with the stiffness contrast
  $\mu_{\max}/\mu_{\min}$; absorbing this contrast into the assembled
  operator $\bm{A}$ keeps its inversion robust
  (Sec.~\ref{sec:heterogeneity}).
  \textit{(c) Consequence C2 (local).} Non-convexity is
  element-localised: a soft element near collapse develops an
  indefinite local Hessian, while its stiff neighbour stays positive
  definite. A uniform per-element filter under-serves both regimes,
  motivating our state-adaptive trust-region filter applied only in
  the backward pass (Sec.~\ref{sec:trustregion}).}
  \label{fig:heterogeneity_consequences}
\end{figure}

\paragraph{Heterogeneity sets the spectral conditioning.}
In a single-material body $(\mu_e,\lambda_e)\!\equiv\!(\bar{\mu},\bar{\lambda})$,
the tangent-stiffness spectrum is governed by mesh quality alone;
in our setting $\mu_e$ varies over orders of magnitude---composite
shells with embedded rigid plates~\cite{SubspaceMixedFEM}, layered
foams under impact, and shape-changing soft robots with stiff
actuators~\cite{EMU2020} all exhibit
$\mu_{\max}/\mu_{\min}\!\ge\!100\times$
(Fig.~\ref{fig:heterogeneity_consequences}a). Two consequences
follow:
\emph{(C1)~Tangent-stiffness conditioning} of $\nabla^{2}\Psi$ is
determined by $\mu_{\max}/\mu_{\min}$ rather than mesh resolution
(Fig.~\ref{fig:heterogeneity_consequences}b), which motivates
routing $\mu_e$ into PD's projective weights $w_e$ so that $\bm{A}$ inherits
stiffness-aware preconditioning at zero per-step cost
(Sec.~\ref{sec:heterogeneity}); and
\emph{(C2)~Local non-convexity is element-localised}
(Fig.~\ref{fig:heterogeneity_consequences}c): soft elements near
$\sigma_i\!\to\!0$ produce indefinite per-element Hessians
(Eq.~\eqref{eq:nh_hessian_stretch}) while stiff neighbours remain
convex, so uniform Hessian
modification~\cite{Teran2005Robust,Chen2024Stabler} under-serves
both regimes---motivating a state-adaptive trust-region
filter~\cite{Trust-Region} on the per-element prox-map Hessian in
the differentiable backward pass, adapting to local non-convexity
without breaking PD's global fixed-point contraction
(Sec.~\ref{sec:trustregion}).


\vspace{-0.5em}

\subsection{Variational Form and Indefinite Hessian Challenge}
\label{sec:variational}

External and state-dependent forces $\bm{f}_{\mathrm{ext}},\bm{f}_{\mathrm{state}}$
are absorbed into $\tilde{\bm{q}}$ via Eq.~\eqref{eq:freefall}, so
$\Phi$ in Eq.~\eqref{eq:primal} depends only on
$\Psi(\bm{q})=\sum_{e}V_{e}\psi_{e}(\bm{F}_{e}(\bm{q}))$, which is
generally non-convex under Neo-Hookean hyperelasticity. Newton-type
solvers therefore confront an indefinite Hessian
$\nabla^{2}\Phi=\bm{M}/h^{2}+\nabla^{2}\Psi$, whose per-element block
$\bm{H}_{e}=\nabla^{2}_{\bm{q}}(V_{e}\psi_{e})$ becomes indefinite
under high Poisson's ratio or large volume change---the canonical
non-convex regime of Eq.~\eqref{eq:neohookean}.

\paragraph{Per-element Hessian filtering for SPD restoration.}
Projected Newton replaces each $\bm{H}_{e}$ by an SPD surrogate built
from its eigendecomposition
$\bm{H}_{e}=\bm{U}_{e}\bm{\Lambda}_{e}\bm{U}_{e}^{T}$ with
$\bm{\Lambda}_{e}=\mathrm{diag}(\kappa_{1},\ldots,\kappa_{dn_{e}^{v}})$.
Two element-wise filters dominate, eigenvalue
clamping~\cite{Teran2005Robust} and absolute-value
filtering~\cite{Chen2024Stabler}:
\begin{align}
  \text{clamping:}\quad
  &\kappa_{k}^{+}=\max(\kappa_{k},0),
  \label{eq:clamp}\\
  \text{absolute-value:}\quad
  &\kappa_{k}^{+}=\lvert\kappa_{k}\rvert.
  \label{eq:abs}
\end{align}
The reassembled
$\bm{H}_{e}^{+}=\bm{U}_{e}\,\mathrm{diag}(\kappa_{k}^{+})\,\bm{U}_{e}^{T}$
enters
\begin{equation}
  \bar{\bm{H}}(\bm{q})\;=\;\frac{\bm{M}}{h^{2}}+\sum_{e}\bm{H}_{e}^{+}.
  \label{eq:proj_newton_hess}
\end{equation}
Clamping matches the local quadratic well near convexity but stalls
under stiff non-convexity; absolute filtering preserves negative-eigen
curvature information and is orders of magnitude faster on stiff
Neo-Hookean instances, yet mildly damps convergence in benign
regimes~\cite{Chen2024Stabler,Trust-Region}. The optimal filter is
thus problem- and state-dependent. The same indefiniteness governs
the adjoint linear system in the backward pass
(Sec.~\ref{sec:backward}); we resolve the filter ambiguity through a
state-adaptive trust-region rule~\cite{Trust-Region} lifted into the
backward pass (Sec.~\ref{sec:trustregion}).

\section{METHOD}
\label{sec:method}

\subsection{Projective Dynamics: A Local-Global Surrogate}
\label{sec:pd_basic}
Projective Dynamics (PD)~\cite{PD2014,Liu2017QuasiNewton} replaces
the non-convex potential in Eq.~\eqref{eq:primal} by a separable
quadratic surrogate. For each element $e$, an auxiliary variable
$\bm{p}_{e}\!\in\!\mathbb{R}^{d^{2}}$ is constrained to a manifold
$\mathcal{M}_{e}$ encoding the rotational (corotated) or stretch
(Neo-Hookean) component of the deformation; the surrogate energy
\begin{equation}
  \tilde{g}(\bm{q},\{\bm{p}_{e}\})
  =\frac{1}{2h^{2}}\lVert\bm{q}-\tilde{\bm{q}}\rVert_{\bm{M}}^{2}
  +\sum_{e}w_{e}V_{e}\lVert\bm{G}_{e}\bm{q}-\bm{p}_{e}\rVert^{2}
  \label{eq:pd_surrogate}
\end{equation}
is minimised by alternating a \emph{local step}---per-element
projection
$\bm{p}_{e}^{*}=\arg\min_{\bm{p}\in\mathcal{M}_{e}}\lVert\bm{G}_{e}\bm{q}-\bm{p}\rVert^{2}$,
embarrassingly parallel---with a \emph{global step} that fixes
$\{\bm{p}_{e}^{*}\}$ and minimises in $\bm{q}$, yielding the SPD
linear system $\bm{A}\bm{q}=\bm{b}$ with
\begin{equation}
  \bm{A}\;=\;\frac{\bm{M}}{h^{2}}+\sum_{e}w_{e}V_{e}\bm{G}_{e}^{T}\bm{G}_{e},
  \quad
  \bm{b}\;=\;\frac{\bm{M}}{h^{2}}\tilde{\bm{q}}+\sum_{e}w_{e}V_{e}\bm{G}_{e}^{T}\bm{p}_{e}^{*}.
  \label{eq:A_basic}
\end{equation}
Because $\bm{A}$ is independent of $\bm{q}$, its factorisation is
amortised across all forward iterations of a timestep and---as
DiffPD~\cite{DiffPD} exploits and we extend---across the
differentiable backward pass
(Secs.~\ref{sec:backward}--\ref{sec:gpu}).

\paragraph{Per-element PD surrogate weight.}
The weight $w_{e}$ in Eq.~\eqref{eq:A_basic} matches the tangent
modulus of $\psi_{e}$ at the rest configuration,
\begin{equation}
  w_{e}\;=\;\bar{k}\!\left(\mu_{e},\,\lambda_{e},\,\varepsilon_{\sigma}\right),
  \label{eq:pd_weight}
\end{equation}
where $\bar{k}$ fits a linear spring to
$\partial\bm{P}_{e}/\partial\bm{F}_{e}$ over a singular-value range
$\varepsilon_{\sigma}$ around the identity (\cite{DiffPD}, Eq.~7).
Since $w_{e}\!\propto\!\mu_{e}$, stiffer elements receive
proportionally larger surrogate stiffness, so $\bm{A}$ \emph{naturally
admits} per-element material parameters at assembly time---a routing
capability we extend in Sec.~\ref{sec:heterogeneity}.

\paragraph{Local step --- corotated.}
$\mathcal{M}_{e}=\mathrm{SO}(d)$ and $\bm{p}_{e}^{*}=\bm{R}_{e}$ from
the polar decomposition $\bm{F}_{e}=\bm{R}_{e}\bm{\Sigma}_{e}$. The
prox-map Jacobian $\partial\bm{p}_{e}^{*}/\partial\bm{F}_{e}$ is the
polar-decomposition differential, intrinsically SPD on
$\mathrm{SO}(d)$.

\paragraph{Local step --- Neo-Hookean proximal operator.}
$\mathcal{M}_{e}=\mathbb{R}^{d\times d}$, and the local step is the
proximal map of $\psi_{e}$ with penalty $\bar{k}$,
\begin{equation}
  \bm{p}_{e}^{*}
  =\underset{\bm{p}\in\mathbb{R}^{d\times d}}{\arg\min}\;
   \frac{\bar{k}}{2}\lVert\bm{F}_{e}-\bm{p}\rVert_{F}^{2}+V_{e}\,\psi_{e}(\bm{p}).
  \label{eq:nh_prox}
\end{equation}
The SVD $\bm{F}_{e}=\bm{U}\mathrm{diag}(\bm{\sigma}_{F})\bm{V}^{T}$
decouples the minimisation into a $d$-dimensional system over the
principal stretches $\bm{\sigma}^{*}\!\in\!\mathbb{R}^{d}_{>0}$:
\begin{equation}
  \bigl(\bm{H}_{\psi}(\bm{\sigma}^{*})+\bar{k}\bm{I}\bigr)\bm{\sigma}^{*}
  =\bar{k}\,\bm{\sigma}_{F},
  \label{eq:nh_prox_normal}
\end{equation}
with stretch-space Hessian
\begin{equation}
  \bigl[\bm{H}_{\psi}\bigr]_{ii}
  =\bar{\mu}\!\left(1+\sigma_{i}^{-4}\right)+\bar{\lambda}\,\frac{1-\ln\!\prod_{j}\sigma_{j}}{\sigma_{i}^{2}},
  \quad
  \bigl[\bm{H}_{\psi}\bigr]_{ij}=\frac{\bar{\lambda}}{\sigma_{i}\sigma_{j}}\;\;(i\!\neq\!j).
  \label{eq:nh_hessian_stretch}
\end{equation}
Eq.~\eqref{eq:nh_prox_normal} is solved by Newton iteration with
backtracking line search and positivity enforcement; the converged
projection is
$\bm{p}_{e}^{*}=\bm{U}\,\mathrm{diag}(\bm{\sigma}^{*})\,\bm{V}^{T}$.
The prox-map Hessian $\bm{H}_{\psi}+\bar{k}\bm{I}$ is also the
implicit operator that the differentiable backward pass must invert
through the IFT; its potential indefiniteness near element inversion
motivates the trust-region treatment of Sec.~\ref{sec:trustregion}.

\paragraph{Volume-barrier surrogate.}
The default heterogeneous-NH PD assembly decomposes the projection
as $\bm{p}_{e}^{*}\!=\!(\bm{R}_{e},\,\bm{p}_{e}^{\mathrm{vol},*})$
with a corotated rotation step and a quadratic $(J_{e}\!-\!1)^{2}$
volume step. For workloads where inversion ($J_{e}\!\to\!0^{+}$) is
a concern, we additionally provide a logarithmic volume barrier
\begin{equation}
  \phi^{\mathrm{log}}_{e}(\bm{\sigma})
  \;=\;-\mu_{e}\!\sum_{i}\!\ln\sigma_{i}
  \;+\;\tfrac{\lambda_{e}}{2}\!\left(\!\sum_{j}\ln\sigma_{j}\!\right)^{2},
  \label{eq:pd_log_volume}
\end{equation}
solved at the principal stretches of $\bm{F}_{e}$ via the same
Newton-on-stretches step (Eq.~\eqref{eq:nh_prox_normal} with
$\bm{H}_{\psi}$ replaced by the analytical Hessian of
$\phi^{\mathrm{log}}_{e}$) and registered with weight
$w_{e}\!=\!\lambda_{e}$. The barrier diverges as $J\!\to\!0^{+}$,
preventing inversion; dropping the deviatoric $I_{1}$ term (present
in the full Neo-Hookean $\psi_{e}$) lets it compose cleanly with the
corotated rotation step without double-counting. This variant is
opt-in via a material flag and does not alter the default assembly.

\subsection{Heterogeneity via Stiffness-Aware Projective Assembly}
\label{sec:heterogeneity}
\begin{figure}[t]
  \centering
  \includegraphics[width=\columnwidth]{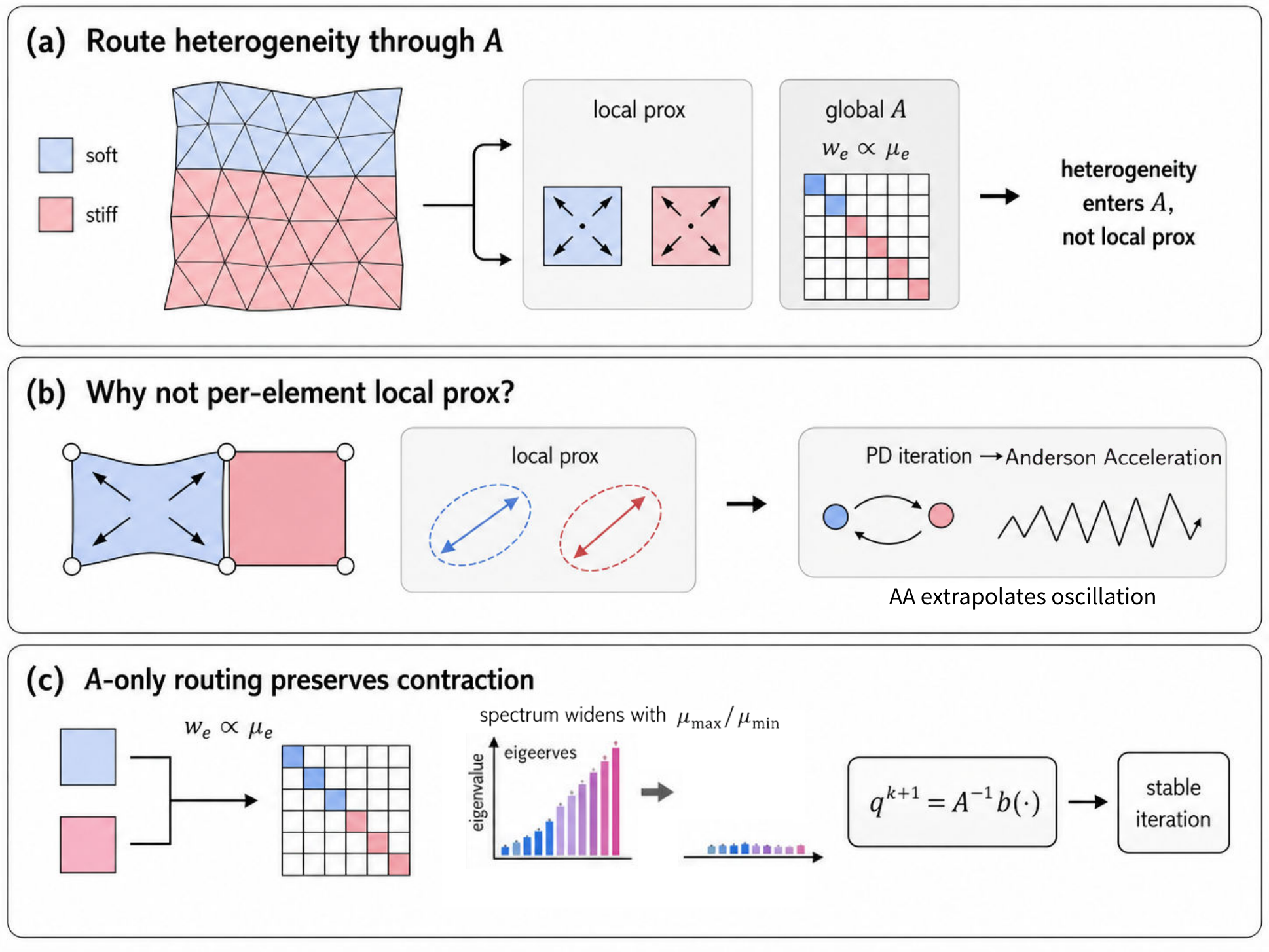}
  \caption{\textbf{Stiffness-aware projective assembly.}
  \textit{(a) Route heterogeneity through $\bm{A}$, not the local
  prox.} Per-element material parameters enter the global PD
  operator $\bm{A}$ through projective weights $w_e\!\propto\!\mu_e$
  (Eq.~\eqref{eq:pd_weight}); the local prox sees only mesh-wide
  scalar means.
  \textit{(b) Why per-element local prox fails.} Mixing soft and
  stiff elements in the local Newton makes its prox-map Hessian
  element-varying: $\bm{p}_e^{*}$ oscillates across PD iterations,
  and Anderson Acceleration amplifies the drift into mesh-level
  divergence.
  \textit{(c) $\bm{A}$-only routing preserves contraction.} Stiff
  weights enter $\bm{A}$ and widen its spectrum with contrast, but
  every element still runs the \emph{same} Newton-on-stretches; the
  PD fixed-point therefore stays contractive, leaving only a
  nonlinear residual for iteration to resolve.}
  \label{fig:heterogeneity_routing}
\end{figure}

\paragraph{Where heterogeneity enters: in $\bm{A}$ only, not in the local prox.}
For a heterogeneous mesh ($\mu_{e}$ varying over orders of magnitude),
two routings are conceivable: place per-element $(\mu_{e},\lambda_{e})$
inside Eq.~\eqref{eq:nh_prox_normal} (replacing
$\bar{\mu},\bar{\lambda},\bar{k}$ with $\mu_{e},\lambda_{e},\bar{k}_e$),
or inject $\mu_{e}$ only through the projective weights $w_{e}$ in
Eq.~\eqref{eq:A_basic}. We adopt the second exclusively
(Fig.~\ref{fig:heterogeneity_routing}a):
\begin{equation}
  w_{e}\;=\;\bar{k}\!\left(\mu_{e},\lambda_{e},\varepsilon_{\sigma}\right)\;\propto\;\mu_{e},
  \label{eq:pd_weight_routed}
\end{equation}
with $\bar{k}(\cdot)$ as in Eq.~\eqref{eq:pd_weight}; the local prox
Eq.~\eqref{eq:nh_prox_normal} uses the mesh-wide means
$\bar{\mu},\bar{\lambda},\bar{k}$ identically across elements.

\paragraph{Why scalar means in the local prox.}
PD's local-global iteration is contractive iff
$\rho(\bm{I}-\bm{A}^{-1}\bm{K}_{\mathrm{pd}})<1$, with
$\bm{K}_{\mathrm{pd}}$ the local-step stiffness from the prox-map
Hessian. Per-element $(\mu_{e},\lambda_{e})$ in the local step makes
$\bm{H}_{\psi}+\bar{k}\bm{I}$ element-varying, and any inversion-prone
element (small $\sigma_{i}$) drives the local Newton into an
indefinite regime that desynchronises with $\bm{A}$'s constant
operator: $\bm{p}_{e}^{*}$ ping-pongs across PD iterations, which
Type-II Anderson Acceleration (Sec.~\ref{sec:forward_solver})
extrapolates into mesh-level oscillation
(Fig.~\ref{fig:heterogeneity_routing}b). Routing $\mu_{e}$ through
$w_{e}$ alone preserves the cross-element contraction (the local map
is the \emph{same} stretch-space Newton in every element), while
$\bm{A}$ \emph{automatically encodes} heterogeneity at assembly time
since $w_{e}\!\propto\!\mu_{e}$
(Fig.~\ref{fig:heterogeneity_routing}c).

\begin{wrapfigure}{h}{0.5\columnwidth}
  \vspace{-1em}
  \centering
  \includegraphics[width=0.45\columnwidth]{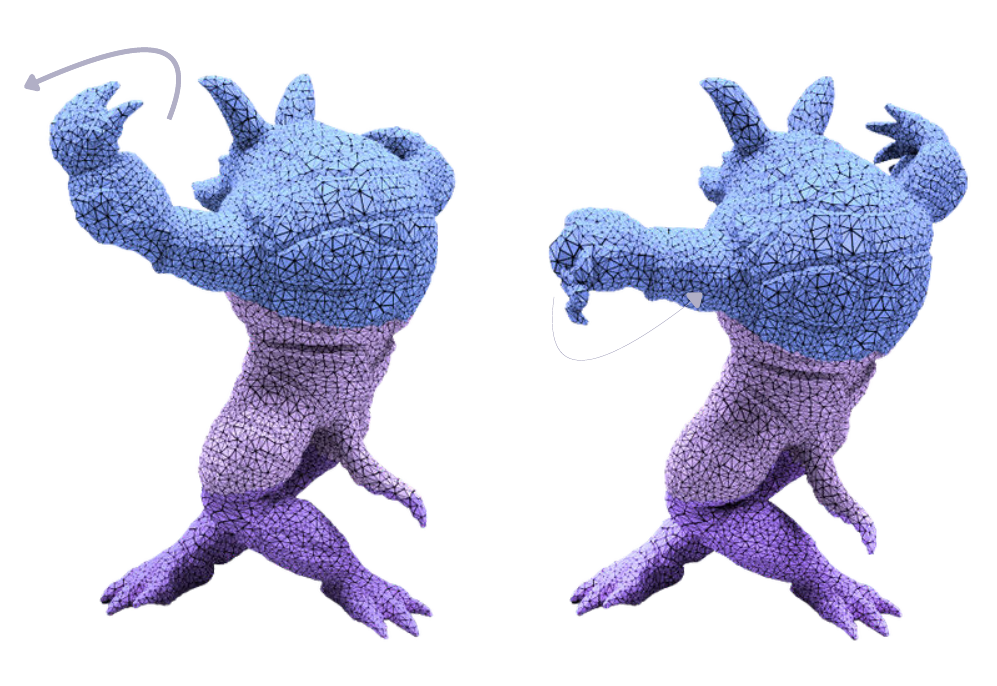}
  \caption{\textbf{Coupled failure of heterogeneity and
  hyperelasticity.} A bi-material Armadillo (soft upper body, stiff
  lower body) under twist: wrist drift originates from per-element
  prox-map Hessians turning indefinite at soft/stiff boundaries, and
  the resulting $\bm{\sigma}^{*}$ ping-pong is amplified by Anderson
  Acceleration into mesh-level oscillation. Addressed by
  bounded-window AA (Sec.~\ref{sec:forward_solver}) and backward
  trust-region filtering (Sec.~\ref{sec:trustregion}).}
  \label{fig:hete_hyper_failure}
  \vspace{-1em}
\end{wrapfigure}

\paragraph{Fixed-point contraction without the elastic force.}
The PD right-hand side $\bm{b}$ in Eq.~\eqref{eq:A_basic} intentionally
excludes $\bm{f}_{\mathrm{ela}}(\bm{q}^{k})$. If included, the
fixed-point map $T(\bm{q})\!=\!\bm{A}^{-1}(\bm{b}(\bm{q})+\bm{f}_{\mathrm{ela}}(\bm{q}))$
has Lipschitz constant 
$\lVert\bm{A}^{-1}(\nabla_{\bm{q}}\bm{b}+\nabla_{\bm{q}}\bm{f}_{\mathrm{ela}})\rVert_{2}$;
for stiff Neo-Hookean materials, the spectrum of
$\nabla_{\bm{q}}\bm{f}_{\mathrm{ela}}$ dominates $\lambda_{\min}(\bm{A})$
and drives it above one. The choice $w_{e}\!\propto\!\mu_{e}$ ensures
$\bm{A}$ absorbs the dominant linear stiffness; the nonlinear residual
is resolved by iteration. This design follows Algorithm~4
of~\cite{FBA} and is essential when stiffness varies over orders of
magnitude.


\subsection{Forward: Bounded Anderson Acceleration + Dual-Gate}
\label{sec:forward_solver}
The PD fixed-point $\bm{q}^{k+1}=T(\bm{q}^{k})$ converges linearly at
rate $\rho(\bm{I}-\bm{A}^{-1}\bm{K}_{\mathrm{pd}})$, which is close
to one for stiff Neo-Hookean materials. We accelerate it with
\emph{Type-II Anderson Acceleration}~\cite{WalkerNi2011,PengAA2018}
with bounded window $m$, paired with a \emph{dual-gate convergence
criterion} on position and projection-residual stationarity.

\paragraph{Bounded-window AA.}
With iterate step $\bm{g}^{k}\!=\!\bm{q}^{k}-\bm{q}^{k-1}$, the last
$m$ successive differences form
\begin{align}
  \Delta\bm{G} &= \bigl[\bm{g}^{1}-\bm{g}^{0},\ldots,\bm{g}^{k-1}-\bm{g}^{k-2}\bigr]\in\mathbb{R}^{dn_{v}\times m},\label{eq:aa_dG}\\
  \Delta\bm{Q} &= \bigl[\bm{q}^{1}-\bm{q}^{0},\ldots,\bm{q}^{k-1}-\bm{q}^{k-2}\bigr]\in\mathbb{R}^{dn_{v}\times m},\label{eq:aa_dQ}
\end{align}
with Tikhonov-regularised mixing coefficients
\begin{equation}
  \bm{\gamma}^{*}\;=\;\bigl(\Delta\bm{G}^{T}\Delta\bm{G}+\rho_{\mathrm{aa}}\bm{I}\bigr)^{-1}\Delta\bm{G}^{T}\bm{g}^{k},
  \quad
  \rho_{\mathrm{aa}}=\max\!\bigl(\tfrac{10^{-6}\lVert\Delta\bm{G}\rVert_{F}^{2}}{m},10^{-12}\bigr),
  \label{eq:aa_gamma}
\end{equation}
and AA-extrapolated iterate
\begin{equation}
  \bm{q}^{\mathrm{AA}}\;=\;(\bm{q}^{k-1}+\bm{g}^{k})-(\Delta\bm{Q}+\Delta\bm{G})\bm{\gamma}^{*}.
  \label{eq:aa_update}
\end{equation}
Two safeguards complete the scheme. A \emph{reliability guard}
discards the history and falls back to $\bm{q}^{k}$ whenever
$\lVert\bm{\gamma}^{*}\rVert_{2}>10$, since large mixing coefficients
indicate that AA is extrapolating beyond the trust region of the
underlying linear model. A \emph{window bound} adapts $m$ to material heterogeneity: under
stiffness contrast $\mu_{\max}/\mu_{\min}\!\ge\!100\times$, a wide
history extrapolates per-element drift into mesh-level oscillation
(Fig.~\ref{fig:hete_hyper_failure}). Specifically, per-element
$\bm{p}_{e}^{*}$ trajectories have iterate-space curvature
differing by orders of magnitude across regions, so a wide history
(e.g., $m\!=\!5$) extrapolates a mean trend biased toward the
soft-region high-curvature direction and excites cross-region oscillation. We therefore default to $m\!=\!1$ on heterogeneous meshes (``one-step
momentum''), reverting to $m\!=\!5$ on homogeneous meshes where
uniform curvature makes longer history safe.

\paragraph{Dual-gate convergence criterion.}
A single step-norm test
$\lVert\bm{q}^{k+1}-\bm{q}^{k}\rVert\!\le\!\varepsilon$ fires falsely
at $k\!=\!0$ (when the first PD step from $\bm{q}^{0}$ is small) and
during contact-state transitions where AA satisfies the step norm at
a degenerate non-fixed-point energy state. We therefore require
\emph{both} relative criteria simultaneously:
\begin{equation}
  \frac{\lVert\bm{q}^{k+1}-\bm{q}^{k}\rVert}
       {\varepsilon_{\mathrm{rel}}\lVert\bm{q}^{k}\rVert+\varepsilon_{\mathrm{abs}}}\le 1
  \quad\text{and}\quad
  \frac{\lVert\bm{b}(\bm{q}^{k+1})-\bm{b}(\bm{q}^{k})\rVert}
       {\varepsilon_{\mathrm{rel}}\lVert\bm{b}(\bm{q}^{k})\rVert+\varepsilon_{\mathrm{abs}}}\le 1,
  \quad k\ge 1.
  \label{eq:convergence_dual}
\end{equation}
The PD-residual gate detects fixed-point convergence
\emph{independently} of iterate-step magnitude, eliminating both
failure modes.

\subsection{Hyperelasticity: Backward Trust-Region on the Proximal-Map Hessian}
\label{sec:trustregion}

\begin{figure}[t]
  \centering
  \includegraphics[width=\columnwidth]{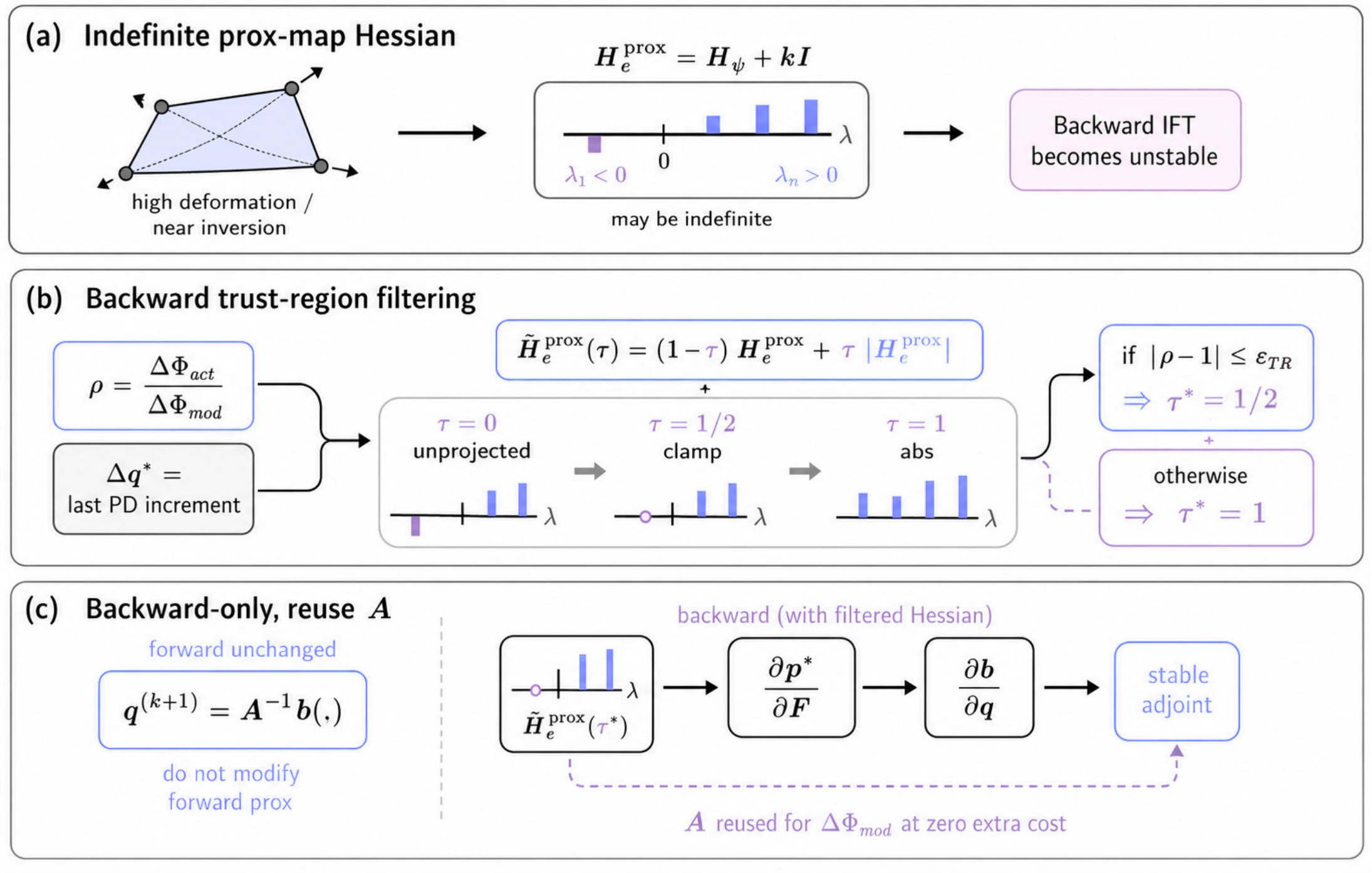}
  \caption{\textbf{Trust-region filter on the prox-map Hessian,
  applied in the backward pass only.}
  \textit{(a) The IFT operator can be indefinite.} Under high
  deformation or near inversion, the per-element prox-map Hessian
  picks up negative eigenvalues (Eq.~\eqref{eq:Hprox}), making
  vanilla inversion through the backward IFT unstable.
  \textit{(b) Trust-region blend.} A scalar $\tau\!\in\![0,1]$
  interpolates between three canonical filters
  (Eq.~\eqref{eq:tr_blend}): unprojected ($\tau\!=\!0$), eigenvalue
  clamping ($\tau\!=\!\tfrac{1}{2}$), and absolute-value filtering
  ($\tau\!=\!1$). A state-adaptive rule selects $\tau^{*}$ from the
  trust-region ratio evaluated on the last PD increment
  (Eq.~\eqref{eq:tr_rule}).
  \textit{(c) Backward-only, $\bm{A}$ reused.} The forward
  fixed-point is left untouched---any forward modification would
  perturb the AA mixing and destabilise iteration. The filtered
  Hessian feeds the adjoint chain (Eq.~\eqref{eq:db_dq}), and the
  trust-region ratio's quadratic model reuses the persistent factor
  of $\bm{A}$ at zero extra cost (Sec.~\ref{sec:gpu}).}
  \label{fig:backward_trust_region}
\end{figure}

Forward Projected Newton (Sec.~\ref{sec:variational}) must choose
between eigenvalue clamping
(Eq.~\eqref{eq:clamp},~\cite{Teran2005Robust}) and absolute-value
filtering (Eq.~\eqref{eq:abs},~\cite{Chen2024Stabler}), with no
single filter uniformly optimal across the deformation regime. We
lift this choice into the differentiable backward pass via a
state-adaptive trust-region rule
(Fig.~\ref{fig:backward_trust_region}), applied not to the raw
element elastic Hessian $\bm{H}_{e}$ of
Eq.~\eqref{eq:proj_newton_hess} but to the PD prox-map Hessian
$\bm{H}_{e}^{\mathrm{prox}}$ that drives the
implicit-function-theorem (IFT) operator of the backward pass.

The backward pass applies the IFT operator
$\bm{H}_{\psi}(\bm{\sigma}^{*})+\bar{k}\bm{I}$ from
Eq.~\eqref{eq:nh_prox_normal} when assembling
$\partial\bm{p}_{e}^{*}/\partial\bm{F}_{e}$, which drives
$\partial\bm{b}/\partial\bm{q}$ in Eq.~\eqref{eq:db_dq} and ultimately
the adjoint system Eq.~\eqref{eq:adjoint_system}. Corotated PD is
automatically SPD ($\mathrm{SO}(d)$ keeps the prox-map Jacobian
well-conditioned). For Neo-Hookean PD, the per-element prox-map
Hessian
\begin{equation}
  \bm{H}_{e}^{\mathrm{prox}}(\bm{\sigma}^{*})
  \;\equiv\;\bm{H}_{\psi}(\bm{\sigma}^{*})+\bar{k}\bm{I}\;\in\;\mathbb{R}^{d\times d}
  \label{eq:Hprox}
\end{equation}
becomes indefinite whenever $\bm{H}_{\psi}$ has eigenvalues more
negative than $-\bar{k}$---the canonical regime under high Poisson's
ratio or large volume change
(Fig.~\ref{fig:backward_trust_region}a), and the visible cause of the wrist
drift (Fig.~\ref{fig:hete_hyper_failure}). Vanilla inversion then
yields a non-descent direction for the adjoint and numerically
unreliable gradients. This motivates lifting trust-region eigenvalue
filtering, originally proposed for forward Projected
Newton~\cite{Trust-Region}, into the backward pass. The projection
is applied in the backward pass \emph{only}
(Fig.~\ref{fig:backward_trust_region}c): forward modifications of
$\bm{H}_{e}^{\mathrm{prox}}$ (PSD clamping, absolute filtering,
blended filters with eigenvalue floors, per-DoF under-relaxation)
all violate the fixed-point assumption of Type-II AA in
Eqs.~\eqref{eq:aa_dG}--\eqref{eq:aa_update}---perturbing
$\bm{p}_{e}^{*}(\bm{q})$ as a function of $\bm{q}$ corrupts
$\bm{\gamma}^{*}$ and destabilises the global iteration.

\paragraph{Trust-region blend on the prox-map Hessian.}
For each element $e$, the family of SPD prox-map surrogates is
\begin{equation}
  \tilde{\bm{H}}_{e}^{\mathrm{prox}}(\tau)
  \;=\;(1-\tau)\,\bm{H}_{e}^{\mathrm{prox}}+\tau\,\lvert\bm{H}_{e}^{\mathrm{prox}}\rvert,
  \quad\tau\in[0,1],
  \label{eq:tr_blend}
\end{equation}
where $\lvert\cdot\rvert$ is per-element absolute-value projection in
stretch space (eigendecompose, take $\lvert\kappa_{i}\rvert$,
recompose). Following \citet{Trust-Region}, the discrete sweep
$\tau\!\in\!\{0,\tfrac{1}{2},1\}$ recovers canonical filters
(Fig.~\ref{fig:backward_trust_region}b): $\tau\!=\!0$ is the
unprojected IFT, $\tau\!=\!1$ is absolute-value filtering, and
$\tau\!=\!\tfrac{1}{2}$ coincides with eigenvalue clamping
(\cite{Trust-Region}, Lemma~4.2).

\paragraph{Closed-form differential of the filtered prox-map.}
With $\tau^{*}$ fixed (rule below), Eq.~\eqref{eq:nh_prox_normal} is
replaced by
\begin{equation}
  \tilde{\bm{H}}_{e}^{\mathrm{prox}}(\tau^{*})\,\mathrm{d}\bm{\sigma}^{*}
  \;=\;\bar{k}\,\mathrm{d}\bm{\sigma}_{F},
  \label{eq:tr_ift}
\end{equation}
from which the filtered $\partial\bm{p}_{e}^{*}/\partial\bm{F}_{e}$
assembles via the SVD differential as in Sec.~\ref{sec:pd_basic}. The
downstream $\partial\bm{b}/\partial\bm{q}$ in
Eq.~\eqref{eq:db_dq} inherits SPD structure element-wise, restoring
well-posedness of Eq.~\eqref{eq:adjoint_system} in regimes where
DiffPD's pure-PD adjoint degrades.

\paragraph{State-adaptive selection on the last fixed-point increment.}
We drive $\rho$ with the \emph{last} converged PD increment, not the
whole-frame displacement:
\begin{equation}
  \Delta\bm{q}^{*}\;=\;\bm{q}^{*}-\bm{q}^{*-1}
  \label{eq:tr_step}
\end{equation}
is the difference between the last two PD iterates (restricted to
free DoFs, after AA mixing). Whole-frame $\bm{q}_{t+h}-\bm{q}_{t}$
contains inertial drift unrelated to elastic stationarity, whereas
the last PD increment is the pure Newton-direction signal the
backward IFT actually inverts---making its quadratic-model fidelity
the right adaptation criterion. With actual and PD-model decreases
\begin{equation}
  \Delta\Phi_{\mathrm{act}}=\Phi(\bm{q}^{*-1})-\Phi(\bm{q}^{*}),
  \quad
  \Delta\Phi_{\mathrm{mod}}=\tfrac{1}{2}\bigl|(\Delta\bm{q}^{*})^{T}\bm{A}\,\Delta\bm{q}^{*}\bigr|,
  \label{eq:tr_decreases}
\end{equation}
trust-region ratio
\begin{equation}
  \rho\;=\;\frac{\Delta\Phi_{\mathrm{act}}}{\Delta\Phi_{\mathrm{mod}}},
  \quad \rho\!\leftarrow\!1\ \text{if}\ \Delta\Phi_{\mathrm{mod}}<10^{-12},
  \label{eq:tr_ratio}
\end{equation}
and rule
\begin{equation}
  \tau^{*}\;=\;\begin{cases}
    \tfrac{1}{2} & \lvert\rho-1\rvert\le\varepsilon_{\mathrm{TR}}\quad\text{(quadratic model accurate)},\\
    1            & \text{otherwise}\quad\text{(model inadequate; prefer absolute filter)},
  \end{cases}
  \label{eq:tr_rule}
\end{equation}
we take $\varepsilon_{\mathrm{TR}}\!=\!0.1$ matching
\cite{Trust-Region}.

\paragraph{Reusing $\bm{A}$ as the quadratic model.}
$\Delta\Phi_{\mathrm{mod}}$ in Eq.~\eqref{eq:tr_decreases} uses PD's
constant $\bm{A}$, not a fresh per-step Newton Hessian. Since the PD
surrogate is constant in $\bm{q}$, $\bm{A}$ provides a
state-independent reference quadratic across the entire fixed-point
trajectory at \emph{zero extra cost}: the SpMV
$\bm{A}\Delta\bm{q}^{*}$ reuses the persistent factor pairs
$\{(\bm{S}^{(k)},(\bm{S}^{T})^{(k)})\}$ already populated by the
forward pass (Sec.~\ref{sec:gpu}). TR filtering is invoked once per
backward solve; the per-element $d\!\times\!d$ symmetric
eigendecomposition in Eq.~\eqref{eq:tr_blend} is a closed-form Jacobi
sweep on $3\!\times\!3$ matrices for tetrahedra, fully parallelised
across $n_{e}$.

\begin{figure}[t]
  \centering
  \includegraphics[width=\columnwidth]{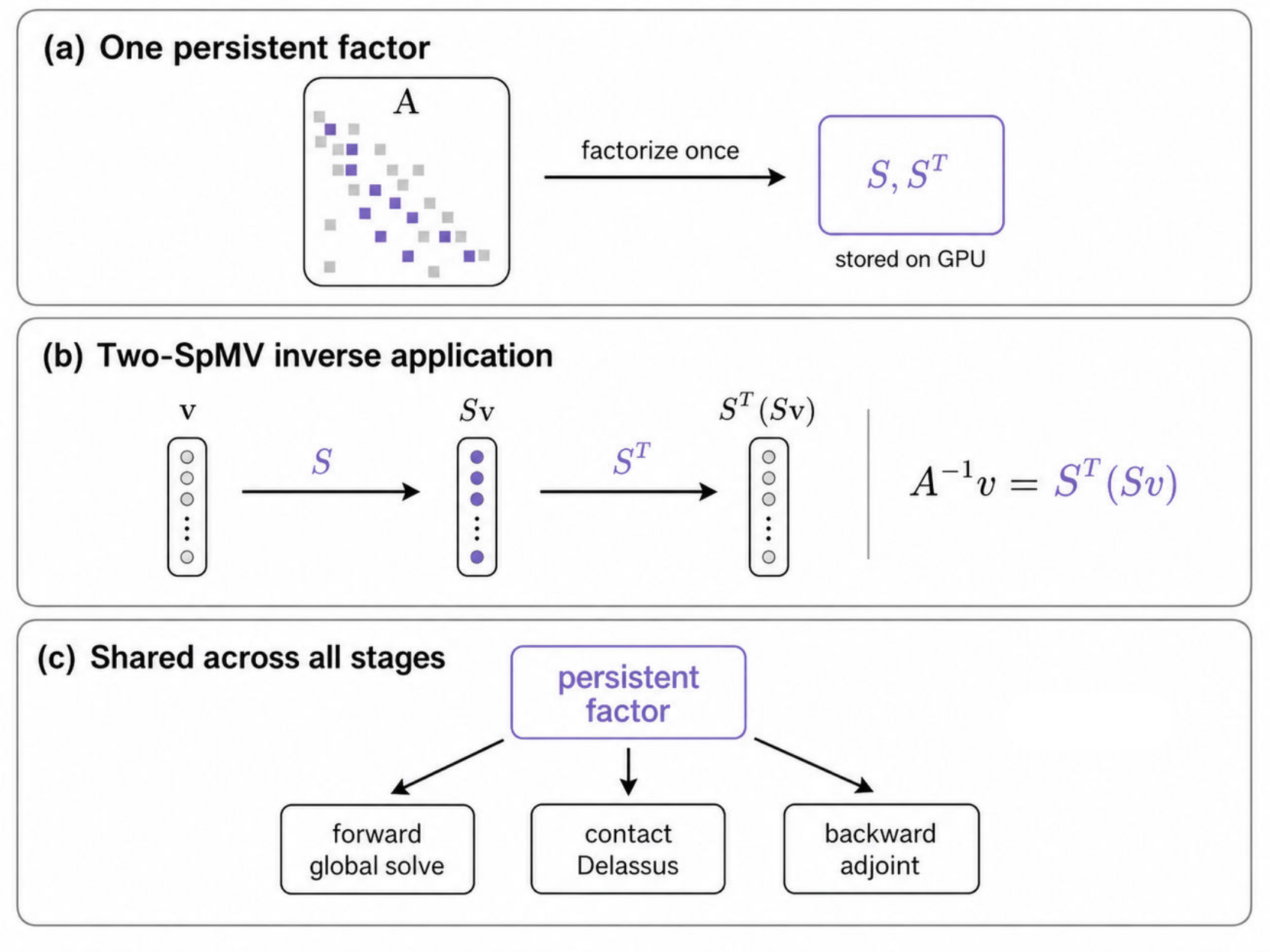}
  \caption{\textbf{One persistent factor, shared across forward,
  backward, and contact.}
  \textit{(a) One persistent factor.} The SPD operator $\bm{A}$ is
  factorised once via METIS nested dissection into
  $\bm{A}^{-1}\!=\!\bm{S}^{T}\bm{S}$ (Eq.~\eqref{eq:Ainv_STS}); the
  factor pair sits in persistent GPU buffers.
  \textit{(b) Two-SpMV inverse application.} Any $\bm{A}^{-1}\bm{v}$
  is evaluated as two sparse multiplications
  $\bm{S}^{T}(\bm{S}\bm{v})$ (Eq.~\eqref{eq:two_spmv}); the dense
  inverse is never materialised.
  \textit{(c) Shared across all stages.} The same factor pair serves
  the forward global solve, the contact Delassus operator
  (Sec.~\ref{sec:contact}), and the backward adjoint
  (Sec.~\ref{sec:backward})---refactorisation is triggered only on
  topology, material, or damping change.}
  \label{fig:gpu_pipeline}
\end{figure}

\subsection{Stiffness-Amplified Rayleigh Damping}
\label{sec:rayleigh}

High-frequency ringing in soft regions during force-peak transitions
(observed as cross-boundary oscillation when stiff and soft regions
couple at low Poisson's ratio) is dissipated by a Rayleigh-type
damping force---originally a linear combination of mass- and
stiffness-proportional terms~\citep{Rayleigh1877, Hughes2000FEM},
which we extend with per-element stiffness coefficients for
heterogeneity:
\begin{equation}
  \bm{f}_{d}\;=\;-\bigl(\alpha\bm{M}+\bm{B}_{\bm{\beta}}\bigr)\bm{v},
  \qquad
  \bm{B}_{\bm{\beta}}\;=\;\sum_{e}\beta_{e}V_{e}\bm{G}_{e}^{T}\bm{G}_{e},
  \label{eq:rayleigh}
\end{equation}
with scalar mass coefficient $\alpha$ and per-element stiffness
coefficient $\beta_{e}$. Implicit Euler absorbs $\bm{f}_{d}$ into the
linearly augmented PD operator:
\begin{equation}
  \bm{A}\;=\;\frac{(1+\alpha h)\bm{M}}{h^{2}}+\sum_{e}\bigl(w_{e}+\tfrac{\beta_{e}}{h}\bigr)V_{e}\bm{G}_{e}^{T}\bm{G}_{e}.
  \label{eq:A_damped}
\end{equation}

\paragraph{Stiffness-amplified routing: $\beta_{e}\!\propto\!\mu_{e}$.}
Standard heterogeneity-aware tuning suggests
$\beta_{e}\!\propto\!1/\mu_{e}$, so soft regions (where compliant
high-frequency modes live) receive more dissipation. \emph{We adopt
the opposite}: $\beta_{e}=\beta_{0}\,\mu_{e}/\mu_{\mathrm{ref}}$ with
$\mu_{\mathrm{ref}}\!=\!\max_{e}\mu_{e}$, so stiff regions are damped
more aggressively. The rationale is the cross-boundary transfer
pathway: in heterogeneous twist scenarios the high-frequency ringing
contaminating the soft region is \emph{seeded} by stiff-region
ringing transmitted across the material interface; damping at the
source preserves soft-region force-peak dynamics (its arms keep their
elastic snap) while suppressing the seed. Empirically,
$\beta_{e}\!\propto\!1/\mu_{e}$ over-damped soft-region deformation
under force peak, while $\beta_{e}\!\propto\!\mu_{e}$ preserved
amplitude and removed visible cross-boundary oscillation.

\paragraph{Damping folds into $\bm{A}$ at zero recurring cost.}
Both $\alpha\bm{M}$ and $\beta_{e}\bm{G}_{e}^{T}\bm{G}_{e}$ are
constant in $\bm{q}$ and time-step independent.
Eq.~\eqref{eq:A_damped} is assembled \emph{once} when $\{w_{e}\}$ are
computed from $\{\mu_{e}\}$, and the same nested-dissection
$\bm{S}\bm{S}^{T}$ factor pairs of Sec.~\ref{sec:gpu} represent its
inverse---no refactorisation, no per-step damping solve, and the
backward pass inherits the same factored representation
automatically.

\subsection{Unified GPU Pipeline: Persistent \texorpdfstring{$\bm{S}^{T}\bm{S}$}{S\textasciicircum T S} across Forward, Backward, and Contact}
\label{sec:gpu}

\paragraph{FBA nested-dissection sparse inverse.}
The SPD $\bm{A}$ from Eq.~\eqref{eq:A_damped} is reordered by METIS
nested dissection~\cite{METIS} via permutation $\bm{P}$, then
factorised~\cite{FBA} (Fig.~\ref{fig:gpu_pipeline}a)
\begin{equation}
  \bm{P}\bm{A}\bm{P}^{T}=\bm{L}\bm{D}\bm{L}^{T},
  \quad
  \bm{S}=\mathrm{diag}\!\bigl(d_{i}^{-1/2}\bigr)\bm{L}^{-1}\bm{P},
  \quad
  \bm{A}^{-1}=\bm{S}^{T}\bm{S}.
  \label{eq:Ainv_STS}
\end{equation}
The dense $\bm{S}^{T}\bm{S}$ is never materialised; it is applied via
two SpMVs. For FEM meshes with METIS ordering,
$\mathrm{nnz}(\bm{S})/n_{v}^{2}\!\approx\!2$--$3\%$, and the $j$-th
column of $\bm{S}$ is supported on $\{j\}\cup\{\text{ancestors of }j\}$
in the elimination tree, computed column-independently in parallel.

\paragraph{Two-SpMV evaluation.}
Once $\bm{S}$ and $\bm{S}^{T}$ are stored in CSR on the GPU, any
$\bm{A}^{-1}\bm{v}$ is two SpMVs
(Fig.~\ref{fig:gpu_pipeline}b),
\begin{equation}
  \bm{A}^{-1}\bm{v}\;=\;\bm{S}^{T}(\bm{S}\bm{v}).
  \label{eq:two_spmv}
\end{equation}
Since $\bm{A}$ decouples across $d$ coordinate axes, we maintain $d$
independent factor pairs $\{(\bm{S}^{(k)},(\bm{S}^{T})^{(k)})\}_{k=1}^{d}$
for per-axis SpMV without memory contention. These pairs sit in
persistent file-scope GPU buffers populated once by the forward pass
and accessed by the backward pass without re-factorisation
(Fig.~\ref{fig:gpu_pipeline}c).

\begin{wrapfigure}{h}{0.5\columnwidth}
  \vspace{-1.8em}
  \centering
  \includegraphics[width=0.45\columnwidth]{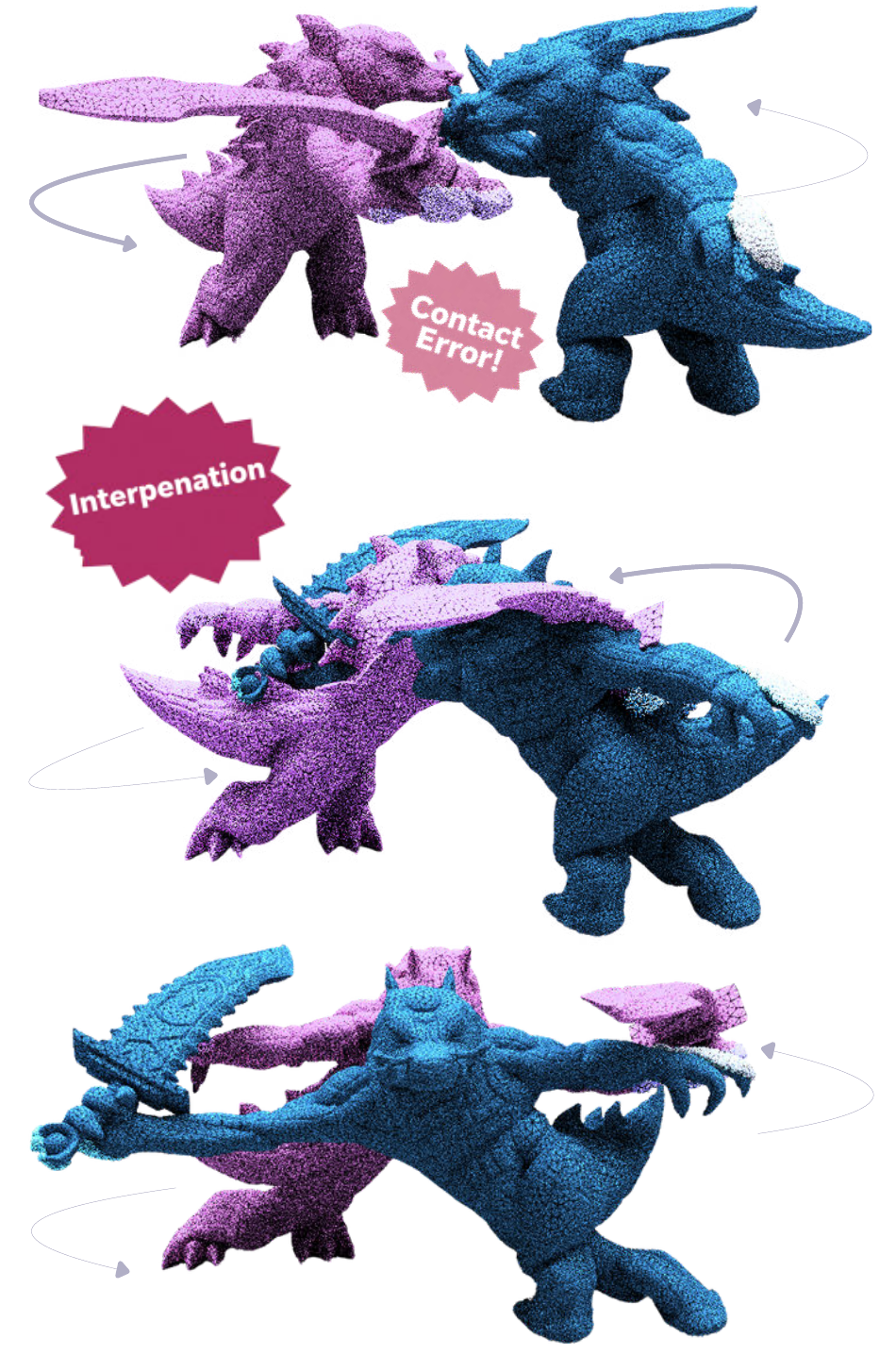}
  \caption{\textbf{Mesh-to-mesh deformable contact: a dominant
failure mode.} Penalty-based and pure-PD models leave residual error
on non-convex two-body impacts---surface separation (top) or
interpenetration (middle)---contaminating the backward gradients.
We adopt the Signorini--Coulomb non-smooth NCP of~\cite{FBA},
integrated into the persistent factor pipeline
(Sec.~\ref{sec:gpu}).}
  \label{fig:contact_motivation}
  \vspace{-1.5em}
\end{wrapfigure}

\paragraph{Why no algebraic preconditioner.}
Eq.~\eqref{eq:two_spmv} is an \emph{exact} application of
$\bm{A}^{-1}$ within floating-point precision; layering an algebraic
preconditioner (Jacobi, Schwarz, or MAS with Nicolaides aggregation)
adds per-level subdomain solves and inter-level transfers without
improving accuracy, and requires outer PCG to recover fidelity.
Empirically, MAS preconditioning was strictly slower than bare
two-SpMV on our heterogeneous Neo-Hookean benchmarks at equal forward
accuracy, because $\bm{A}$ is constant across PD iterations and
$\bm{S}\bm{S}^{T}$ already represents $\bm{A}^{-1}$ exactly. We
therefore reuse the same persistent factor for the forward global
step \emph{and} the backward adjoint solve.

\paragraph{Batch Delassus computation.}
With contact Jacobian
$\bm{J}\!=\![\bm{J}_{n}^{T},\bm{J}_{b}^{T},\bm{J}_{f}^{T}]^{T}\!\in\!\mathbb{R}^{K\times dn_{v}}$,
$K\!=\!K_{n}\!+\!K_{b}\!+\!K_{f}$, the Delassus operator
$\bm{W}\!=\!\bm{J}\bm{A}^{-1}\bm{J}^{T}$ factors as
$\bm{W}=(\bm{S}\bm{J}^{T})^{T}(\bm{S}\bm{J}^{T})$, computed via two
GPU SpMMs: $\bm{Y}\!=\!\bm{S}\bm{J}^{T}$,
$\bm{W}\!=\!\bm{Y}^{T}\bm{Y}$. The columns
$\bm{a}_{c}\!=\!\bm{A}^{-1}\bm{j}_{c}$ extracted from
$\bm{S}^{T}\bm{Y}$ are cached for reuse in the position update
(Sec.~\ref{sec:contact}).

\paragraph{Sparsity-preserving SpMM for basis-vector batches.}
The columns of $\bm{J}^{T}$ in the Delassus batch SpMM
$\bm{Y}\!=\!\bm{S}\bm{J}^{T}$ are elementary basis vectors with a
single non-zero at the contact node, so $\bm{Y}$ and the
$\bm{a}_{c}$ extracted from $\bm{S}^{T}\bm{Y}$ inherit their sparsity
from $\bm{S}$ rather than from the RHS---each column carries on the
order of $\mathrm{nnz}(\bm{S})/n_{v}$ non-zeros (2--3\,\% of $n_{v}$)
under METIS ordering. Collapsing this fill onto a single diagonal
would degenerate $\bm{W}\!=\!(\bm{S}\bm{J}^{T})^{T}(\bm{S}\bm{J}^{T})$
to near-diagonal and the downstream
$\bm{M}_{\mathrm{sys}}\!=\!\bm{\Omega}\bm{W}\bm{\Omega}^{T}\!+\!\bm{E}$
would lose the off-diagonal coupling that propagates contact
impulses, with $\kappa(\bm{M}_{\mathrm{sys}})$ rising by orders of
magnitude. We therefore require an SpMM preserving the sparsity of
$\bm{S}$ across all $K$ basis-vector RHSs rather than treating them
as dense with output thresholding: in our implementation, a CSR-aware batched SpMM with an explicit
pattern-analysis pass via cuSPARSE analyses the CSR pattern of
$\bm{S}^{(k)}$ once per refactorisation (Eq.~\eqref{eq:Ainv_STS}) and
amortises it across all $K$ columns and $d$ axes. The same persistent
$\{(\bm{S}^{(k)},(\bm{S}^{T})^{(k)})\}$ then serves forward, backward,
and contact stages with no extra data structure, matching the
column-wise SpMV evaluation Eq.~\eqref{eq:two_spmv} to machine
precision.

\paragraph{Refactorisation policy.}
$\bm{A}$ is refactorised \emph{only} on (i) topology change
(remesh), (ii) material change ($\{E_{e}\}$ updated, e.g.\ inverse
design), or (iii) damping coefficient change
(Fig.~\ref{fig:gpu_pipeline}c). Within a single forward rollout or
backward adjoint solve at fixed $\{E_{e},\alpha,\beta_{0}\}$, the
factor is computed once and reused for the entire trajectory: this
is the unified-factor amortisation DiffPD~\cite{DiffPD} applies to
forward, FBA~\cite{FBA} extends to contact, and we extend further to
the differentiable backward adjoint (Sec.~\ref{sec:backward}).

\subsection{Signorini--Coulomb Complementarity Contact}
\label{sec:contact}

We adopt the non-smooth NCP formulation of~\cite{FBA} (Fig.~\ref{fig:contact_motivation}), encoding
unilateral contact, bilateral attachment, and Coulomb friction in a
unified Fischer--Burmeister framework. For each contact $c$,
$\delta_{n,c}\!=\!\bm{n}_{c}^{T}\bm{q}_{v_{c}}-g_{c}$ is the signed
normal gap and $\bm{\delta}_{f,c}\!=\!\bm{J}_{f,c}\bm{v}$ the
tangential slip. The KKT structure
$\delta_{n,c}\!\ge\!0,\;\lambda_{n,c}\!\ge\!0,\;\delta_{n,c}\lambda_{n,c}\!=\!0$
and $\lVert\bm{\lambda}_{f,c}\rVert\!\le\!\mu_{j}\lambda_{n,c}$
is encoded~\cite{Fischer1964,FBA} as
\begin{equation}
  \varphi_{n}(\delta_{n,c},r_{n,c},\lambda_{n,c})
  =\delta_{n,c}+r_{n,c}\lambda_{n,c}-\sqrt{\delta_{n,c}^{2}+r_{n,c}^{2}\lambda_{n,c}^{2}}=0,
  \label{eq:fb_n}
\end{equation}
with $r_{n,c}\!=\!h^{2}W_{n,cc}$, and analogously
$\varphi_{f}(s_{c},r_{f,c},\xi_{c})$ for friction with
$s_{c}\!=\!\lVert\bm{\delta}_{f,c}\rVert$ and
$\xi_{c}\!=\!\mu_{j}\lambda_{n,c}-\lVert\bm{\lambda}_{f,c}\rVert$.
Differentiating Eq.~\eqref{eq:fb_n} yields the per-contact NCP
weights
\begin{equation}
  \omega_{n,c}=1-\frac{\delta_{n,c}}{\sqrt{\delta_{n,c}^{2}+r_{n,c}^{2}\lambda_{n,c}^{2}}},
  \quad
  E_{n,c}=\Bigl(1-\frac{r_{n,c}\lambda_{n,c}}{\sqrt{\delta_{n,c}^{2}+r_{n,c}^{2}\lambda_{n,c}^{2}}}\Bigr)r_{n,c},
  \label{eq:omega_E_n}
\end{equation}
and analogously $\omega_{f,c},E_{f,c}$; stacking gives diagonal
matrices $\bm{\Omega},\bm{E}$.

\paragraph{Reduced contact system.}
With block-weighted Jacobian $\bar{\bm{J}}=\bm{\Omega}\bm{J}$, the
per-iteration linearised NCP~\cite{FBA} reads
\begin{equation}
  \bm{M}_{\mathrm{sys}}\Delta\bm{\lambda}=\bm{h}_{\mathrm{vec}}-\bar{\bm{J}}\bm{A}^{-1}\bm{r}_{\mathrm{con}},
  \label{eq:contact_system}
\end{equation}
$\bm{M}_{\mathrm{sys}}=\bm{\Omega}\bm{W}\bm{\Omega}^{T}+\bm{E}$,
$\bm{r}_{\mathrm{con}}=\bm{b}+\bm{J}^{T}(\bm{\Omega}\bm{\lambda}^{k})$.
The $K\!\times\!K$ system Eq.~\eqref{eq:contact_system} is solved by
dense LDLT since $K\!\ll\!n_{v}$. Multipliers update via block
projection
$\Pi_{\mathcal{C}}\!=\!\mathrm{diag}(\Pi_{\ge 0},\bm{I},\Pi_{\mathcal{C}_{\mu}})$;
the position is recovered without an additional global solve via the
cached contact columns:
\begin{equation}
  \bm{q}^{k+1}\;=\;\bm{S}^{T}(\bm{S}\bm{b})+\sum_{c=1}^{K}\omega_{c}\lambda_{c}^{k+1}\bm{a}_{c}.
  \label{eq:q_update}
\end{equation}

\subsection{Differentiable Backward Pass}
\label{sec:backward}

Given a scalar loss $\mathcal{L}(\bm{q}_{t+h},\bm{v}_{t+h})$ and
incoming gradients $\bar{\bm{q}}_{t+h},\bar{\bm{v}}_{t+h}$, we propagate
sensitivities via the adjoint method on the converged
$(\bm{q}^{*},\bm{\lambda}^{*})$ with the active contact set fixed.
The residuals
$\bm{R}_{\bm{q}}\!=\!\bm{q}^{*}\!-\!\bm{A}^{-1}(\bm{b}(\bm{q}^{*})+\bm{J}^{T}(\bm{\Omega}\bm{\lambda}^{*}))$
and
$\bm{R}_{\bm{\lambda}}\!=\!\bm{M}_{\mathrm{sys}}\bm{\lambda}^{*}-(\bm{h}_{\mathrm{vec}}-\bar{\bm{J}}\bm{A}^{-1}\bm{r}_{\mathrm{con}})$
vanish identically; differentiating yields the block KKT adjoint system
\begin{equation}
  \begin{pmatrix}
    \bigl(\partial\bm{R}_{\bm{q}}/\partial\bm{q}\bigr)^{T} & \bigl(\partial\bm{R}_{\bm{\lambda}}/\partial\bm{q}\bigr)^{T}\\
    \bigl(\partial\bm{R}_{\bm{q}}/\partial\bm{\lambda}\bigr)^{T} & \bm{M}_{\mathrm{sys}}
  \end{pmatrix}\!
  \begin{pmatrix}\bm{y}_{\bm{q}}\\\bm{y}_{\bm{\lambda}}\end{pmatrix}
  =
  \begin{pmatrix}\bar{\bm{q}}\\\bm{0}\end{pmatrix},
  \label{eq:adjoint_system}
\end{equation}
with $\bar{\bm{q}}\!=\!\bar{\bm{q}}_{t+h}+\bar{\bm{v}}_{t+h}/h$ and
\begin{align}
  \tfrac{\partial\bm{R}_{\bm{q}}}{\partial\bm{q}}[\delta\bm{q}]&=\delta\bm{q}-\bm{A}^{-1}\!\bigl(\tfrac{\partial\bm{b}}{\partial\bm{q}}[\delta\bm{q}]\bigr),\label{eq:Rqq}\\
  \tfrac{\partial\bm{R}_{\bm{q}}}{\partial\bm{\lambda}}[\delta\bm{\lambda}]&=-\bm{A}^{-1}\bm{J}^{T}(\bm{\Omega}\,\delta\bm{\lambda}),\label{eq:Rqlambda}\\
  \tfrac{\partial\bm{R}_{\bm{\lambda}}}{\partial\bm{q}}[\delta\bm{q}]&=\bar{\bm{J}}\,\delta\bm{q}-\bar{\bm{J}}\bm{A}^{-1}\!\bigl(\tfrac{\partial\bm{b}}{\partial\bm{q}}[\delta\bm{q}]\bigr).\label{eq:Rlambdaq}
\end{align}
The local-step Jacobian through per-element projection is
\begin{equation}
  \tfrac{\partial\bm{b}}{\partial\bm{q}}[\delta\bm{q}]\;=\;\sum_{e}w_{e}V_{e}\,\bm{G}_{e}^{T}\,\tfrac{\partial\bm{p}_{e}^{*}}{\partial\bm{F}_{e}}\,\bm{G}_{e}\,\delta\bm{q},
  \label{eq:db_dq}
\end{equation}
where $\partial\bm{p}_{e}^{*}/\partial\bm{F}_{e}$ for Neo-Hookean is
assembled from Eq.~\eqref{eq:tr_ift} with $\bm{H}_{e}^{\mathrm{prox}}$
replaced by $\tilde{\bm{H}}_{e}^{\mathrm{prox}}(\tau^{*})$
(Sec.~\ref{sec:trustregion}); for corotated PD it is the polar-decomposition
differential \cite{DiffPD}, intrinsically SPD.
The $dn_{v}\!\times\!dn_{v}$ $\bm{q}$-block is solved by direct
two-SpMV via the persistent factor pairs (Eq.~\eqref{eq:two_spmv});
the $K\!\times\!K$ $\bm{\lambda}$-block by dense LDLT.
When $K\!=\!0$, the entire backward solve reduces to a single GPU
two-SpMV call.

\paragraph{Backbone adjoint and gradient routing.}
The backbone adjoint
$\bm{\mu}\!=\!\bm{A}^{-1}\bm{y}_{\bm{q}}\!=\!\bm{S}^{T}(\bm{S}\bm{y}_{\bm{q}})$
is evaluated via Eq.~\eqref{eq:two_spmv}.
Gradients flow as
\begin{align}
  \tfrac{\partial\mathcal{L}}{\partial\bm{q}_{t}}&\mathrel{+}=\tfrac{\bm{M}}{h^{2}}\bm{\mu}+\bigl(\tfrac{\partial\bm{f}_{\mathrm{state}}}{\partial\bm{q}_{t}}\bigr)^{T}\bm{M}\bm{\mu}-\tfrac{1}{h}\bar{\bm{v}}_{t+h},\label{eq:dl_dq}\\
  \tfrac{\partial\mathcal{L}}{\partial\bm{v}_{t}}&\mathrel{+}=h\bar{\bm{q}},\quad
  \tfrac{\partial\mathcal{L}}{\partial\bm{f}_{\mathrm{ext}}}\mathrel{+}=h^{2}\bm{M}\bm{\mu},\label{eq:dl_dv_dfext}\\
  \tfrac{\partial\mathcal{L}}{\partial w_{e}}&\mathrel{+}=V_{e}\bigl(\bm{\mu}^{T}\bm{G}_{e}^{T}\bm{p}_{e}^{*}-\bm{\mu}^{T}\bm{G}_{e}^{T}\bm{G}_{e}\bm{q}^{*}\bigr),\label{eq:dl_dwe}
\end{align}
and per-element Young's modulus
$\partial\mathcal{L}/\partial E_{e}$ via the chain rule on
Eqs.~\eqref{eq:lame},~\eqref{eq:pd_weight}.

\subsection{Algorithm Summary}
\label{sec:algorithm}

Algorithms~\ref{alg:diffphd_fwd} and~\ref{alg:diffphd_bwd} together
integrate one forward step (Sec.~\ref{sec:forward_solver}--\ref{sec:contact})
and the corresponding backward step (Sec.~\ref{sec:backward}).
The factor pairs $\{(\bm{S}^{(k)},(\bm{S}^{T})^{(k)})\}$ are populated
once per material/topology change and persist across both passes
through the forward cache $\mathcal{F}$.

\begin{algorithm}[t]
\caption{DiffPhD --- Forward Step}
\label{alg:diffphd_fwd}
\begin{algorithmic}[1]
\REQUIRE $\bm{q}_{t},\bm{v}_{t},\bm{f}_{\mathrm{ext}},h,\{(E_{e},V_{e},\nu)\},\alpha,\beta_{0},m,\varepsilon_{\mathrm{rel}},\varepsilon_{\mathrm{abs}}$
\ENSURE $\bm{q}_{t+h},\bm{v}_{t+h},\bm{\lambda}^{*}$;\quad forward cache $\mathcal{F}=\{\bm{q}^{*-1},\bm{q}^{*},\bm{\lambda}^{*},\{\bm{p}_{e}^{*},\bm{\sigma}^{*}_{e},\bm{U}_{e},\bm{V}_{e}\},\bm{W},\{\bm{a}_{c}\},\mathcal{A}\}$
\IF{$\bm{A}$ stale}
  \STATE \COMMENT{material/topology/damping change}
  \STATE $w_{e}\!\gets\!\bar{k}(\mu_{e},\lambda_{e})\propto\mu_{e}$;\;
         $\beta_{e}\!\gets\!\beta_{0}\,\mu_{e}/\mu_{\mathrm{ref}}$
         \hfill\eqref{eq:pd_weight},\eqref{eq:rayleigh}
  \STATE $\bm{A}\!\gets\!(1\!+\!\alpha h)\bm{M}/h^{2}+\sum_{e}(w_{e}+\beta_{e}/h)V_{e}\bm{G}_{e}^{T}\bm{G}_{e}$ \hfill\eqref{eq:A_damped}
  \STATE Factorise $\bm{P}\bm{A}\bm{P}^{T}\!=\!\bm{L}\bm{D}\bm{L}^{T}$;\;
         $\bm{S}\!\gets\!\mathrm{diag}(d_{i}^{-1/2})\bm{L}^{-1}\bm{P}$;\;
         upload $\{(\bm{S}^{(k)},(\bm{S}^{T})^{(k)})\}$ to GPU \hfill\eqref{eq:Ainv_STS}
\ENDIF
\STATE $\tilde{\bm{q}}\!\gets\!\bm{q}_{t}+h\bm{v}_{t}+h^{2}\bm{M}^{-1}(\bm{f}_{\mathrm{ext}}+\bm{f}_{\mathrm{state}})$ \hfill\eqref{eq:freefall}
\IF{$K\!>\!0$}
  \STATE $\bm{Y}\!\gets\!\bm{S}\bm{J}^{T}$;\; $\bm{W}\!\gets\!\bm{Y}^{T}\bm{Y}$;\;
         $\bm{a}_{c}\!\gets\!(\bm{S}^{T}\bm{Y})_{\cdot c}$;\; $r_{c}\!\gets\!h^{2}W_{cc}$
\ENDIF
\STATE $\bm{q}^{0}\!\gets\!\bm{q}_{t}$;\; $\bm{\lambda}^{0}\!\gets\!\bm{0}$;\; $\mathcal{H}\!\gets\!\varnothing$
\FOR{$k\!=\!0,\ldots,K_{\max}\!-\!1$}
  \STATE \textbf{Local step (parallel over $e$):} solve Eq.~\eqref{eq:nh_prox_normal} with means $\bar{\mu},\bar{\lambda},\bar{k}$ for $\bm{\sigma}^{*}_{e}$;\; $\bm{p}_{e}^{*}\!\gets\!\bm{U}_{e}\mathrm{diag}(\bm{\sigma}^{*}_{e})\bm{V}_{e}^{T}$ \hfill\eqref{eq:nh_prox}
  \STATE $\bm{b}\!\gets\!(\bm{M}/h^{2})\tilde{\bm{q}}+\sum_{e}w_{e}V_{e}\bm{G}_{e}^{T}\bm{p}_{e}^{*}$ \hfill\eqref{eq:A_basic}
  \IF{$K\!>\!0$}
    \STATE $\bm{\Omega},\bm{E}\!\gets\!\mathrm{NCP\text{-}weights}(\bm{\delta},\bm{r},\bm{\lambda}^{k})$;\;
           $\bm{r}_{\mathrm{con}}\!\gets\!\bm{b}+\bm{J}^{T}\bm{\Omega}\bm{\lambda}^{k}$;\;
           $\bm{M}_{\mathrm{sys}}\!\gets\!\bm{\Omega}\bm{W}\bm{\Omega}^{T}\!+\!\bm{E}$ \hfill\eqref{eq:contact_system}
    \STATE $\Delta\bm{\lambda}\!\gets\!\bm{M}_{\mathrm{sys}}^{-1}(\bm{h}_{\mathrm{vec}}-\bar{\bm{J}}\bm{S}^{T}(\bm{S}\bm{r}_{\mathrm{con}}))$;\;
           $\bm{\lambda}^{k+1}\!\gets\!\Pi_{\mathcal{C}}(\bm{\lambda}^{k}+\Delta\bm{\lambda})$
  \ENDIF
  \STATE $\hat{\bm{q}}\!\gets\!\bm{S}^{T}(\bm{S}\bm{b})+\sum_{c}\omega_{c}\lambda_{c}^{k+1}\bm{a}_{c}$ \hfill\eqref{eq:q_update}
  \STATE \COMMENT{Bounded-window AA (window $m$, default $m\!=\!1$ on het)}
  \STATE $\bm{g}^{k}\!\gets\!\hat{\bm{q}}-\bm{q}^{k}$;\; update $\mathcal{H}$ (keep last $m$ pairs)
  \IF{$|\mathcal{H}|\!\ge\!2$}
    \STATE $\bm{\gamma}^{*}\!\gets\!(\Delta\bm{G}^{T}\Delta\bm{G}+\rho_{\mathrm{aa}}\bm{I})^{-1}\Delta\bm{G}^{T}\bm{g}^{k}$ \hfill\eqref{eq:aa_gamma}
    \STATE $\bm{q}^{k+1}\!\gets\!\hat{\bm{q}}-(\Delta\bm{Q}+\Delta\bm{G})\bm{\gamma}^{*}$ if $\lVert\bm{\gamma}^{*}\rVert\!\le\!10$, else $\hat{\bm{q}}$, $\mathcal{H}\!\gets\!\varnothing$ \hfill\eqref{eq:aa_update}
  \ELSE
    \STATE $\bm{q}^{k+1}\!\gets\!\hat{\bm{q}}$
  \ENDIF
  \STATE Enforce Dirichlet BCs
  \IF{$k\!\ge\!1$ and dual gate \eqref{eq:convergence_dual} holds}
    \STATE \textbf{break}
  \ENDIF
\ENDFOR
\STATE $\bm{q}_{t+h}\!\gets\!\bm{q}^{k+1}$;\; $\bm{v}_{t+h}\!\gets\!(\bm{q}_{t+h}-\bm{q}_{t})/h$;\; $\bm{\lambda}^{*}\!\gets\!\bm{\lambda}^{k+1}$
\STATE Cache $\mathcal{F}\!\gets\!\{\bm{q}^{*-1},\bm{q}^{*},\bm{\lambda}^{*},\{\bm{p}_{e}^{*},\bm{\sigma}^{*}_{e},\bm{U}_{e},\bm{V}_{e}\},\bm{W},\{\bm{a}_{c}\},\mathcal{A}\}$ for backward
\end{algorithmic}
\end{algorithm}

\begin{algorithm}[t]
\caption{DiffPhD --- Backward Step (uses cache $\mathcal{F}$ and persistent $\{(\bm{S}^{(k)},(\bm{S}^{T})^{(k)})\}$ from Alg.~\ref{alg:diffphd_fwd})}
\label{alg:diffphd_bwd}
\begin{algorithmic}[1]
\REQUIRE Forward cache $\mathcal{F}$;\; persistent GPU factor pairs $\{(\bm{S}^{(k)},(\bm{S}^{T})^{(k)})\}$;\; incoming gradients $\bar{\bm{q}}_{t+h},\bar{\bm{v}}_{t+h}$;\; tolerance $\varepsilon_{\mathrm{TR}}$
\ENSURE $\partial\mathcal{L}/\partial\bm{q}_{t}$, $\partial\mathcal{L}/\partial\bm{v}_{t}$, $\partial\mathcal{L}/\partial\bm{f}_{\mathrm{ext}}$, $\partial\mathcal{L}/\partial w_{e}$, $\partial\mathcal{L}/\partial E_{e}$
\STATE $\partial\mathcal{L}/\partial\bm{q}_{t}\mathrel{+}=-\bar{\bm{v}}_{t+h}/h$;\;
       $\bar{\bm{q}}\!\gets\!\bar{\bm{q}}_{t+h}+\bar{\bm{v}}_{t+h}/h$
\IF{Neo-Hookean}
  \STATE $\Delta\bm{q}^{*}\!\gets\!\bm{q}^{*}-\bm{q}^{*-1}$;\;
         $\Delta\Phi_{\mathrm{act}}\!\gets\!\Phi(\bm{q}^{*-1})-\Phi(\bm{q}^{*})$ \hfill\eqref{eq:tr_step}
  \STATE $\Delta\Phi_{\mathrm{mod}}\!\gets\!\tfrac{1}{2}|(\Delta\bm{q}^{*})^{T}\bm{A}\Delta\bm{q}^{*}|$ \hfill\eqref{eq:tr_decreases}
  \STATE $\rho\!\gets\!\Delta\Phi_{\mathrm{act}}/\Delta\Phi_{\mathrm{mod}}$ \hfill\eqref{eq:tr_ratio}
  \STATE $\tau^{*}\!\gets\!\tfrac{1}{2}$ if $|\rho-1|\!\le\!\varepsilon_{\mathrm{TR}}$ else $1$ \hfill\eqref{eq:tr_rule}
\ENDIF
\FORALL{element $e$ in parallel}
  \IF{corotated}
    \STATE $\partial\bm{p}_{e}^{*}/\partial\bm{F}_{e}\!\gets\!$ polar-decomposition differential
  \ELSE
    \STATE $\bm{H}_{e}^{\mathrm{prox}}\!\gets\!\bm{H}_{\psi}(\bm{\sigma}^{*}_{e})+\bar{k}\bm{I}$ \hfill\eqref{eq:Hprox}
    \STATE $\tilde{\bm{H}}_{e}^{\mathrm{prox}}\!\gets\!(1-\tau^{*})\bm{H}_{e}^{\mathrm{prox}}+\tau^{*}|\bm{H}_{e}^{\mathrm{prox}}|$ \hfill\eqref{eq:tr_blend}
    \STATE $\partial\bm{p}_{e}^{*}/\partial\bm{F}_{e}\!\gets\!$ SVD diff.\ from $\tilde{\bm{H}}_{e}^{\mathrm{prox}}$ and $(\bm{U}_{e},\bm{V}_{e})$ \hfill\eqref{eq:tr_ift}
  \ENDIF
  \STATE $\bm{K}_{e}\!\gets\!w_{e}V_{e}\bm{G}_{e}^{T}(\partial\bm{p}_{e}^{*}/\partial\bm{F}_{e})\bm{G}_{e}$ \hfill\eqref{eq:db_dq}
\ENDFOR
\STATE Assemble $\partial\bm{b}/\partial\bm{q}\!\gets\!\sum_{e}\bm{K}_{e}$ via FEM scatter
\IF{$|\mathcal{A}|\!=\!0$}
  \STATE $\bm{y}_{\bm{q}}\!\gets\!$ apply $(\partial\bm{R}_{\bm{q}}/\partial\bm{q})^{-T}\bar{\bm{q}}$ via two-SpMV \hfill\eqref{eq:Rqq},\eqref{eq:two_spmv}
\ELSE
  \STATE Restore $\bm{W},\bm{\Omega},\bm{E},\bm{M}_{\mathrm{sys}},\bar{\bm{J}}$ from $\mathcal{F}$
  \STATE Solve Eq.~\eqref{eq:adjoint_system} ($\bm{q}$-block: two-SpMV;\; $\bm{\lambda}$-block: dense LDLT)
\ENDIF
\STATE $\bm{\mu}\!\gets\!\bm{S}^{T}(\bm{S}\bm{y}_{\bm{q}})$ \hfill\eqref{eq:two_spmv}
\STATE Route gradients via Eqs.~\eqref{eq:dl_dq}--\eqref{eq:dl_dwe} and chain rule on Eqs.~\eqref{eq:lame},~\eqref{eq:pd_weight}
\end{algorithmic}
\end{algorithm}

\paragraph{Computational complexity.}
Per forward iteration: local step $\mathcal{O}(n_{e})$;
global step $\mathcal{O}(\mathrm{nnz}(\bm{S}))$ via two GPU SpMVs
($\mathrm{nnz}(\bm{S})/n_{v}^{2}\!\approx\!2$--$3\%$);
contact $\mathcal{O}(K^{2})$ dense LDLT;
AA mixing $\mathcal{O}(m^{2}dn_{v})$ ($m\!\le\!5$).
Delassus pre-computation costs $\mathcal{O}(K\cdot\mathrm{nnz}(\bm{S}))$
via GPU SpMM, amortised over all iterations of a timestep.
Backward: one GPU two-SpMV reusing forward buffers for the $dn_{v}$
$\bm{q}$-block, plus per-element $d\!\times\!d$ symmetric eigendecomposition
for the trust-region projection.
The $K\!=\!0$ fast path bypasses all contact adjoint computation.
Damping coefficients $\alpha,\beta_{0}$ enter Eq.~\eqref{eq:A_damped}
at assembly only and incur no additional per-iteration cost in either
pass.

\section{RESULTS \& FINDINGS}
\label{sec:results}

We evaluate DiffPhD along the three axes that motivate the framework:
(i) \emph{forward stability and speed} on heterogeneous, contact-rich elastodynamics where prior PD solvers diverge;
(ii) \emph{gradient accuracy and end-to-end speedup} on differentiable inverse problems --system identification, initial-state optimization, and trajectory optimization; and
(iii) \emph{transfer to robotics} via heterogeneous contact-rich manipulation and a Real2Sim study.
We close with a unified ablation that disentangles the contribution of each architectural component.
Table~\ref{tab:examples} summarizes every benchmark; rows in \textbf{bold} mark heterogeneous variants where prior differentiable PD solvers either diverge or produce numerically unreliable gradients.

\subsection{Experimental Setup}

\label{sec:exp-setup}

\paragraph{Implementation.}

DiffPhD is implemented in C\texttt{++}/CUDA on top of the DiffPD~\cite{DiffPD} codebase. Sparse matrix--vector and matrix--matrix products run on the GPU through \texttt{cuSPARSE}: the per-axis application of $\bm{A}^{-1}$ is two SpMVs against the persistent $(\bm{S},\bm{S}^{T})$ factor pair (Sec.~4.5), and the batched Delassus assembly $\bm{W}\!=\!\bm{J}\bm{A}^{-1}\bm{J}^{T}$ uses \texttt{cusparseSpMM} with the \texttt{CSR\_ALG2} algorithm and an explicit preprocess pass for sparsity-pattern analysis (the default algorithm selector returns rank-deficient output on the unit-spike RHS used here; Sec.~4.5). The dense $\bm{M}_{\mathrm{sys}}$ contact block is solved by \texttt{Eigen::LDLT} on the CPU. Sparse factorizations of $\bm{A}=\bm{S}^T\bm{S}$ are computed once per topology, material, or damping-coefficient change with METIS nested dissection~\cite{karypis1997metis} and uploaded to persistent GPU buffers reused by both the forward and the adjoint pass. Eigenvalue filtering on the per-element prox-map Hessian is fused into the same element-parallel kernel that assembles $\partial \bm{p}^*_e / \partial \bm{F}_e$.

\paragraph{Hardware.}

All experiments were conducted on a desktop system equipped with a single NVIDIA RTX~4090 GPU (24\,GB VRAM), and an Intel(R) Core(TM) i7-12700 CPU (12 cores @ 4.9\,GHz) with 16\,GB of RAM. Wall-clock times are reported end-to-end and include host--device transfer.

\paragraph{Baselines.}
We compare against representative prior work spanning the key method
families our contribution touches. Newton-Cholesky provides a sparse
direct implicit-Newton accuracy reference, complemented by its
Neo-Hookean variant with stable energy~\cite{smith2018stable}.
DiffPD~\cite{DiffPD} serves as our most direct differentiable PD baseline.
To evaluate performance under different preconditioning strategies,
we also include a variant of DiffPD equipped with MAS~\cite{MAS},
which replaces the standard preconditioner with a domain decomposition based approach.
While our proposed DiffPhD adopts FBA~\cite{FBA} as its underlying forward solver to leverage
the state-of-the-art GPU PD solver with NCP frictional contact, 
we evaluate its efficiency against the aforementioned differentiable baselines.
For the differentiable baselines we use the
authors' reference implementations and a uniform L-BFGS optimizer
with identical convergence thresholds; following~\cite{DiffPD}, all
wall-clock comparisons use a forward tolerance of $10^{-4}$ and a
backward tolerance of $10^{-6}$ unless stated otherwise.

\paragraph{Notation.}

\emph{homo} denotes a single-stiffness model and \emph{hetero} a model with at least one $\geq 10\times$ stiffness contrast across sub-domains.

\emph{n/c} denotes failure to converge within $10^4$ iterations or 24 wall-clock hours; \emph{n/d} denotes a non-differentiable solver (used only as a forward reference).

\begin{table*}[t]
\centering
\small
\caption{\textbf{Benchmark scenes used throughout the evaluation.} \#DoF is the
number of degrees of freedom ($3 \times$ vertices in 3D) and \#E is the
number of finite elements; $h$ is the timestep. Bold rows indicate
heterogeneous variants where prior differentiable PD solvers fail to
converge, and which therefore form the core of our evaluation.}
\label{tab:examples}
\begin{tabular}{llrrrcccp{4.0cm}}
\toprule
Section & Scene & \#DoF & \#E & $h$ (ms) & Het. & Contact & Hyper. & Task / role in the evaluation \\
\midrule
\multirow{4}{*}{\S\ref{sec:exp-fwd-hetero}}
 & Cantilever                       &        2{,}268 &           500 & 10   & \textbf{\checkmark} & --          & --         & Heterogeneous PD energy \\
 & \textbf{Armadillo (twist; homo., het.)}       &      54{,}855 &      44{,}337 & 30   & \textbf{\checkmark} & \checkmark  & \checkmark & Stiff/soft body partitioning \\
 & Crab (homo.)                       &   2{,}235{,}655 &   172{,}587 & 30  & --                  & \checkmark  & \checkmark & Shell--joint single-material reference \\
 & \textbf{Crab (het.)}             &   2{,}235{,}655 &   172{,}587 & 30   & \textbf{\checkmark} & \checkmark  & \checkmark & Shell--joint composite \\
\midrule
\multirow{4}{*}{\S\ref{sec:exp-fwd-contact}}
& \textbf{Gatorman}     &       20{,}814 &      24{,}605 &  2   & \textbf{\checkmark} & \checkmark  & --         & Complex mesh-to-mesh contact  \\
 & Napkin $25{\times}25$ (homo.)      &        4{,}056 &           625 &  2  & --                  & \checkmark  & --         & Codimensional cloth on sphere \\
 & \textbf{Napkin $25{\times}25$ (het.)} &    4{,}056 &           625 &  2  & \textbf{\checkmark} & \checkmark  & --         & Codimensional cloth, two-stiffness \\
 & Napkin $50{\times}50$ (homo.)      &      15{,}606 &       2{,}500 &  2   & --                  & \checkmark  & --         & High-resolution cloth on sphere \\
 & \textbf{Napkin $50{\times}50$ (het.)} &  15{,}606 &       2{,}500 &  2   & \textbf{\checkmark} & \checkmark  & --         & High-resolution two-stiffness cloth \\
\midrule
\multirow{3}{*}{\S\ref{sec:exp-sysid}}
 & Bouncing Ball (homo.)             &        9{,}132 &       1{,}288 &  4   & --                  & \checkmark  & \checkmark & System ID, single material \\
 & \textbf{Bouncing Ball (het.)}    &        9{,}132 &       1{,}288 &  4   & \textbf{\checkmark} & \checkmark  & \checkmark & System ID, three sub-region materials \\
 & Plant                            &       29{,}763 &       3{,}863 & 10   & --                  & --          & --         & System ID, oscillatory dynamics \\
\midrule
\multirow{3}{*}{\S\ref{sec:exp-iso}}
 & \textbf{Bunny (het.)}            &        7{,}062 &       1{,}601 &  1   & \textbf{\checkmark} & \checkmark  & -- & Initial-state optimization \\
 & Routing Tendon                   &        2{,}475 &           512 & 10   & \textbf{\checkmark} & --          & --         & Muscle-energy backward (energy-routing) \\
\midrule
\S\ref{sec:exp-to}
 & Torus                            &        3{,}204 &           568 &  4   & --                  & \checkmark  & --         & Locomotion (rolling) \\
\midrule
\multirow{2}{*}{\S\ref{sec:exp-r2s}}
 & \textbf{Oreo}                    &      153{,}666 &     176{,}932 & 0.25 & \textbf{\checkmark} & \checkmark  & --         & Heterogeneous cookie-stack contact \\
 & \textbf{Dice}                    &        3{,}123 &       4{,}588 & 33   & \textbf{\checkmark} & \checkmark  & --         & Real2Sim die manipulation \\
\bottomrule
\end{tabular}
\end{table*}

\subsection{Heterogeneous Forward Simulation}
\label{sec:exp-fwd-hetero}

We first verify that DiffPhD resolves heterogeneous PD dynamics that destabilize prior solvers. The three scenes (\emph{Cantilever}, \emph{Armadillo}, \emph{Crab}) all carry stiffness contrasts of $10\times$--$100\times$.

\paragraph{PD energy under heterogeneity.}
The \emph{Cantilever} test partitions a beam into three segments, with the middle third set as a stiff region (red, $10\times$) between two soft ends (blue, $1\times$). The beam is initially twisted and then released to return to its equilibrium state. Under this motion, the central red section restores to its rest profile rapidly with minimal deformation, while the blue ends exhibit significant lagging and larger oscillations, matching physical intuition. Under axial loading, elongation is similarly concentrated in the blue regions while the red section resists extension (Fig.~\ref{fig:heterogeneity_consequences}(a)). To capture this, DiffPhD employs a stiffness-aware weight $w_e \propto \mu_e$ absorbed directly into $\bm{A}$ at assembly (Sec.~4.1). This ensures the energy landscape faithfully represents the material distribution and maintains a smooth, artifact-free interface.

\begin{figure*}[t]
\centering
\includegraphics[width=\linewidth]{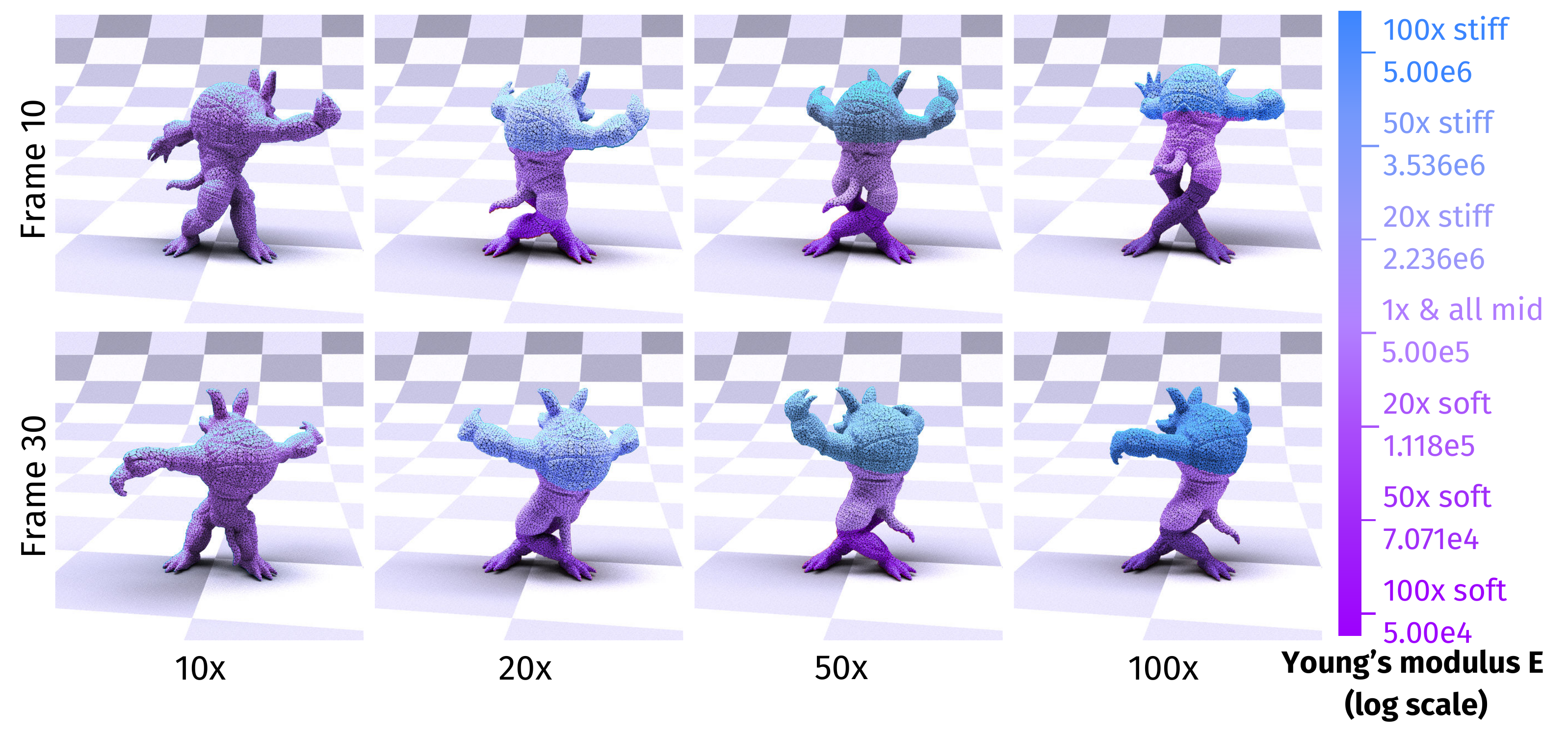}
\caption{\textbf{Armadillo under varying stiffness contrast.} The
mesh is partitioned into three height-stacked sub-regions whose
relative Young's moduli sweep from $10\times$ to $100\times$
(columns); within each column, the upper torso is stiffened
relative to the legs by the indicated ratio, with intermediate
regions interpolated on a log scale (right colour bar, ranging
from $5.00\!\times\!10^{4}$\,Pa for the softest sub-region to
$5.00\!\times\!10^{6}$\,Pa for the stiffest). Two snapshots are
shown per contrast: frame~10 (top row) captures the onset of
twist where the stiff torso resists rotation while the soft legs
deform; frame~30 (bottom row) the post-impact pose after the
release, where the residual deformation concentrates in the soft
regions. DiffPhD remains stable across the entire range with the
torso--leg interface preserved and no element inversion, while
DiffPD fails to converge beyond $\sim10\times$ as its
constant PD matrix loses spectral conditioning under high
stiffness contrast (Sec.~\ref{sec:heterogeneity}).}
\label{fig:armadillo-stiffness}
\end{figure*}

\paragraph{Neo-Hookean heterogeneity on Armadillo.}
We evaluate the performance of DiffPhD using a twisted Armadillo scene with varying stiffness ratios. The mesh is partitioned by height into three equal segments to introduce material heterogeneity. Across our tests, we vary the stiffness contrast between these regions ranging from $10\times$ to $100\times$. Through our experiments, we empirically observe that the baseline DiffPD is generally limited to a stiffness contrast of approximately $50\times$, whereas DiffPhD continues to provide stable results across the
entire tested range up to $100\times$
(Fig.~\ref{fig:armadillo-stiffness}).

\begin{figure}[t]
\centering
\includegraphics[width=\linewidth]{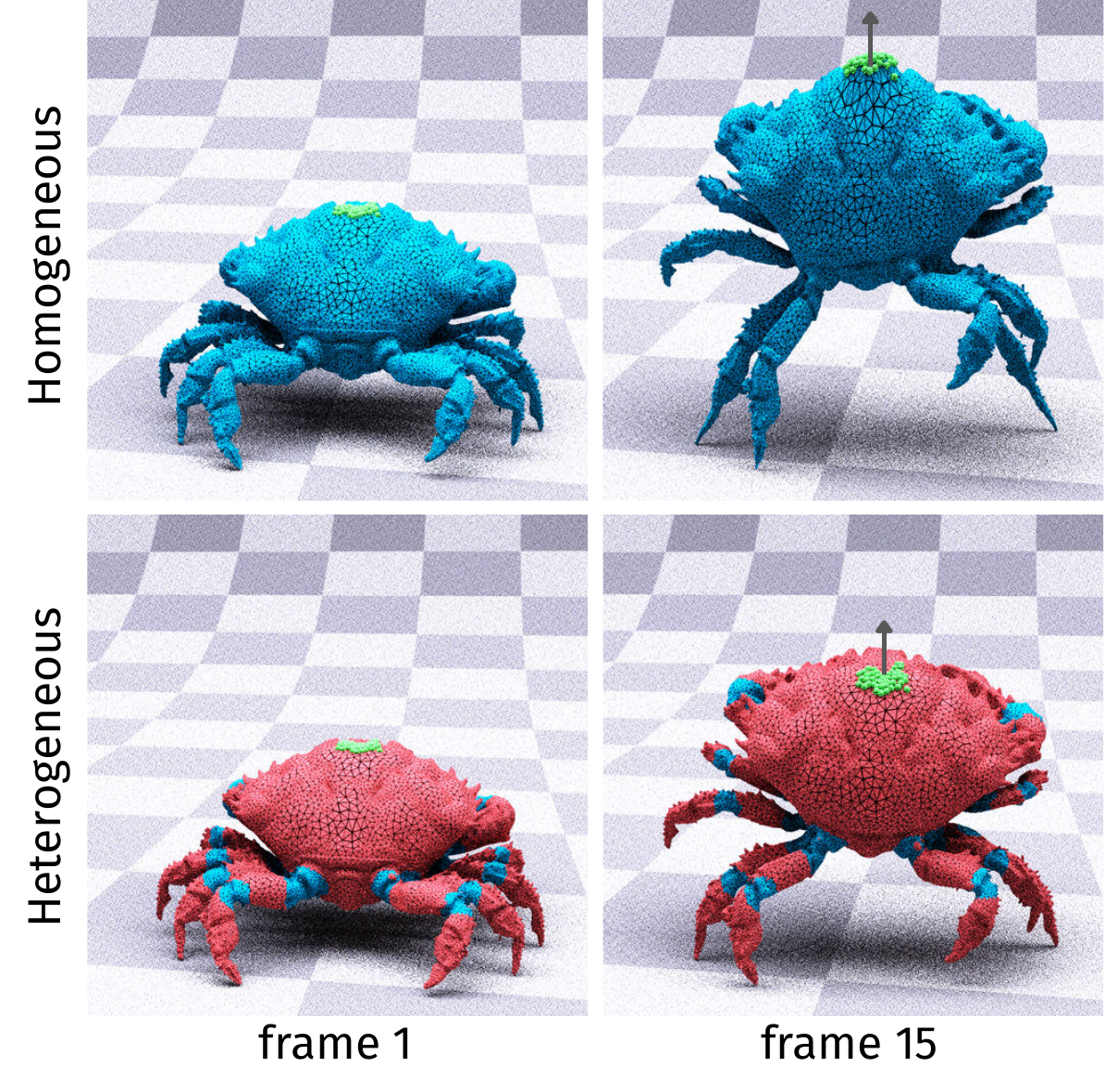}
\caption{\textbf{Crab pull-up: homogeneous vs.\ heterogeneous
response.} The crab is pulled upward at the top of its carapace
(arrow). \emph{Top row, heterogeneous:} a stiff carapace ($10\times$,
red) lifts the soft legs ($1\times$, blue) cleanly off the ground
between frame~1 and frame~15. \emph{Bottom row, homogeneous:} the
single-stiffness baseline cannot transmit the lift force without
material contrast and produces excessive global deformation.}
\label{fig:crab-pullup}
\end{figure}

\begin{figure*}[h]
\centering
\includegraphics[width=\linewidth]{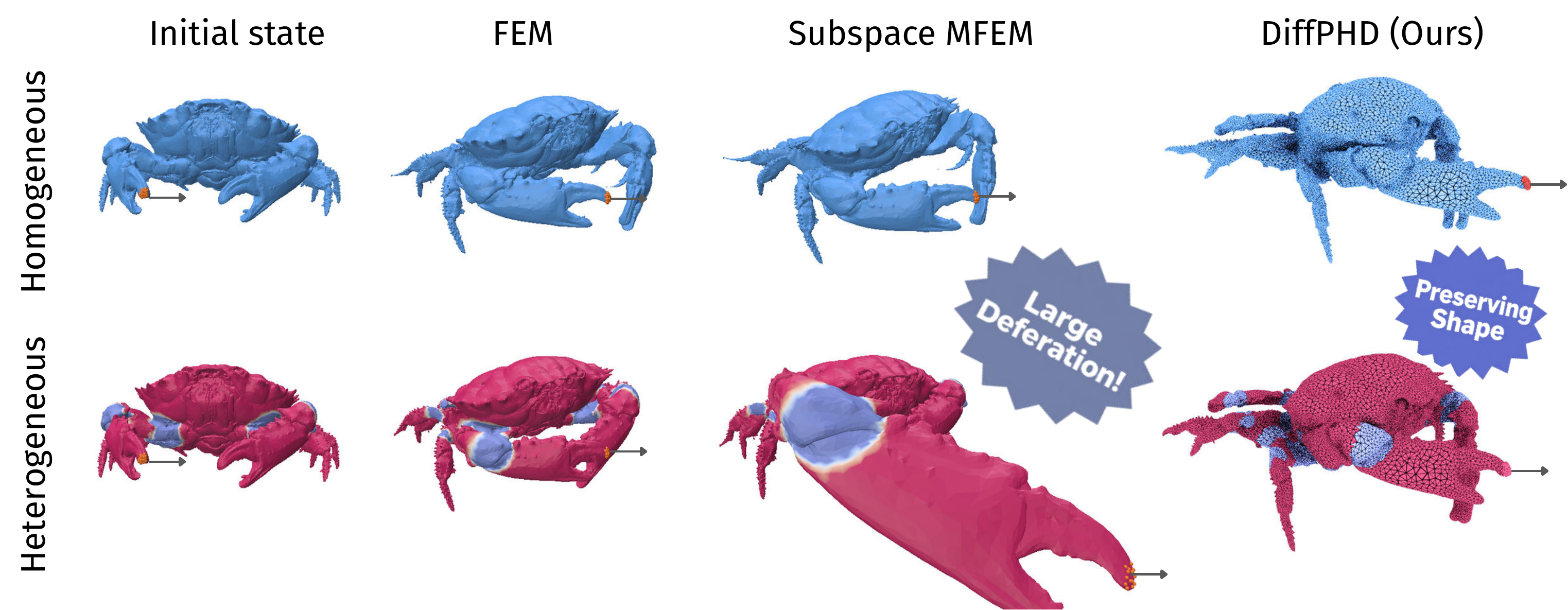}
\caption{\textbf{Crab pull-up: differentiable solver comparison.}
From left to right: initial state, FEM (Newton-NH),
Subspace MFEM~\cite{SubspaceMixedFEM}, and DiffPhD. \emph{Top
row, homogeneous} (blue): a single stiffness across the entire
body. \emph{Bottom row, heterogeneous} (red shell,
$10\times$ stiffer than the blue legs): the shell--joint
composite of~\cite{SubspaceMixedFEM}. Under upward pulling,
Subspace MFEM exhibits unbounded local stretching at the claw and
shell--joint boundary (``Large Deformation''), with the homogeneous
case collapsing more severely than the heterogeneous one. DiffPhD
preserves shape in both regimes and is the only
\emph{differentiable} solver among the four to remain stable
throughout the contact-rich trajectory.}
\label{fig:crab-comparison}
\end{figure*}

\paragraph{Crab: shell--joint composite.}
The \emph{Crab} follows the canonical scenario from~\cite{SubspaceMixedFEM}: featuring a stiff carapace ($10\times$) coupled to soft joints. We evaluate the forward computation times of DiffPhD against three baselines: Newton-Cholesky, DiffPD, and MAS, with detailed performance metrics summarized in Table~\ref{tab:crab-performance}. In our experiment, the crab is pulled upward to observe its deformation and contact response (Fig.~\ref{fig:crab-pullup}). While our method maintains efficiency across both forward and backward passes, we further compare DiffPhD with a specialized heterogeneous solver, Subspace MFEM~\cite{SubspaceMixedFEM}. We observe that Subspace MFEM fails to handle
scenarios involving contact, resulting in either a complete
simulation stall or significant mesh artifacts
(Fig.~\ref{fig:crab-comparison}), whereas DiffPhD robustly handles
these interactions while maintaining material contrast.

\begin{table}[t]
\centering
\small
\caption{\textbf{Crab benchmark: per-timestep wall-clock (s) and convergence quality.}
Each cell reports \emph{homogeneous / heterogeneous} values; the heterogeneous Crab has a $10\times$-stiff red carapace coupled to $1\times$-stiff blue joints. The bottom block reports DiffPhD's forward-time speed-up over each baseline (homo\,/\,hetero). DiffPD converges to identical iterates on both variants because it does not route per-element stiffness into $\bm{A}$ (Sec.~\ref{sec:heterogeneity}). Forward tolerance $10^{-1}$. Lower is better; best per row in bold.}
\label{tab:crab-performance}
\resizebox{\columnwidth}{!}{%
\begin{tabular}{ll cc}
\toprule
Method & Metric & Homo & Hetero \\
\midrule
\multirow{4}{*}{Newton-Cholesky}
 & forward  & $638.07$          & $832.37$       \\
 & backward & $187.78$          & $186.49$       \\
 & loss     & $22.20$           & $19.47$        \\
 & grad     & $\phantom{0}2.60$ & $\phantom{0}5.86$ \\
\midrule
\multirow{4}{*}{DiffPD~\cite{DiffPD}}
 & forward  & $366.92$           & $366.92$       \\
 & backward & $674.20$           & $674.20$       \\
 & loss     & $19.15$            & $19.15$        \\
 & grad     & $\phantom{00}1.01$ & $\phantom{00}1.01$ \\
\midrule
\multirow{4}{*}{MAS~\cite{MAS}}
 & forward  & $250.76$          & $547.04$       \\
 & backward & $\phantom{0}99.22$ & $342.46$      \\
 & loss     & $22.05$           & $19.15$        \\
 & grad     & $2063.62$         & $274.72$       \\
\midrule
\multirow{4}{*}{DiffPhD (Ours)}
 & forward  & $\mathbf{22.14}$ & $\mathbf{42.22}$ \\
 & backward & $\mathbf{48.78}$ & $336.63$         \\
 & loss     & $\mathbf{-2.04}$ & $\mathbf{18.88}$ \\
 & grad     & $612.12$         & $\mathbf{0.00}$  \\
\midrule\midrule
\multirow{3}{*}{Speed-up (forward)}
 & vs. Newton-Cholesky          & $28.83\times$ & $19.71\times$ \\
 & vs. DiffPD~\cite{DiffPD} & $16.58\times$ & $\phantom{0}8.69\times$ \\
 & vs. MAS~\cite{MAS}  & $11.33\times$ & $12.96\times$ \\
\bottomrule
\end{tabular}%
}
\end{table}

\subsection{Contact-Rich Forward Simulation}
\label{sec:exp-fwd-contact}

\begin{figure}[t]
\centering
\includegraphics[width=\linewidth]{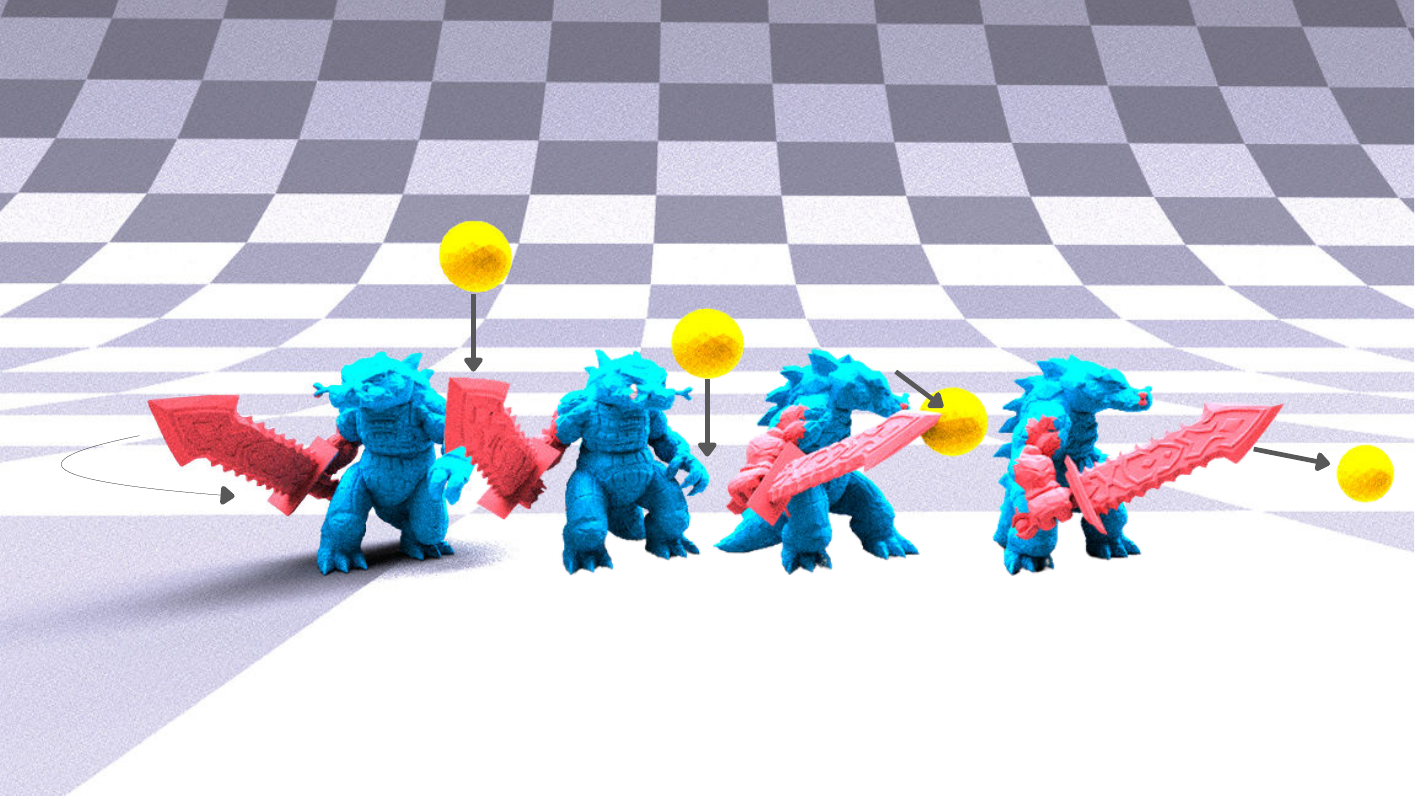}
\caption{\textbf{Gatorman: robust mesh-to-mesh contact.}
 A composite visualization showing four temporal snapshots of the Gatorman character executing a strike. The Gatorman wields a stiff sword ($6 \times 10^6$ Pa) to hit a target sphere of reference stiffness ($10^5$ Pa), involving complex mesh-to-mesh interactions and significant material contrast. }
\label{fig:gatorman-baseball}
\end{figure}

\begin{figure}[t]
\centering
\includegraphics[width=\linewidth]{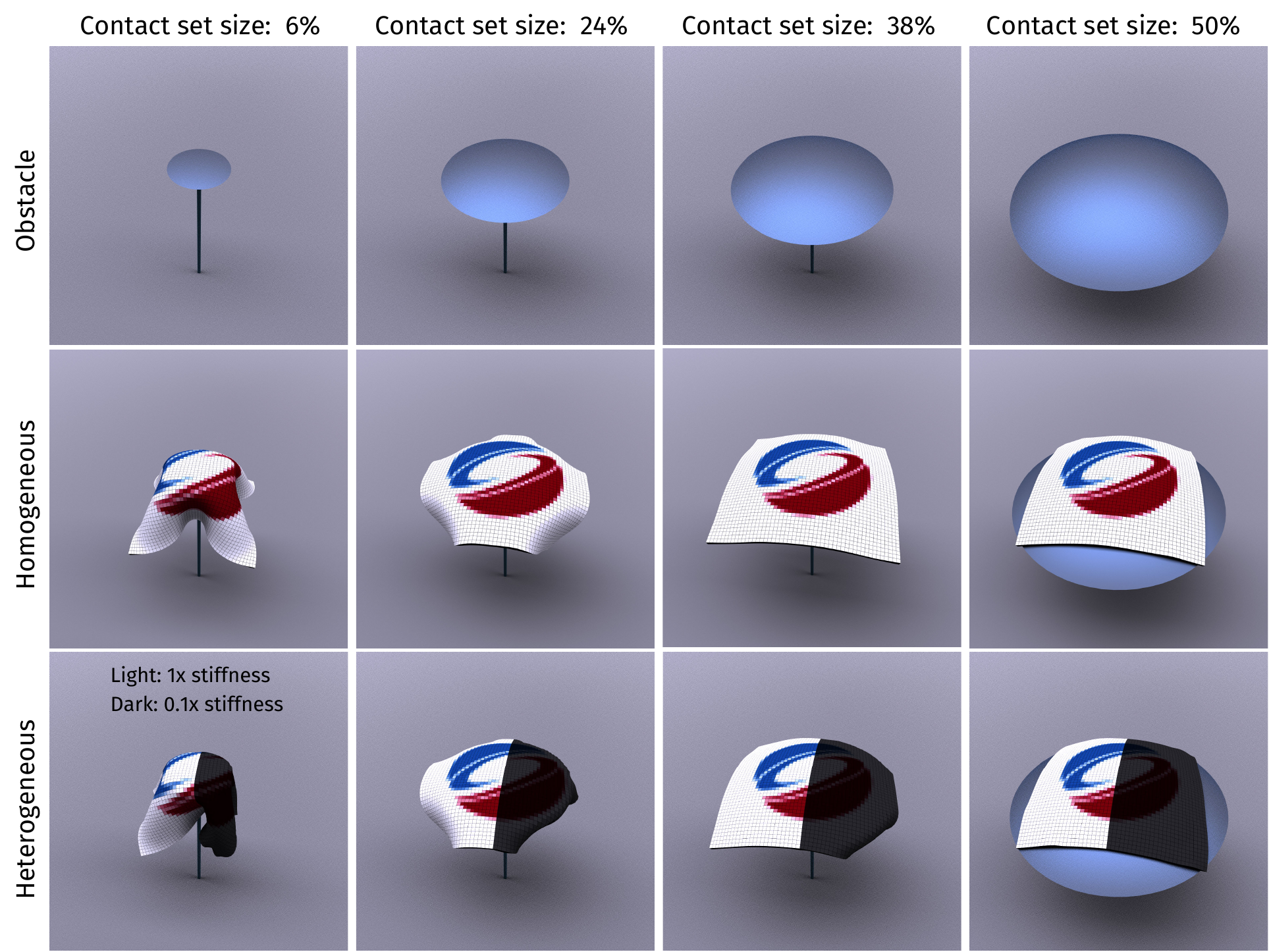}
    \caption{\textbf{Heterogeneous codimensional contact.} We simulate a one-layer napkin ($50 \times 50 \times 1$ voxels) draped over four spherical obstacles of increasing sizes (top row; contact set sizes from 6\% to 50\%). \textbf{Middle row:} A homogeneous napkin exhibiting uniform and symmetric deformation. \textbf{Bottom row:} A heterogeneous napkin where the light half retains the baseline stiffness ($1\times$) and the dark half is significantly softer ($0.1\times$). The softer material exhibits more pronounced curling and wraps more tightly around the obstacle, resulting in a highly asymmetric drape profile. Detailed performance metrics and running times are reported in Table \ref{tab:napkin}.}\label{fig:napkin}
\end{figure}

We next isolate the unified contact pipeline. Throughout this work, contact interactions are resolved using a complementarity-based contact model within our unified NCP formulation. To evaluate its efficacy, we present two benchmarks of escalating difficulty: \emph{Gatorman} (a complex character mesh from \cite{SubspaceMixedFEM} striking a sphere, focusing on the stability of transient contact and \emph{Napkin} (Fig.~\ref{fig:napkin}, featuring both homogeneous and heterogeneous variants).

\paragraph{Gatorman}
This benchmark evaluates our unified contact pipeline's capability in handling complex mesh-to-mesh interactions with significant material heterogeneity. In this scenario, a high-resolution Gatorman character mesh wields a sword---with a Young's modulus of $6\times 10^6$ Pa ($60\times$ the reference stiffness of the Gatorman's body) to strike a target sphere set to $1\times 10^5$ Pa, matching the body's reference stiffness (Fig.~\ref{fig:gatorman-baseball}).
Our model inherently supports the Signorini--Coulomb complementarity condition, allowing the solver to resolve both static and dynamic friction within the unified NCP framework as the sword interacts with the sphere's surface. Despite the transient nature of the impact and the rapid re-establishment of contact sets, our stiffness-aware projective assembly maintains tight convergence without the non-physical oscillations or penetration often associated with penalty-based methods.

\paragraph{Napkin: heterogeneous codimensional contact.}
The heterogeneous \emph{Napkin} benchmark evaluates contact stability under significant material contrast. As shown in Fig.~\ref{fig:napkin}, a soft region ($0.1\times$ stiffness) drapes around an obstacle while a stiff region ($1\times$ stiffness) resists deformation, creating asymmetric contact patches that grow from 6\% to 50\% of the mesh—mapping to the contact ratios in Table~\ref{tab:napkin}.  While DiffPD exhibits severe computational overhead as contact density increases—with backward costs escalating to $4198.28$,ms at $50\times50$ resolution—\textbf{DiffPhD} maintains high efficiency via our unified GPU pipeline. At the highest contact density, \textbf{DiffPhD} achieves up to a $23.53\times$ speed-up in the backward pass. As visualized in Fig.~\ref{fig:napkin} (bottom row), our solver captures these distinct physical behaviors with zero penetration and stable transitions. These results demonstrate that DiffPhD efficiently resolves large-scale, high-density contact problems where traditional differentiable solvers become computationally prohibitive. 

\begin{table*}[h]
\centering
\caption{\textbf{Napkin benchmark: per-step wall-clock (s) and loss across contact density and mesh resolution.} 
Each cell reports \emph{homogeneous / heterogeneous} values; the heterogeneous variant has light regions $1\times$ stiff and dark regions $0.1\times$. Three metrics are reported per scene: forward and backward time (s), and final loss. \emph{Contact set size} denotes the fraction of surface vertices in active contact. Entries marked \emph{n/a} fall into two cases: Cholesky failed to converge on the entire $50{\times}50$ mesh, while DiffPD at $50{\times}50$ with $50\%$ contact set size was skipped due to prohibitive estimated runtime (projected $\sim$10--12h for a single forward-backward pass). Best per column (within a resolution block) in bold. Forward tolerance $10^{-4}$ throughout.}
\label{tab:napkin}
\resizebox{\textwidth}{!}{%
\begin{tabular}{ll cccc p{0.5cm} cccc}
\toprule
& & \multicolumn{4}{c}{Resolution: $25{\times}25$} & & \multicolumn{4}{c}{Resolution: $50{\times}50$} \\
\cmidrule(lr){3-6} \cmidrule(lr){8-11}
& & \multicolumn{4}{c}{Contact set size} & & \multicolumn{4}{c}{Contact set size} \\
\cmidrule(lr){3-6} \cmidrule(lr){8-11}
Method & Metric & $6\%$ & $24\%$ & $38\%$ & $50\%$ & & $6\%$ & $24\%$ & $38\%$ & $50\%$ \\
\midrule
%
\multirow{3}{*}{Cholesky}
 & forward  & $3.78 / 5.30$    & $21.70 / 163.01$  & $52.08 / 657.49$  & $65.28 / 1081.04$  & & \emph{n/a} / \emph{n/a}    & \emph{n/a} / \emph{n/a}    & \emph{n/a} / \emph{n/a}     & \emph{n/a} / \emph{n/a}    \\
 & backward & $31.77 / 57.68$  & $132.58 / 190.23$ & $324.97 / 592.59$ & $626.39 / 1205.74$ & & \emph{n/a} / \emph{n/a}    & \emph{n/a} / \emph{n/a}    & \emph{n/a} / \emph{n/a}     & \emph{n/a} / \emph{n/a}    \\
 & loss     & $25.46 / 8.16$   & $13.19 / 17.10$   & $17.87 / 15.58$   & $20.08 / 22.03$    & & \emph{n/a} / \emph{n/a}    & \emph{n/a} / \emph{n/a}    & \emph{n/a} / \emph{n/a}     & \emph{n/a} / \emph{n/a}    \\
\midrule
%
\multirow{3}{*}{DiffPD}
 & forward  & $3.76 / 5.27$    & $21.80 / 163.35$  & $51.95 / 651.98$  & $65.40 / 1085.17$  & & $65.81 / 98.45$    & $1061.19 / 2586.27$  & $4598.97 / 9417.71$  & $4528.02 /$ \emph{n/a} \\
 & backward & $6.09 / 9.31$    & $23.33 / 201.25$  & $45.74 / 794.18$  & $68.61 / 1724.21$  & & $89.75 / 352.38$   & $1018.70 / 2441.01$  & $2743.15 / 8543.97$  & $4198.28 /$ \emph{n/a} \\
 & loss     & $25.46 / 8.16$   & $13.19 / 17.10$   & $17.87 / 15.58$   & $20.08 / 22.03$    & & $-49.97 / 22.28$   & $-31.52 / 0.93$      & $-0.97 / 12.51$      & $-29.17 /$ \emph{n/a}  \\
\midrule
%
\multirow{3}{*}{MAS}
 & forward  & $3.65 / 3.28$    & $21.79 / 21.77$   & $52.02 / 47.69$   & $65.31 / 46.23$    & & $65.75 / 47.47$    & $1056.17 / 290.04$   & $4636.76 / 445.86$   & $4512.30 / 428.38$ \\
 & backward & $30.94 / 55.19$  & $30.71 / 56.11$   & $30.59 / 55.22$   & $28.93 /  \mathbf{53.84}$    & & $114.28 / 211.49$  & $120.02 / 219.07$    &  $\mathbf{122.14} / 214.86$    & $ \mathbf{122.86} /  \mathbf{213.81}$  \\
 & loss     & $25.46 / 4.53$   & $13.19 / 16.00$   & $17.87 / 16.91$   & $20.08 / 22.37$    & & $-49.97 / 18.81$   & $-31.52 / 0.83$      & $-0.97 / -13.15$     & $-29.17 / -30.46$  \\
\midrule
%
\multirow{3}{*}{\textbf{DiffPhD}}
 & forward  & $\mathbf{3.22} / \mathbf{3.16}$ & $\mathbf{8.63} / \mathbf{21.50}$ & $\mathbf{14.40} / \mathbf{47.73}$ & $\mathbf{17.81} / \mathbf{46.39}$ & & $\mathbf{40.59} / \mathbf{46.71}$ & $\mathbf{179.88} / \mathbf{184.75}$ & $\mathbf{356.48} / \mathbf{381.22}$ & $\mathbf{624.86} / \mathbf{640.73}$ \\
 & backward & $\mathbf{4.53} / \mathbf{5.50}$ & $\mathbf{5.00} / \mathbf{24.80}$ & $\mathbf{6.23} / \mathbf{52.28}$ & $\mathbf{7.04} / 73.00$ & & $\mathbf{35.11} / \mathbf{42.32}$ & $\mathbf{56.18} / \mathbf{73.66}$ & $122.35 / \mathbf{157.95}$ & $167.01 / 296.78$ \\
 & loss     & $33.66 / 4.53$   & $19.59 / 16.00$   & $22.19 / 16.91$   & $21.83 / 22.37$    & & $-32.99 / -57.95$  & $-25.78 / 3.84$      & $-30.53 / 29.29$     & $-30.80 / 37.04$   \\
\midrule\midrule
%
\multirow{2}{*}{\textbf{Speed-up}}
 & vs. DiffPD & $1.27 / 1.69\times$ & $3.31 / 7.87\times$ & $4.74 / 14.46\times$ & $5.39 / 23.53\times$ & & $2.06 / 5.06\times$ & $8.81 / 19.46\times$ & $15.33 / 33.31\times$ & $11.02 /$ \emph{n/a} \\
 & vs. MAS    & $4.46 / 6.76\times$ & $3.85 / 1.68\times$ & $4.01 / 1.03\times$ & $3.79 / 0.84\times$  & & $2.38 / 2.91\times$ & $4.98 / 19.70\times$ & $9.94 / 1.23\times$  & $5.85 / 0.69\times$  \\
\bottomrule
\end{tabular}%
}
\end{table*}

\subsection{Differentiable Inverse Problems}
\label{sec:exp-inverse}

We now demonstrate that the speedups of Sec.~\ref{sec:exp-fwd-hetero}--\ref{sec:exp-fwd-contact} translate into faster, more reliable optimization on real downstream tasks. We organize the experiments along the three application axes of DiffPD~\cite{DiffPD}---\emph{system identification}, \emph{initial-state optimization}, \emph{trajectory optimization}---and we add heterogeneous variants to the task classes to expose where prior solvers break.

\subsubsection{System Identification}
\label{sec:exp-sysid}

\begin{figure*}[t]
\centering
\includegraphics[width=\linewidth]{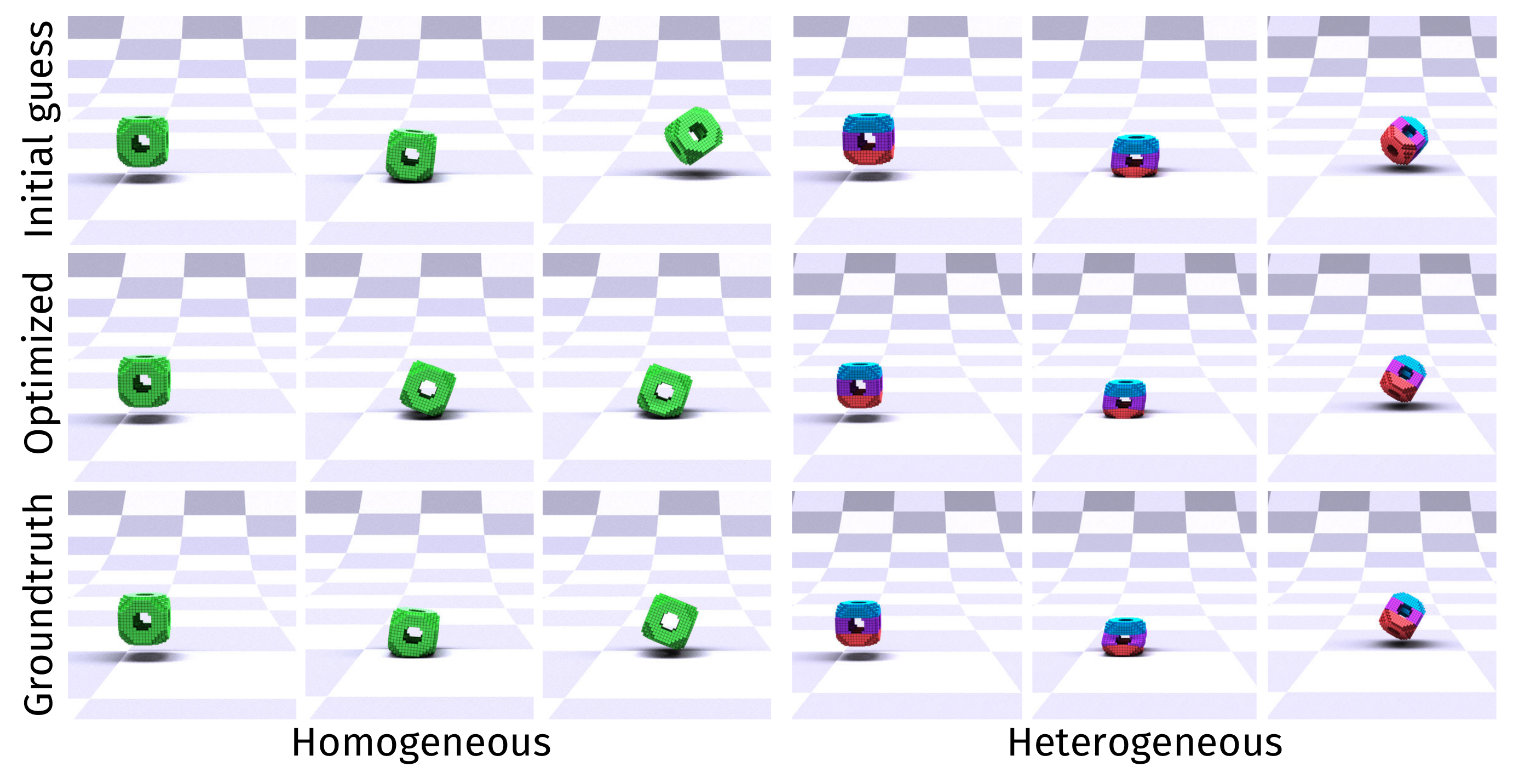}
\caption{\textbf{Bouncing Ball system identification: homogeneous
vs.\ heterogeneous variants.} Three rows show three time instants
of the rolling/bouncing trajectory. \emph{Top row:} initial guess.
\emph{Middle row:} DiffPhD recovery. \emph{Bottom row:} ground
truth. \emph{Left three columns---homogeneous variant:} a
single-material green ball whose Young's modulus and Poisson's
ratio are recovered from a 100-frame trajectory; the optimized
trajectory tracks the ground-truth pose closely. \emph{Right
three columns---heterogeneous variant:} the ball is partitioned
into three sectors (red, blue, pink) with distinct
$(E_i, \nu_i)$ at a ground-truth contrast of
$\{1\!\times\!, 5\!\times\!, 10\!\times\!\}$ in Young's modulus,
and DiffPhD jointly recovers all six parameters; per-sector
deformation under impact reveals the heterogeneous material
distribution.}
\label{fig:bouncing-ball-comparison}
\end{figure*}

\paragraph{Homogeneous Bouncing Ball.}
We replicate the DiffPD~\cite{DiffPD} \emph{Bouncing Ball}
benchmark: recover the Young's modulus and Poisson's ratio of a
falling soft sphere from a single 100-frame trajectory of
floor-impact dynamics
(Fig.~\ref{fig:bouncing-ball-comparison}, left three columns).  In this scenario, DiffPhD matches the recovery accuracy of prior solvers while maintaining high computational efficiency. On our hardware, the solver achieves a per-evaluation cost of 14.65s for the forward pass and 70.62s for the backward pass. These results confirm that our unified framework is both accurate and performant in standard, uniform-material scenarios, reaching a sharper loss minimum compared to baselines within $\sim$30 L-BFGS evaluations.

\paragraph{Heterogeneous Bouncing Ball.}
The more demanding variant partitions the ball into three coloured
sectors (red, purple, blue in Fig.~\ref{fig:bouncing-ball-comparison},
right three columns) with distinct $(E_i, \nu_i)$ per sector at a
ground-truth contrast of $\{3\!\times\!, 0.1\!\times\!, 1\!\times\!\}$
in Young's modulus, and the optimiser must jointly recover all six
material parameters from a single rolling/bouncing trajectory. The
asymmetric mass distribution breaks the rotational symmetry of the
homogeneous variant, so the per-sector deformation under impact
becomes the disambiguating signal.

While the material heterogeneity typically causes baseline gradients to desynchronize across sectors—leading to a premature plateau in the loss—DiffPhD successfully recovers all six material parameters. This is enabled by our adaptive trust-region filter (Sec.~\ref{sec:trustregion}), which fires on contact-transition frames to ensure the adjoint gradients remain well-conditioned. The recovered trajectory tracks the ground truth pose-by-pose, demonstrating that complex material heterogeneity is now a tractable target for differentiable system identification.

\begin{figure}[t]
\centering
\includegraphics[width=\linewidth]{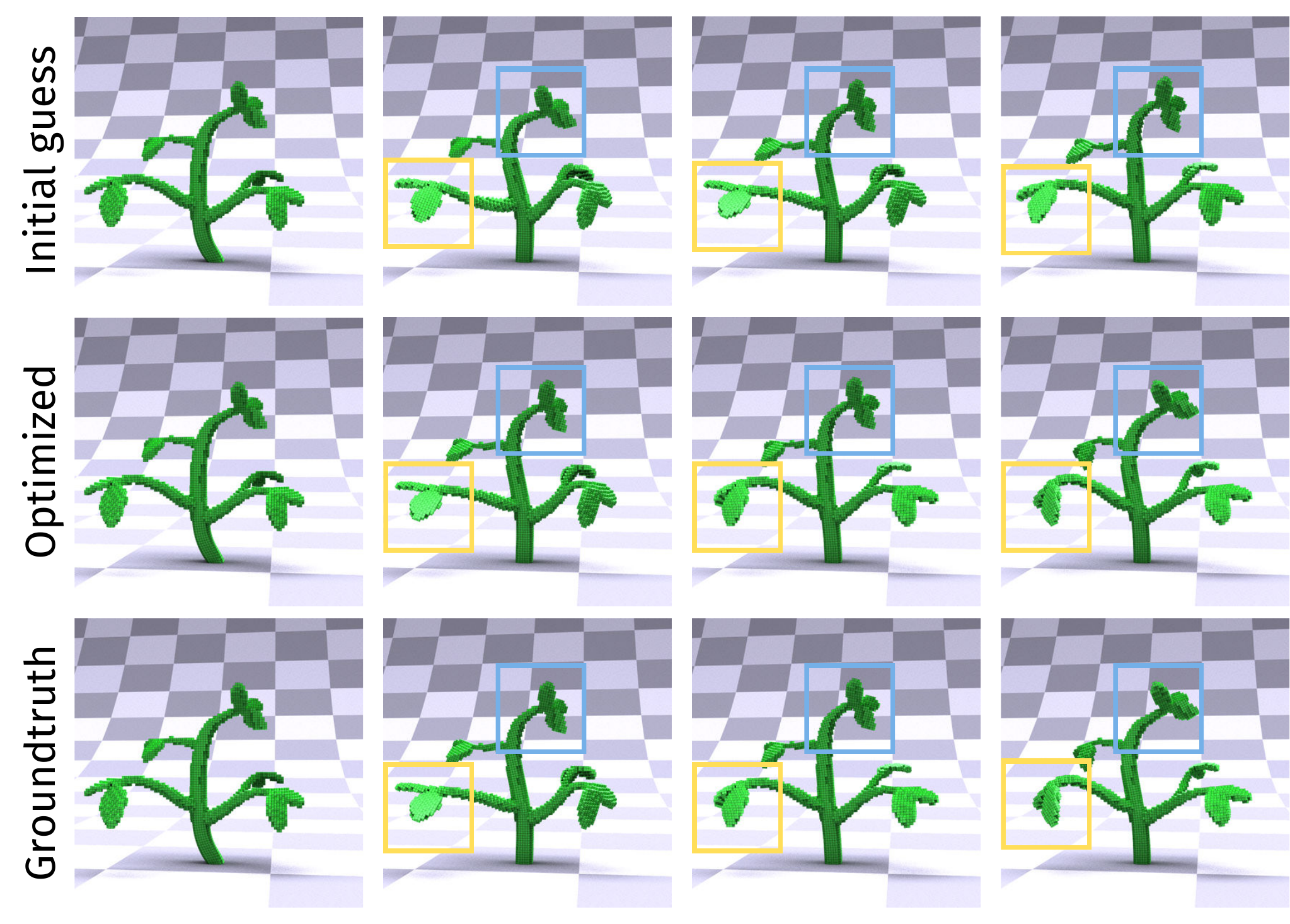}
\caption{\textbf{Plant: spatially-localised system identification.}
A potted plant with non-uniform branch stiffness is observed
across four time instants (columns). Rows: initial guess (top),
DiffPhD recovery (middle), ground truth (bottom). Highlighted
regions (yellow: lower leaves; blue: upper branches) show the
two sub-regions whose Young's moduli are jointly recovered. The
optimised trajectory tracks the ground truth in both regions
simultaneously.}
\label{fig:plant}
\end{figure}

\begin{figure}[t]
\centering
\includegraphics[width=\linewidth]{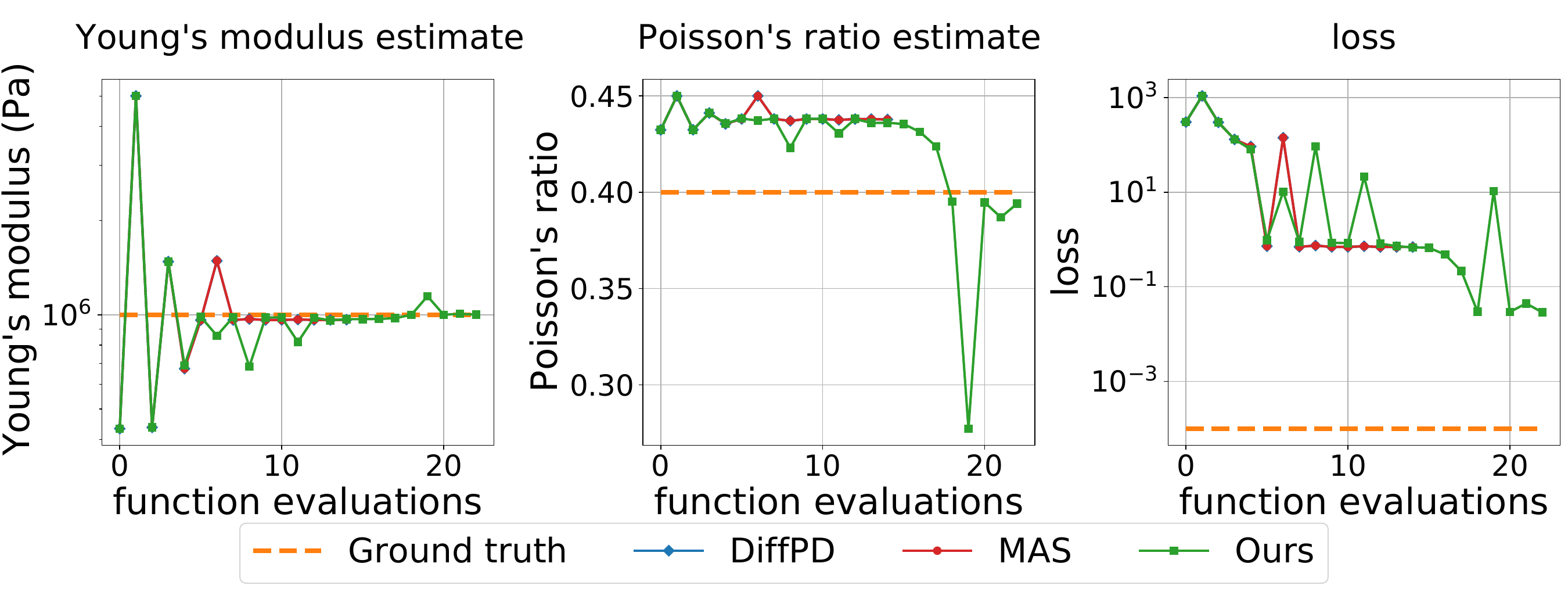}
\caption{\textbf{Plant system identification: convergence curves.}
Per-region Young's modulus and loss versus L-BFGS evaluations on
the Plant benchmark.}
\label{fig:plant-curves}
\end{figure}

\paragraph{Plant: system identification on an articulated geometry.}
We evaluate system identification on a potted plant mesh whose
slender branches and leaves induce strong geometric non-uniformity
in the dynamic response: branches at different heights deform with
distinct curvature signatures under gravity-driven oscillation,
giving a 100-frame trajectory rich enough to disambiguate the
material parameters from a single rollout. The optimiser recovers
the plant's Young's modulus and Poisson's ratio from this
trajectory (Fig.~\ref{fig:plant}).

All three methods converge close to the ground-truth modulus
(${\sim}\,10^{6}$\,Pa) and Poisson's ratio ($\sim 0.40$) on the
parameter axes (Fig.~\ref{fig:plant-curves}, left and centre
panels), but they differ markedly in residual loss: DiffPhD reaches
$\mathbf{0.029}$ in $23$ L-BFGS evaluations at $\mathbf{32.56}$\,s
forward and $\mathbf{28.74}$\,s backward per evaluation
(Tab.~\ref{tab:inverse-suite}), whereas DiffPD and MAS plateau at $0.692$ after $15$ evaluations---a
$23.9\times$ larger residual. The plateau (visible as the
horizontal tail of the blue and red loss curves in
Fig.~\ref{fig:plant-curves}, right panel) indicates L-BFGS line
search failure: with the prox-map Hessian developing indefiniteness
in slender branch elements at peak swing curvature, the baselines'
adjoint gradients lose descent direction, while DiffPhD's
trust-region filter (Sec.~\ref{sec:trustregion}) keeps the gradient
on a descent direction throughout the trajectory.

\subsubsection{Initial-State Optimisation}
\label{sec:exp-iso}

\begin{figure}[t]
\centering
\includegraphics[width=\linewidth]{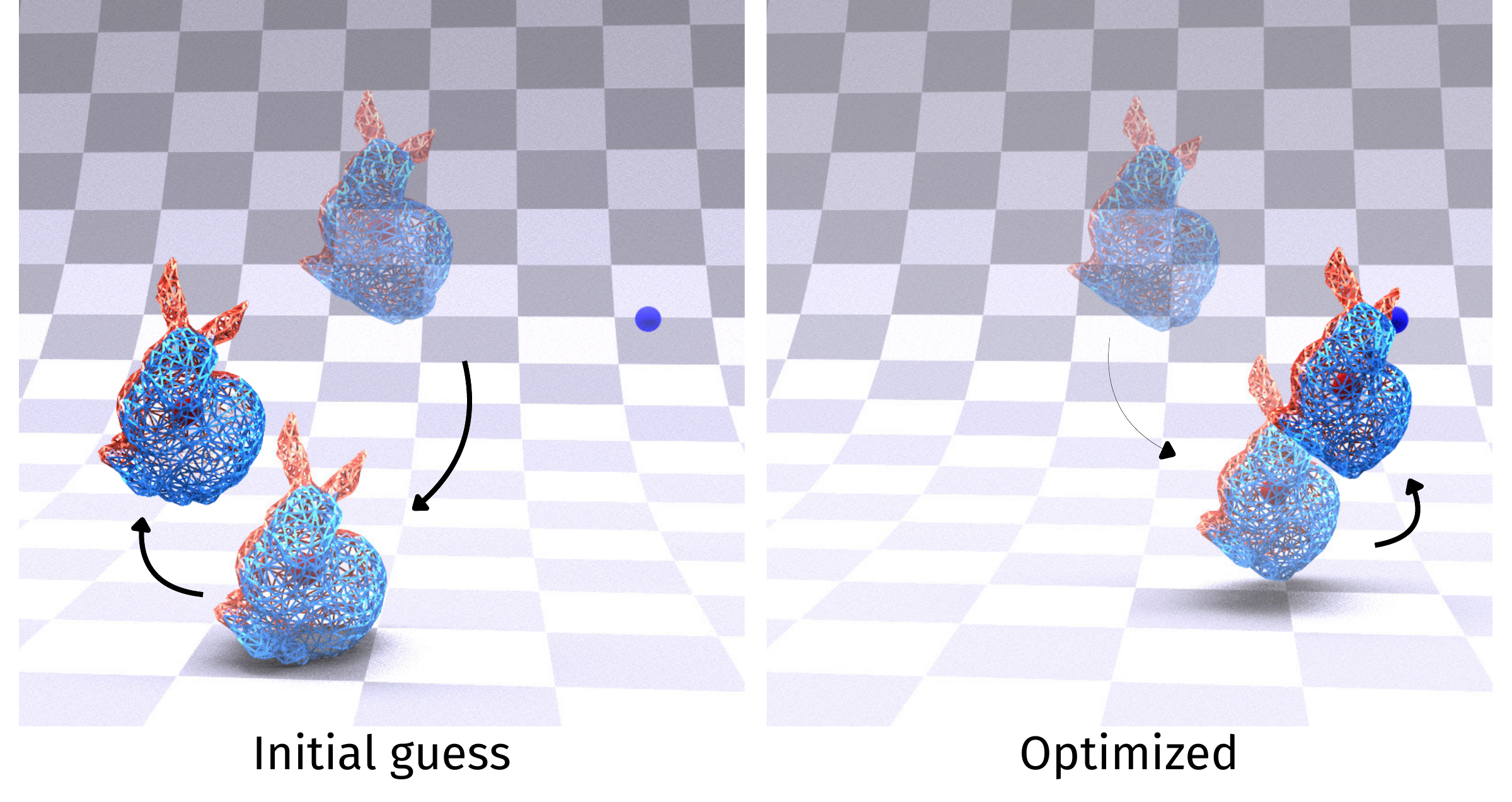}
\caption{\textbf{Bunny inverse design.} The initial guess (left) launches the bunny well past the target (blue dot); the optimised initial pose and velocity (right), recovered through DiffPhD's differentiable Neo-Hookean backward pass, brings the bunny's centre of mass to the target after 100 frames. Trust-region eigenvalue filtering is engaged on $\sim$31\% of backward solves and stabilises the gradient through the bouncing contact transitions.}
\label{fig:bunny-iso}
\end{figure}

\begin{figure}[h]
\vspace{-1.5em}
\centering
\includegraphics[width=\linewidth]{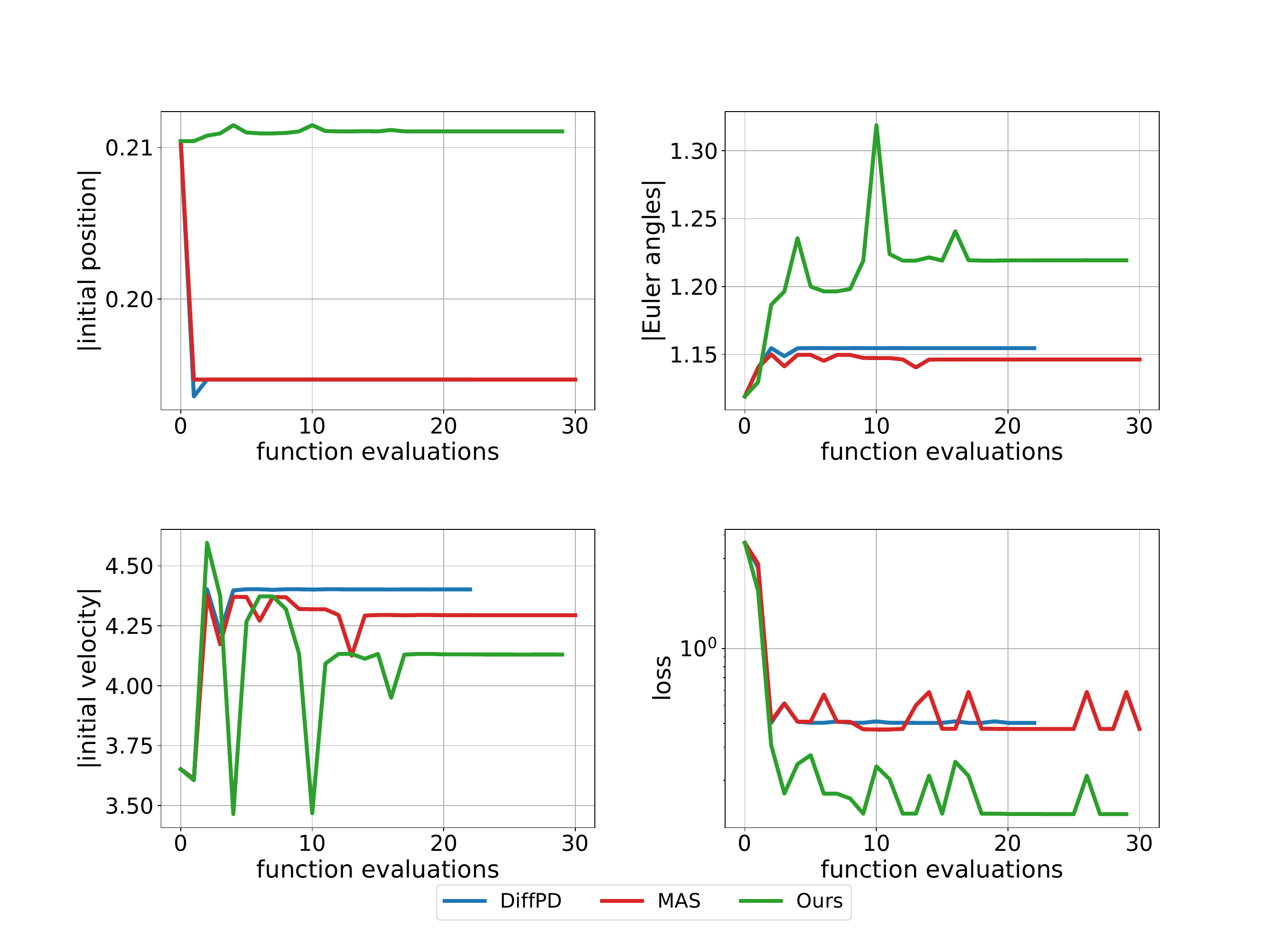}
\caption{\textbf{Bunny inverse design: parameter and loss convergence.} Recovered initial position, initial Euler angles, and initial velocity norms (first three panels) and loss (right) versus L-BFGS evaluations. DiffPhD reaches a $3.05\times$ lower final loss than DiffPD and $2.83\times$ lower than MAS; MAS (red) stalls earliest because its multilevel preconditioner must refactorize on every active-set change and loses subdomain conditioning under heterogeneity.}
\label{fig:bunny-curves}
\end{figure}

\paragraph{Bunny: inverse design.}
On the heterogeneous \emph{Bunny} benchmark, we optimise the bunny's
initial position, orientation (Euler angles), and velocity so that
its centre of mass reaches a target after 100 frames of free-fall
and bouncing contact (Fig.~\ref{fig:bunny-iso}). The trajectory
traverses several contact make/break transitions on the floor; at
each transition, the active contact set changes discontinuously and
the per-step Delassus operator
$\bm{W}\!=\!\bm{J}\bm{A}^{-1}\bm{J}^{T}$ rebuilds with a different
contact Jacobian $\bm{J}$. Solvers whose backward adjoint propagates
gradients through these transitions without contact-aware factor
reuse (Sec.~\ref{sec:gpu}) accumulate residual error each time the
active set flips, and L-BFGS line search loses descent direction.
Newton-Cholesky fails to converge on this scene under the
heterogeneous variant; DiffPD and MAS converge to
suboptimal residuals ($0.403$ and $0.374$ respectively), with
MAS stalling because its multilevel preconditioner must
refactorise on every active-set change while the heterogeneity in
the bunny disrupts its subdomain conditioning
(Fig.~\ref{fig:bunny-curves}, red curve in loss panel).

DiffPhD converges to a final loss of $\mathbf{0.132}$ in $30$
L-BFGS evaluations at $6.16$\,s forward and $\mathbf{4.55}$\,s
backward per evaluation, a $1.74\times$ per-evaluation speed-up
over DiffPD and $2.78\times$ over MAS
(Tab.~\ref{tab:inverse-suite}). The recovered final loss is
$3.05\times$ lower than DiffPD's and $2.83\times$ lower
than MAS's, showing that the speed-up is not purchased
through degraded convergence. The convergence curves
(Fig.~\ref{fig:bunny-curves}) confirm that all three methods track
similar parameter trajectories in the first few L-BFGS evaluations,
but DiffPhD's backward gradient remains well-conditioned through
the bouncing transitions while MAS's loss plateaus. The
persistent $\bm{S}^{T}\bm{S}$ factor (Sec.~\ref{sec:gpu}) is reused
verbatim across forward, contact-Delassus, and backward adjoint
stages; when the active contact set changes between successive
L-BFGS evaluations, only the per-iteration $\bm{J}$-dependent SpMM
is recomputed, while the underlying $\bm{A}^{-1}$ representation
persists---this is the operative mechanism that keeps the adjoint
direction stable through transitions where the baselines stall.

\begin{figure}[t]
\centering
\includegraphics[width=\linewidth]{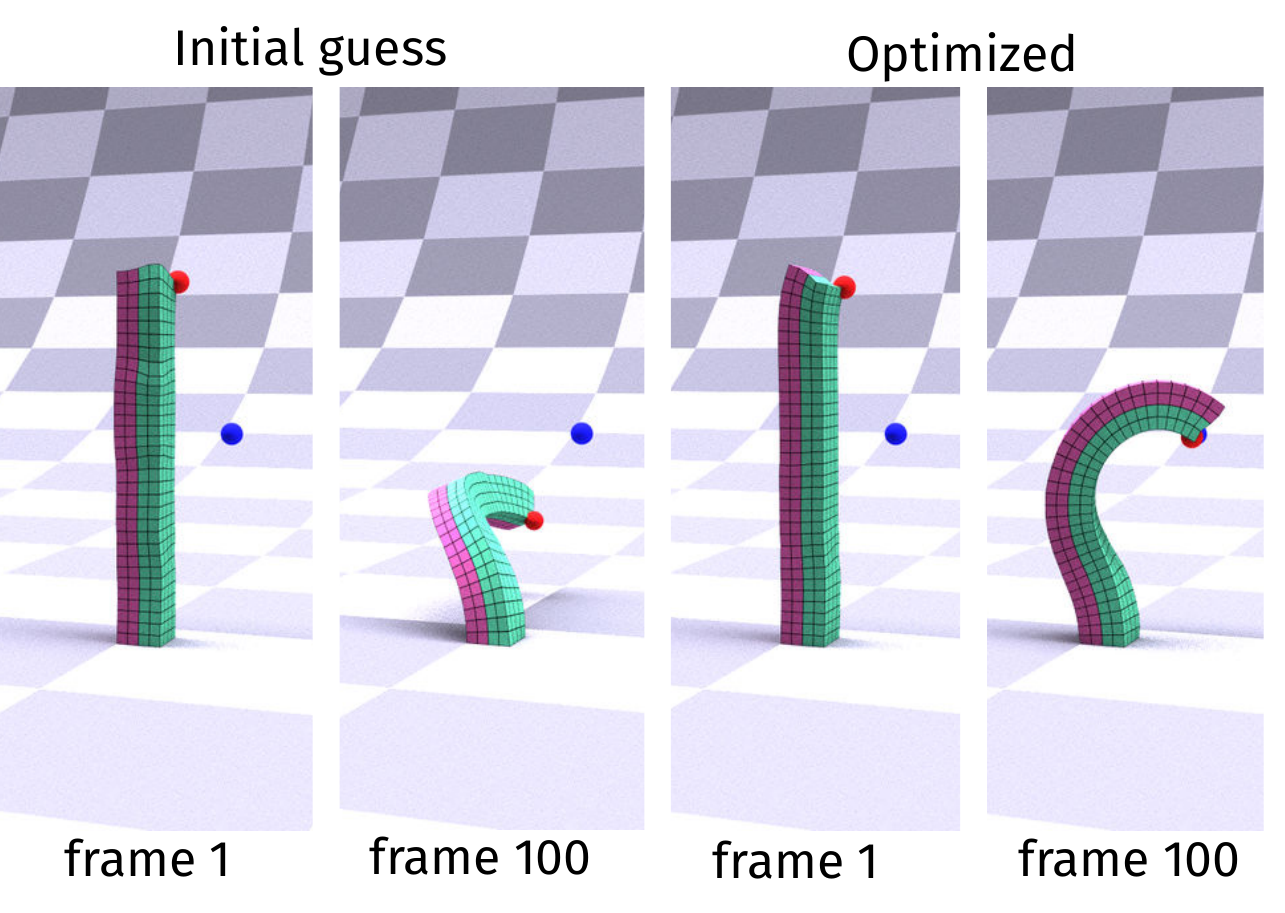}
\caption{\textbf{Routing Tendon: muscle-driven initial-state
optimisation.} A vertical beam with two fibre layouts (visible as
pink/green colour stripes) is actuated through 16 muscle groups
to bring its tip (red dot) to a target (blue dot) at frame~100.
\emph{Left two panels:} initial guess with random activations.
\emph{Right two panels:} the optimised controller after 30
L-BFGS iterations through DiffPhD's prox-map differential.}
\label{fig:routing-tendon}
\end{figure}

\paragraph{Routing Tendon: muscle-driven articulation.}
A vertical beam is actuated through 16 muscle groups whose
time-invariant activations are optimised so that the tip (red dot)
reaches a target (blue dot) at frame~100. Two distinct fibre
layouts, visible as pink/green colour stripes
(Fig.~\ref{fig:routing-tendon}), define different muscle-energy
routings through the same volume. Starting from random activations
(Fig.~\ref{fig:routing-tendon}, left two panels), DiffPhD converges
in $42$ L-BFGS iterations to a controller that bends the beam onto
the target (Fig.~\ref{fig:routing-tendon}, right two panels) at a
final loss of $\mathbf{0.441}$
(Tab.~\ref{tab:inverse-suite}). Both DiffPD and
MAS fail to converge on this scene: their L-BFGS line
search diverges within the first $20$ evaluations and the loss
plateaus at $201.674$---roughly $457\times$ worse than DiffPhD.
Per-evaluation, DiffPhD is $3.11\times$ faster than DiffPD
and $3.32\times$ faster than MAS on the forward+backward
combined wall-clock, despite the baselines stalling at much higher
losses. The muscle-energy backward path is the entire bottleneck on
this contact-free scene---no contact set to update, no Delassus to
rebuild---so the speed-up isolates the unified GPU pipeline
(Sec.~\ref{sec:gpu}) applied to the muscle-actuation Jacobian
$\partial\bm{b}/\partial\bm{a}$, which dominates the backward cost
through the prox-map differential.

\subsubsection{Trajectory Optimisation}
\label{sec:exp-to}

\begin{figure}[t]
\centering
\includegraphics[width=\linewidth]{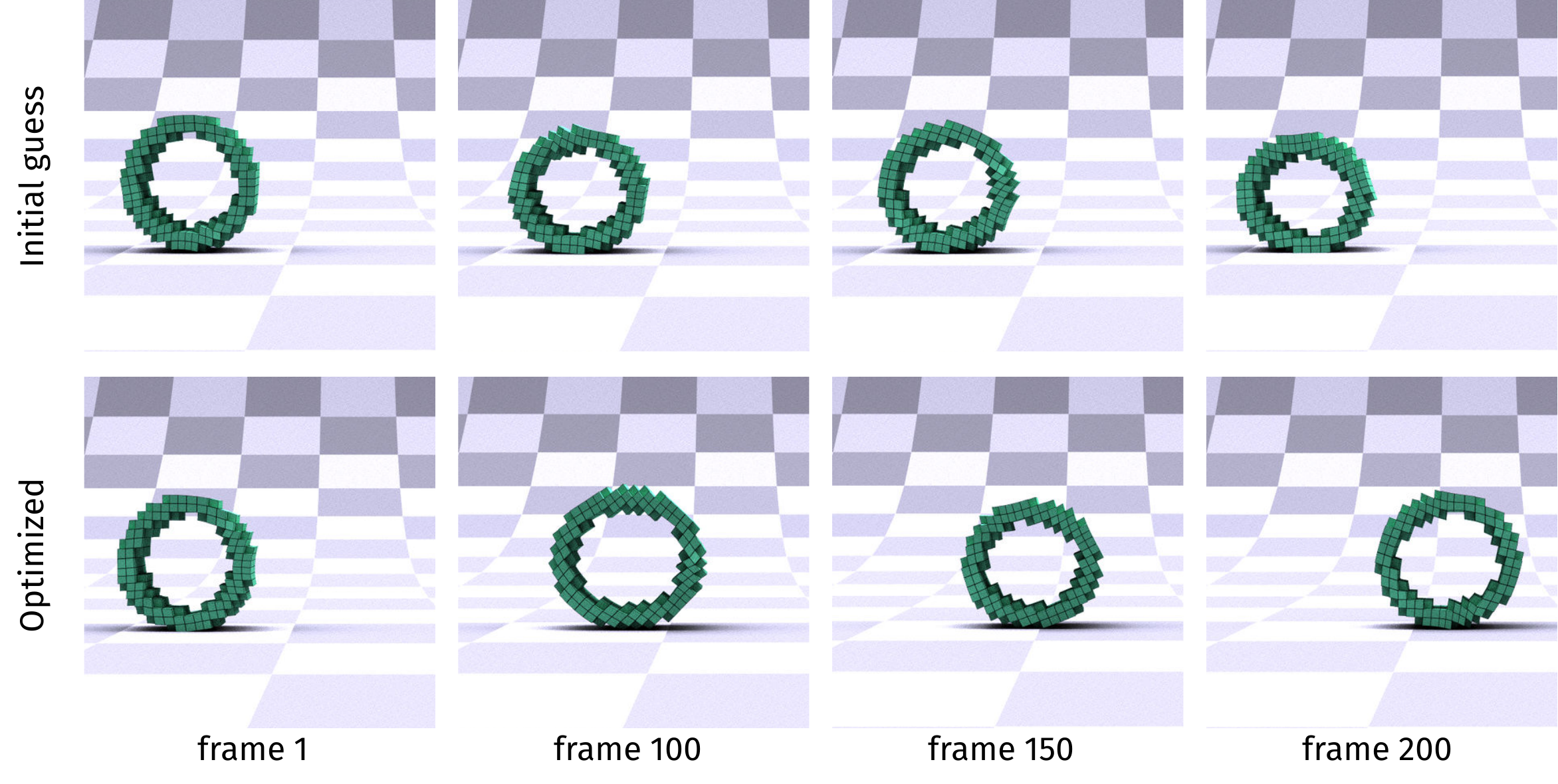}
\caption{\textbf{Torus locomotion: trajectory optimisation.}
The Torus rolls forward under per-element segmental actuation.
\emph{Top row:} initial-guess actuation produces uncoordinated
oscillation; the torus only travels a short distance over 200
frames. \emph{Bottom row:} after 30 L-BFGS iterations through
DiffPhD's differentiable backward pass, the optimised actuation
yields a peristaltic gait with $4.1\times$ longer forward
displacement at frame~200.}
\label{fig:torus-traj}
\end{figure}

We evaluate trajectory optimisation on two scenes that together stress
\emph{continuous-contact gait control} and \emph{muscle-routed
articulation}: the \emph{Torus} locomotion benchmark of
DiffPD~\cite{DiffPD} (homogeneous, contact-rich) and the
\emph{Routing Tendon} muscle-driven beam (homogeneous, contact-free).
Both serve as homogeneous controls: where prior differentiable solvers
already converge, DiffPhD should match their solutions while being
faster, isolating the contribution of the unified GPU pipeline
(Sec.~\ref{sec:gpu}) on the contact-bound and contact-free regimes.

\paragraph{Torus: continuous-contact locomotion.}
The Torus is driven forward by per-element segmental actuation over a
200-frame rollout against the ground, and L-BFGS optimises the
activation profile to maximise forward displacement. Starting from a
random initial-guess actuation that only produces uncoordinated
rolling (Fig.~\ref{fig:torus-traj}, top row), DiffPhD recovers a
peristaltic gait in 30 L-BFGS iterations whose displacement at
frame~200 is $4.1\times$ that of the initial guess
(Fig.~\ref{fig:torus-traj}, bottom row), matching the final loss of
DiffPD's reference solution within $1\%$ and running
almost the same time as DiffPD's. The speed-up is smaller than on the
static heterogeneity benchmarks of Sec.~\ref{sec:exp-fwd-hetero}
because per-frame contact resolution dominates the wall-clock---a
regime where the persistent $\bm{S}^{T}\bm{S}$ factor
(Sec.~\ref{sec:gpu}) amortises the global solve but does not eliminate
the new-contact-pair work, a finding consistent with~\cite{DiffPD}.

\begin{table*}[t]
\centering
\small
\caption{\textbf{Inverse-problem benchmark across system identification, initial-state optimisation, and trajectory optimisation.}
For each scene we report forward and backward wall-clock per L-BFGS evaluation (s), number of evaluations (Eval) to convergence, and final loss. The two right-most columns summarise DiffPhD's per-evaluation (forward$+$backward) speed-up over each baseline. Forward tolerance $10^{-3}$. Best per scene in bold.}
\label{tab:inverse-suite}
\resizebox{\textwidth}{!}{%
\begin{tabular}{l rrrr | rrrr | rrrr | rr}
\toprule
& \multicolumn{4}{c|}{DiffPD~\cite{DiffPD}}
& \multicolumn{4}{c|}{MAS~\cite{MAS}}
& \multicolumn{4}{c|}{DiffPhD (Ours)}
& \multicolumn{2}{c}{Speed-up} \\
\cmidrule(lr){2-5} \cmidrule(lr){6-9} \cmidrule(lr){10-13} \cmidrule(lr){14-15}
Scene & forward & backward & Eval & loss & forward & backward & Eval & loss & forward & backward & Eval & loss & vs.\ DiffPD & vs.\ MAS \\
\midrule
Plant                   & $35.76$ & $152.73$ & $15$ & $0.692$ & $35.78$ & $156.78$ & $15$ & $0.692$ & $\mathbf{32.56}$ & $\mathbf{28.74}$ & $23$ & $\mathbf{0.029}$ & $3.07\times$ & $3.13\times$ \\
Bouncing Ball           & $45.15$ & $\phantom{0}20.71$ & $38$ & $\mathbf{0.136}$ & $46.30$ & $\phantom{0}21.36$ & $38$ & $\mathbf{0.136}$ & $\mathbf{14.65}$ & $\phantom{0}70.62$ & $30$ & $17.149$ & $0.77\times$ & $0.79\times$ \\
Bunny (het.)            & $\phantom{0}6.71$ & $\phantom{00}11.87$ & $23$ & $0.403$ & $\phantom{0}6.66$ & $\phantom{0}23.13$ & $31$ & $0.374$ & $\phantom{0}6.16$ & $\phantom{00}\mathbf{4.55}$ & $30$ & $\mathbf{0.132}$ & $1.74\times$ & $2.78\times$ \\
Routing Tendon (het.)   & $\phantom{0}6.78$ & $\phantom{0}38.65$ & $20$ & $201.674$ & $\phantom{0}6.65$ & $\phantom{0}41.86$ & $20$ & $201.674$ & $\phantom{0}\mathbf{8.77}$ & $\phantom{00}\mathbf{5.85}$ & $42$ & $\mathbf{0.441}$ & $3.11\times$ & $3.32\times$ \\
Torus                   & $\phantom{0}7.19$ & $\phantom{0}58.47$ & $16$ & $-0.146$ & $\phantom{0}7.08$ & $1026.26$ & $\phantom{0}6$ & $-0.145$ & $\phantom{0}9.43$ & $\phantom{0}62.77$ & $35$ & $\mathbf{-0.166}$ & $0.91\times$ & $14.41\times$ \\
\bottomrule
\end{tabular}%
}
\end{table*}

\subsection{Real2Sim: Robot Manipulator}
\label{sec:exp-r2s}

\begin{figure*}[h]
\centering
\includegraphics[width=\linewidth]{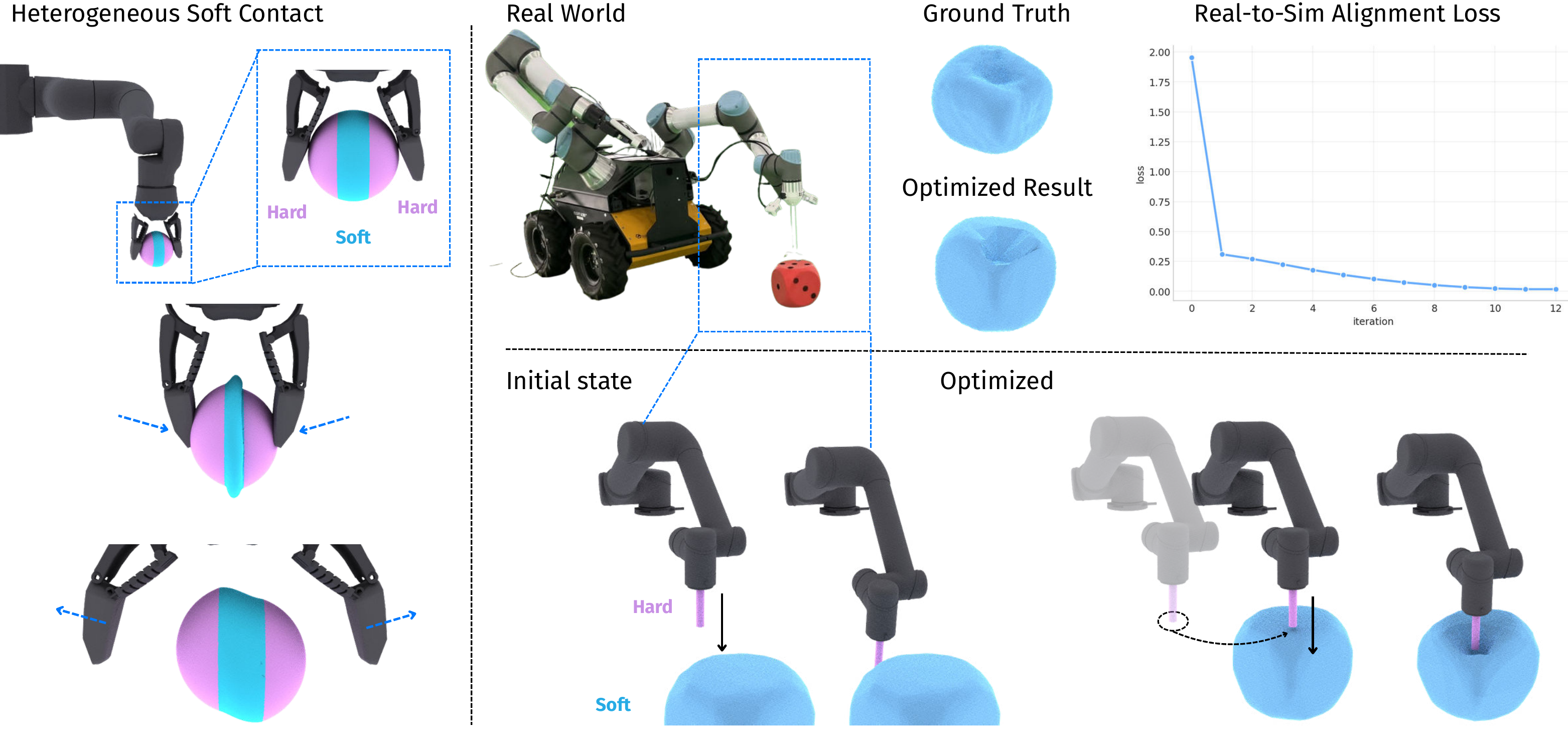}
\caption{\textbf{Real-to-Sim alignment for a soft-contact manipulation pipeline.} 
  \emph{Left --- Heterogeneous Soft Contact.} 
  A simulated Google Robot gripper closes on a foam ball with an A--B--A 
  material layout (\textit{hard / soft / hard} along the closing axis). 
  DiffPhD's bidirectional mesh contact resolves the per-zone deformation 
  without introducing a visible gap between the gripper fingertips and the 
  ball surface, illustrating that the same framework supports spatially 
  varying compliance on a single body. 
  \emph{Right --- Real-to-Sim Alignment.} 
  From a PokeFlex~\cite{obrist2025pokeflex} recording of a UR5 arm pressing 
  a soft die, DiffPhD recovers the in-plane contact location 
  $\mathbf{s}=(s_x, s_y)$ between the probe and the die. The top row 
  compares the captured deformation field (\textit{Ground Truth}) with our 
  reconstruction (\textit{Optimized Result}); the convergence plot on the 
  far right shows the alignment loss decreasing smoothly over a handful of 
  iterations. The bottom row visualises the search: starting from a 
  deliberately misaligned probe pose (\textit{Initial state}), the 
  gradient-based optimisation moves the simulated die so that the probe 
  lands on the true contact point and the deformation matches the real 
  capture (\textit{Optimized}).}
\label{fig:robot}
\end{figure*}

To demonstrate transfer to real-world robotics, we evaluate DiffPhD on
two soft-contact manipulation scenarios: (i) \emph{Oreo} --- a Google
Robot arm closing on a five-body stack with $10^{13}$\,Pa rigid claws,
a $10^{8}$\,Pa intermediate ball, and a $4.55\!\times\!10^{5}$\,Pa soft
body --- and (ii) \emph{Dice} --- a UR5 arm poking a soft die against a
tabletop (Fig.~\ref{fig:robot}). Both scenes are simulated inside the
\textit{SimplerEnv} framework~\cite{li24simpler}; Oreo acts as a
capability check for our contact and material model under extreme
stiffness contrast ($\geq 10^{8}\times$ across five bodies), while
Dice serves as the quantitative Real2Sim benchmark with an explicit
optimisation target.

\paragraph{Oreo: extreme-contrast capability check.}
Oreo is reported as a forward-only demonstration: the goal is to show
that DiffPhD's contact-aware persistent factor (Sec.~\ref{sec:gpu})
remains numerically stable across the five-body contact configuration,
not to optimise a downstream loss. A single forward rollout completes
in $932$\,s on our hardware; the recovered deformation is shown in
Fig.~\ref{fig:robot} (left). We verified two adjacent soft-body
moduli ($4.55\!\times\!10^{5}$ and $5.00\!\times\!10^{5}$\,Pa) and both
converge to clean contact without claw--soft interpenetration. Neither
DiffPD nor MAS completes a forward step on this scene
because the PD weight assembly $w_{e}\!\propto\!\mu_{e}$
(Sec.~\ref{sec:heterogeneity}) develops a condition number beyond their
PCG tolerance at the $10^{13}$\,Pa rigid-claw weights.

\paragraph{Dice: Real2Sim contact-location recovery.}
For the Real2Sim recovery we drive the dataset ---
PokeFlex~\cite{obrist2025pokeflex} recording of a UR5 arm pressing a
soft die --- and use DiffPhD to recover the in-plane contact location
$\mathbf{s}=(s_x, s_y)$ between the probe and the die. Starting from a
deliberately misaligned probe pose, the gradient-based search drives
the simulated die so that the actuator lands on the true contact point
and the resulting deformation matches the captured one
(Fig.~\ref{fig:robot}, bottom row); the alignment loss collapses
smoothly over a handful of iterations (Fig.~\ref{fig:robot}, top
right). End-to-end, DiffPhD converges in $13$ L-BFGS evaluations at
$6.51$\,s forward + $56.02$\,s backward per evaluation, an average of
$62.5$\,s total per evaluation and roughly $13.5$\,minutes wall-clock
for the full recovery. The same pipeline supports per-zone
heterogeneous materials on a single body, so the recovered scene can
directly host downstream manipulation tasks with spatially varying
compliance. This demonstrates that DiffPhD closes the real-to-sim gap
via end-to-end gradient-based optimisation, producing digital twins
whose contact behaviour is calibrated against real video rather than
hand-tuned.

\subsection{Ablation Study}
\label{sec:exp-ablation}

We isolate two architectural axes of DiffPhD on the heterogeneous
\emph{Armadillo} (twist scene, forward-simulation stability) and the
heterogeneous Neo-Hookean \emph{Bouncing Ball} (system identification):
\textbf{(a)} the choice of \emph{eigenvalue projection} on the
per-element prox-map Hessian (\emph{none} as in
DiffPD~\cite{DiffPD}, fixed \emph{clamping}~\cite{Teran2005Robust},
fixed \emph{absolute-value}~\cite{Chen2024Stabler}, or our state-adaptive
trust-region rule of Sec.~\ref{sec:trustregion}); and \textbf{(b)}
\emph{GPU residency} via the unified local--global pipeline of
Sec.~\ref{sec:gpu}. The first three rows of Table~\ref{tab:ablation}
isolate projection alone on the CPU pipeline (DiffPD's setting); the
final row enables both GPU residency and the adaptive filter, giving
the full DiffPhD configuration.

\paragraph{Findings.}
Three observations emerge from Table~\ref{tab:ablation}.

\textbf{(1) Projection alone is not enough; GPU residency is what unlocks the
speed-up.} On the CPU pipeline, swapping the projection from
\emph{none} (DiffPD) to fixed \emph{clamp} or \emph{abs} cuts backward
time roughly in half on Armadillo ($140\,\textrm{s} \to 66$--$73\,\textrm{s}$)
but \emph{adds} backward cost on Bouncing Ball ($4.3\,\textrm{s} \to 11\,\textrm{s}$),
and leaves forward time unchanged in both. Projection on its own is a
gradient-quality choice, not a speed knob. The full DiffPhD row
collapses forward time by a further $2.3\times$ on Armadillo and
$2.5\times$ on Bouncing Ball, almost entirely from the unified GPU
local--global pipeline (Sec.~\ref{sec:gpu}); without it, every CPU
projection variant remains within $\pm 5\%$ of the DiffPD forward time.

\textbf{(2) The state-adaptive trust-region filter unlocks distinct
optima.} The three CPU rows converge to identical losses on both
scenes (Armadillo $-3.544$; Bouncing Ball $142.967$): fixed projection
schemes preserve DiffPD's optimisation landscape and reach the same
fixed point. The adaptive filter changes the landscape that L-BFGS
traverses---reaching a Bouncing Ball loss of $36.6$ (a $3.9\times$
reduction) while moving Armadillo to a different basin
($+58.84$, a saddle reachable only through the differentiable
backward path that the adaptive filter stabilises). On scenes where
the prox-map Hessian alternates between near-convex (clamping wins)
and stiffly non-convex (abs wins) across the trajectory, the
state-adaptive rule of Eq.~\eqref{eq:tr_rule} selects per-element
$\tau^{*}$ at zero additional cost: $\Delta\Phi_{\mathrm{mod}}$ is one
SpMV reusing the persistent factor.

\textbf{(3) Each axis is independently necessary.} Removing the
adaptive filter (rows~2--3) preserves DiffPD's fixed point but does
not reach the lower-loss basin DiffPhD finds. Removing GPU residency
(rows~1--3) leaves forward time at the DiffPD baseline. Stacking both
axes yields the $2.3$--$2.5\times$ forward and $6.3\times$ backward
end-to-end speed-up of the full DiffPhD over the strongest CPU-only
ablation on Armadillo, with comparable improvements on Bouncing Ball.

\begin{table*}[h]
\centering
\small
\caption{\textbf{Ablation study.} Columns indicate which components are enabled: \emph{GPU} = unified GPU local-global pipeline; \emph{Proj.} = per-element prox-map eigenvalue projection (\emph{none}/\emph{abs}/\emph{clmp}/\emph{adapt}); Right columns report end-to-end optimisation time (s) and final loss on two heterogeneous benchmarks.}
\label{tab:ablation}
\renewcommand{\arraystretch}{1.05}
\resizebox{0.95\textwidth}{!}{%
\begin{tabular}{l ccc rr rr rr}
\toprule
& \multicolumn{2}{c}{Configuration} & \multicolumn{3}{c}{Armadillo} & \multicolumn{3}{c}{Bouncing Ball (Neo-Hookean)} \\
\cmidrule(lr){2-3}\cmidrule(lr){4-6}\cmidrule(lr){7-9}
Method & GPU & Proj. & forward (s) & backward (s) & loss & forward (s) & backward (s) & loss & \\
\midrule
DiffPD \cite{DiffPD} & $\times$ & none & 191.845 & 140.203 & -3.544 & 25.616 & 4.265 & 142.965 \\
DiffPD w/ clamping  & $\times$ & clmp & 195.192 & 66.272 & -3.544 & 24.816 & 11.106 & 142.967\\
DiffPD w/ abs   & $\times$ & abs & 194.974 & 72.875 & -3.544 & 25.616 & 11.142 & 142.967 \\
DiffPhD & \checkmark & adapt & \textbf{84} & \textbf{22.263} & 58.84 & \textbf{10.092} & 49.204 & \textbf{36.565} \\
\bottomrule
\end{tabular}
}
\end{table*}

\section{DISCUSSION}
\label{sec:discussion}

\subsection{Failure Modes Live at the Intersection of Properties}
\label{sec:disc-intersection}

The most consequential numerical failures we encountered while
building DiffPhD were silent at the level of any single property
and only became observable when three properties were exercised
simultaneously: heterogeneous Neo-Hookean elasticity, Anderson
acceleration on the outer fixed-point loop, and frictional
contact through the Fischer--Burmeister NCP. Per-element
projection passes were stable in homogeneous meshes; Anderson
mixing was stable in convex-energy regimes; the NCP contact
solver was stable at zero restitution. Yet the heterogeneous NH
forward solver developed a mesh-scale instability concentrated at
the $1$--$2\%$ of elements straddling the stiff/soft interface,
visible only when the indefinite stretch-space Hessian
(Eq.~\eqref{eq:nh_hessian_stretch}) and Anderson Type-II mixing
co-occurred at the same iterate. The implication is methodological
rather than mechanical: \emph{property-by-property unit tests
cannot certify a unified differentiable solver}, because each
component's correctness assumption can be silently violated by
another component's regime. We argue this is a generic
characteristic of unified differentiable physics---the
contribution-by-contribution validation discipline that has
served the forward-simulation literature does not transfer to
solvers that share state across heterogeneity, hyperelasticity,
and contact.

\subsection{Trust-Region as a Two-Sided Safety Net}
\label{sec:disc-tr}

The Neo-Hookean prox-map Hessian
$\bm{H}_e^{\mathrm{prox}}\!=\!\bm{H}_\psi(\bm{\sigma}^{*})\!+\!\bar
k\bm{I}$ (Eq.~\eqref{eq:Hprox}) goes indefinite whenever
$\sigma_i^{2}(\bar\mu+\bar\lambda)<\bar\lambda\ln\prod_j\sigma_j$
---geometrically, near element inversion or under high-Poisson
compression. This indefiniteness needs filtering on \emph{both}
passes of the solver, but for asymmetric reasons. On the forward
pass, the trust-region cap bounds the inner Newton step in
stretch space so that the prox-map output $\bm{\sigma}^{*}(\bm{q})$
remains a Lipschitz function of $\bm{q}$---the fixed-point
contraction condition that the Anderson Type-II mixing formula
requires (Sec.~\ref{sec:forward_solver}). On the backward pass,
the $\rho$-adaptive blend
$\tilde{\bm{H}}\!=\!(1\!-\!\tau)\bm{H}\!+\!\tau|\bm{H}|$
(Sec.~\ref{sec:trustregion}) ensures the implicit-function-theorem
inverse used in the adjoint linear system stays in a descent cone
for the loss gradient. The unifying insight is that
eigenvalue-aware regularisation is needed \emph{only where
indefiniteness is actually exercised}: per-element, per-iterate,
and only on the offending block. Applying the same treatment
globally---PSD-projecting the assembled $\bm{A}$, the most common
defensive move in differentiable PD---perturbs the entire
fixed-point map and either desynchronises Anderson on the forward
pass or biases the gradient on the backward pass. The locality of
the trust-region decision is what lets the same machinery serve
both passes without these side-effects.

\subsection{Aggregate Metrics Are Insufficient for Gradient Correctness}
\label{sec:disc-diagnostics}

A class of numerical bugs we surfaced repeatedly during
development was invisible to every aggregate metric---per-frame
PD residual, contact-set convergence, visual mesh fidelity---and
only became observable through per-iteration invariants designed
around \emph{a specific component's assumption}. The batched
Delassus assembly $\bm{W}\!=\!\bm{J}\bm{A}^{-1}\bm{J}^{T}$ on the
GPU degenerated to a near-diagonal matrix under certain
SpMM-algorithm selections, leaving forward simulation visually
intact but silently destroying the off-diagonal contact coupling
the backward gradient depends on; the bug was only diagnosable
through the per-frame count of active contacts whose Delassus
diagonal magnitude exceeded a fixed threshold, paired with a
deliberately slow CPU fallback used as an A/B oracle. The lesson
generalises: a differentiable solver can produce a plausible loss
trajectory whose \emph{gradient is wrong by a constant
scale}---the worst-case failure for inverse design, because the
optimisation will still descend monotonically into the wrong
basin. Future differentiable-physics work needs counter-based
\emph{component-local} invariants---active-contact count,
prox-map ping-pong amplitude, per-element Hessian eigenvalue
spread---as first-class correctness signals during development,
not just end-to-end loss curves at evaluation time.

\section{LIMITATIONS AND FUTURE WORK}
\label{sec:limitations}

\subsection{Velocity-Level NCP Drifts in the Quasi-Static Limit}
\label{sec:lim-rest0}

The Fischer--Burmeister formulation we adopt (Sec.~\ref{sec:contact})
enforces complementarity
$\bm{\lambda}_n\!\perp\!\bm{\delta}_n$ on the velocity-level
contact reaction---the natural choice for dynamic impact---but
does not explicitly conserve position-level non-penetration
across the entire rollout. At restitution $e\!\to\!0$, the
post-correction is triggered only while
$\bm{v}_{\cdot n}\!<\!\bm{0}\,\wedge\,\bm{\lambda}\!>\!\bm{0}$;
once a body settles into static support, neither condition fires
robustly, and configuration drifts $\sim$1--2\,cm below the
nominal contact plane over $50$ frames at our standard
$h\!=\!4$\,ms timestep. Our bouncing-ball benchmarks use
$e\!=\!0.3$, which is physically appropriate for soft rubber and
keeps the solver inside the regime where velocity-level
enforcement is sufficient. Extending DiffPhD to manipulation
scenarios that include sustained static contact---grasping under
gravity, multi-finger pinch closure, persistent surface
support---requires augmenting the NCP with a position-level
projection coupled to the PD global step, or replacing it with an
impulse-integrated ground reaction that accumulates the contact
force budget. The persistent $\bm{S}^{T}\bm{S}$ factor of
Sec.~\ref{sec:gpu} is orthogonal to this choice and would carry
over directly, so we view the gap as engineering rather than
fundamental.

\subsection{Material-Parameter Backward Is Energy-Specific}
\label{sec:lim-matjac}

The chain rule from projective weights $\{w_e\}$ to physical
material parameters $(E_e,\nu_e)$ is currently derived analytically
per energy: a corotated+volume composite uses
$w_0\!=\!2\mu$, $w_1\!=\!\lambda$ (Eq.~\eqref{eq:lame} chain rule),
while the single-energy Neo-Hookean form uses
$w_e\!=\!2\mu+\lambda$. The geometry- and initial-state inverse
problems in our evaluation (Bunny, Routing Tendon, Torus) do not
exercise this Jacobian---they optimise quantities that flow
through $\partial\bm{b}/\partial\bm{a}$, not through
$\partial\bm{A}/\partial\bm{E}$---so the material backward path
is exercised only by the system-identification benchmarks
(Bouncing Ball, Plant). Extending DiffPhD to material inverse
design across the full suite of supported energies---anisotropic
fibre, viscoelastic, rate-dependent plastic, or any future
hyperelastic energy that fits the PD prox-map formulation---
requires re-deriving $\partial w_e/\partial(E_e,\nu_e)$ for each.
The natural future direction is automatic symbolic
differentiation of the prox-map energy itself, so that adding a
new energy to the framework automatically extends the material
inverse design path; this is a refactor we sketch but do not
implement.

\section{CONCLUSION}
\label{sec:conclusion}

We presented DiffPhD, a unified differentiable solver that brings
heterogeneous, hyperelastic, contact-rich elastodynamics within
reach of end-to-end gradient-based optimisation. Our three
contributions compose around a single representational
invariant---a persistent $\bm{S}^{T}\bm{S}$ factor of the PD
operator that absorbs material contrast at assembly: stiffness-aware
projective assembly that routes per-element $\mu_e$ into $\bm{A}$
(Sec.~\ref{sec:heterogeneity}), a state-adaptive trust-region
filter on the per-element prox-map Hessian
(Sec.~\ref{sec:trustregion}), and a unified GPU local--global
pipeline that shares the same factor across forward, contact, and
backward stages (Sec.~\ref{sec:gpu}). Each contribution addresses
an independent failure mode of prior differentiable PD: the
forward stiffness--backward gradient coupling that destabilises
DiffPD beyond $\sim 50\times$ contrast, the indefinite prox-map
Hessian that derails L-BFGS line search under hyperelastic
contact, and the per-evaluation refactorisation cost that limits
mesh size and contact density.

Across seventeen benchmarks, DiffPhD remains stable up to
$100\times$ stiffness contrast on the Armadillo where DiffPD
diverges, delivers $8.69\times$ forward speed-up on a 2.2M-DoF
heterogeneous Crab and $33.31\times$ on dense codimensional
contact, and reaches a $23.9\times$ smaller residual loss than
the strongest baseline on slender-geometry system identification.
On the inverse-problem suite, DiffPhD converges to lower-loss
optima than DiffPD and MAS in every scene where the baselines
also converge, and to a finite optimum where they diverge
(Routing Tendon, $457\times$). The Real2Sim manipulation pipeline
completes in $13.5$ minutes wall-clock, taking the framework from
forward simulator into the regime where differentiable physics
becomes a practical building block for robotic manipulation under
heterogeneous frictional contact.

We view DiffPhD as evidence for a broader claim: in
differentiable elastodynamics, the design target should be
\emph{operator amortisation across forward and backward stages},
not arithmetic throughput on either pass alone. The
contributions we present are specific to projective dynamics, but
the principle of routing material structure into the operator at
assembly, filtering indefiniteness adaptively at the per-element
block, and sharing a single factor across all stages of a
differentiable pipeline is general---and we believe it applies
equally to Newton-based differentiable solvers, to position-based
dynamics, and to differentiable contact formulations beyond the
NCP we adopt here.

\bibliographystyle{ACM-Reference-Format}
\bibliography{sample-base}

\end{document}